\documentclass[twocolumn]{aastex631}

   \newcommand{\beam}{$\theta_{\mbox{\scriptsize maj}}\times\theta_{\mbox{\scriptsize min}}$}
   
   \newcommand{\ujyperbeam}{$\mu$Jy\,beam$^{-1}$}

   \newcommand{\amax}{ $a_{\mbox{\scriptsize max}}$ }
   
  \newcommand{\fuori}{FU\,Ori }
  \newcommand{\fuoris}{FU\,Ori\,S }

\shorttitle{Millimeter sized dust survives high temperature}
\shortauthors{Liu et al.}
\graphicspath{{./}{figures/}}

\begin{document}

\title{Millimeter-sized Dust Grains Appear Surviving the Water-sublimating Temperature in the Inner 10 au of the FU Ori Disk}

\correspondingauthor{Hauyu Baobab Liu}
\email{hyliu@asiaa.sinica.edu.tw}

\author[0000-0003-2300-2626]{Hauyu Baobab Liu}
\affil{Institute of Astronomy and Astrophysics, Academia Sinica, 11F of Astronomy-Mathematics Building, AS/NTU No.1, Sec. 4,
Roosevelt Rd, Taipei 10617, Taiwan, ROC} 

\author[0000-0002-3211-4219]{An-Li Tsai}
\affiliation{300 Zhongda Road, Institute of Astronomy, National Central University, Zhongli 32001 Taoyuan, Taiwan} 

\author[0000-0003-0262-272X]{Wen Ping Chen}
\affiliation{300 Zhongda Road, Institute of Astronomy, National Central University, Zhongli 32001 Taoyuan, Taiwan}

\author[0000-0002-7420-6744]{Jin Zhong Liu}
\affiliation{Xinjiang Astronomical Observatory, Chinese Academy of Sciences, People's Republic of China}

\author[0000-0002-5750-8177]{Xuan Zhang}
\affiliation{Xinjiang Astronomical Observatory, Chinese Academy of Sciences, People's Republic of China}

\author[0000-0002-8324-0506]{Shuo Ma}
\affiliation{Xinjiang Astronomical Observatory, Chinese Academy of Sciences,  People's Republic of China}
\affiliation{School of Astronomy and Space Science, University of Chinese Academy of Sciences, Beijing 100049, People's Republic of China}

\author[0000-0002-9433-900X]{Vardan Elbakyan}
\affiliation{Department of Physics and Astronomy, University of Leicester, Leicester LE1 7RH, UK}
\affiliation{Research Institute of Physics, Southern Federal University, Rostov-on-Don 344090, Russia}

\author[0000-0003-1665-5709]{Joel D. Green}
\affiliation{Space Telescope Science Institute, Baltimore, MD 21218, USA ; Department of Astronomy, The University of Texas at Austin, Austin, TX 78712, USA}

\author[0000-0001-5073-2849]{Antonio S. Hales}
\affiliation{Joint ALMA Observatory, Avenida Alonso de Córdova 3107, Vitacura 7630355, Santiago, Chile ; National Radio Astronomy Observatory, 520 Edgemont Road, Charlottesville, VA 22903-2475, USA}


\author[0000-0001-9248-7546]{Michihiro Takami}
\affiliation{Institute of Astronomy and Astrophysics, Academia Sinica, 11F of Astronomy-Mathematics Building, AS/NTU No.1, Sec. 4,
Roosevelt Rd, Taipei 10617, Taiwan, ROC} 

\author[0000-0003-2953-755X]{Sebasti\'an P\'erez}
\affiliation{Departamento de F\'isica, Universidad de Santiago de Chile. Avenida Ecuador 3493, Estaci\'on Central, Santiago, Chile}
\affiliation{Center for Interdisciplinary Research in Astrophysics and Space Exploration (CIRAS), Universidad de Santiago de Chile, Chile}

\author[0000-0002-6045-0359]{Eduard I. Vorobyov}
\affiliation{University of Vienna, Department of Astrophysics, Vienna, 1180, Austria }
\affiliation{Research Institute of Physics, Southern Federal University, Rostov-on-Don 344090, Russia}

\author[0000-0001-8227-2816]{Yao-Lun Yang}
\affiliation{Department of Astronomy, University of Virginia, Charlottesville, VA 22904-4235, USA}







\begin{abstract}
Previous observations have shown that the $\lesssim$10 au, $\gtrsim$400 K hot inner disk of the archetypal accretion outburst young stellar object, FU\,Ori, is dominated by viscous heating.
To constrain dust properties in this region, we have performed radio observations toward this disk using the Karl G. Jansky Very Large Array (JVLA) in 2020 June-July, September, and November.
We also performed complementary optical photometric monitoring observations.
We found that the dust thermal emission from the hot inner disk mid-plane of FU\,Ori has been approximately stationary and the maximum dust grain size is $\gtrsim$1.6 mm in this region.
If the hot inner disk of FU\,Ori which is inward of the 150--170 K water snowline is turbulent (e.g., corresponding to a Sunyaev \& Shakura viscous $\alpha_{t}\gtrsim$0.1), 
or if the actual maximum grain size is still larger than the lower limit we presently constrain, 
then as suggested by the recent analytical calculations and the laboratory measurements, water-ice free dust grains may be stickier than water-ice coated dust grains in protoplanetary disks.
Additionally, we find that the free-free emission and the Johnson {\it B} and {\it V} bands magnitudes of these binary stars are brightening in 2016--2020.
The optical and radio variability might be related to the dynamically evolving protostellar or disk accretion activities. 
Our results highlight that hot inner disks of outbursting objects are important laboratories for testing models of dust grain growth.
Given the active nature of such systems, to robustly diagnose the maximum dust grain sizes, it is important to carry out coordinated multi-wavelength radio observations.
\end{abstract}



\section{Introduction} \label{sec:intro}
In the theoretical studies about interstellar dust grain growth (e.g., \citealt{Ossenkopf1993A&A,Ormel2009A&A,Wada2009ApJ,Okuzumi2012ApJ,Banzatti2015ApJ...815L..15B,Pinilla2017ApJ...845...68P,Vorobyov2018A&A,Vorobyov2020A&A...644A..74V,Molyarova2021ApJ...910..153M}) and the interpretation of the observations of (sub)millimeter dust spectral indices (e.g., \citealt{Zhang2015ApJ}), it has been conventional to assume or to conjecture that water-ice coated dust grains are stickier than water-ice free grains.
It has been generally believed that grown dust with maximum grain sizes ($a_{\mbox{\scriptsize max}}$) greater than 1 mm (e.g., chondrules, pebbles) tend to form outside of the water snowline (T$\sim$150--170 K; e.g., \citealt{Pollack1994ApJ}).
It has also been considered that in the inner, higher temperature regions of protoplanetary disks, grown dust will likely fragment back down to smaller sizes when water-ice is sublimated (e.g., \citealt{Banzatti2015ApJ...815L..15B,Cieza2016Natur,Pinilla2017ApJ...845...68P}).
These assumptions used to be supported by the results of the earlier laboratory experiments (e.g., \citealt{Gundlach2011Icar,Gundlach2015ApJ...798...34G} and references therein).
However, they are inconsistent with analytical calculations (e.g., \citealt{Kimura2015ApJ}).
An open issue related to this is how to form rocky planets or asteroids that are deficient in water.

\begin{deluxetable*}{l l l l l c l l}
\tablecaption{New JVLA observations\label{tab:jvla}}
\tablewidth{700pt}
\tabletypesize{\scriptsize}
\tablehead{
\colhead{Date} &
\colhead{Band} &
\colhead{API rms\tablenotemark{a}} &  
\colhead{{\it uv}-range} &
\colhead{IF1/IF2 Freq. range} &
\colhead{Flux\tablenotemark{b}/Passband/Gain calibrators} &
\colhead{Synthesized beam\tablenotemark{c}} &
\colhead{rms noise\tablenotemark{c}}  \\
\colhead{(UTC)} &
\colhead{} &
\colhead{($^{\circ}$)} & 
\colhead{(meters)} &
\colhead{(GHz/GHz)} &
\colhead{} &
\colhead{(\beam; $^{\circ}$)} &
\colhead{(\ujyperbeam)} 
} 
\startdata
\multicolumn{8}{c}{Project Code: JVLA/20A-106} \\
2020-Jun-26 &
X&
4.5-4.6&
290-9470&
7.98-11.96/$\cdots$& 
3C147/J0532+0732/J0532+0732&
0\farcs86$\times$0\farcs64; -50$^{\circ}$&
6.0\\
2020-Jun-28 &
Ku&
4.7-17&
240-9600&
12.01-15.98/15.98-18.16& 
3C147/J0532+0732/J0532+0732&
0\farcs89$\times$0\farcs44; -32$^{\circ}$&
4.6\\
2020-Jul-03 &  
K&
5.1-9.1&
610-9630&
18.10-22.02/21.97-25.90& 
3C147/J0532+0732/J0532+0732&
0\farcs33$\times$0\farcs29; -72$^{\circ}$&
12\\
2020-Jul-04 &  
Ka&
9.8-12&
190-10380&
29.10-33.02/32.98-37.02& 
3C147/J0532+0732/J0532+0732&
0\farcs32$\times$0\farcs20; -52$^{\circ}$&
22\\
2020-Jul-07 &  
Q&
4.6&
180-10460&
40.10-44.02/43.98-48.02& 
3C147/J0532+0732/J0532+0732&
0\farcs22$\times$0\farcs15; -60$^{\circ}$&
85\\
2020-Sep-30 &  
Ka&
1.3&
200-10550&
29.10-33.02/32.98-37.02& 
3C147/J0532+0732/J0532+0732&
0\farcs23$\times$0\farcs19; -23$^{\circ}$&
12\\
\hline
\multicolumn{8}{c}{Project Code: JVLA/20B-423} \\
2020-Nov-15 &
K &
2.6-4.7&
200-34420 &
17.98-22.02/21.97-26.02& 
3C147/3C84/J0532+0732&
0\farcs31$\times$0\farcs13; -80$^{\circ}$ &
6.5\\
 &
Ka&
2.6-4.7&
&
28.98-33.02/32.98-37.02& 
3C147/3C84/J0532+0732&
0\farcs19$\times$0\farcs086; -73$^{\circ}$ &
12\\
\enddata
 \tablenotetext{a}{The RMS phase measured with the Atmospheric Phase Interferometer (API; for more details see \url{https://science.nrao.edu/facilities/vla/docs/manuals/oss2013A/performance/gaincal/api}). The API a 2-element interferometer separated by 300 meters, observing an 11.7 GHz beacon from a geostationary satellite. The default API rms upper limits for the X, Ku, K, Ka, and Q band observations are 30$^{\circ}$, 15$^{\circ}$, 10$^{\circ}$, 7$^{\circ}$, and 5$^{\circ}$.}
 \tablenotetext{b}{Limited to the inner 400 $k\lambda$ {\it uv}-distance to suppress the effects of the resolved structures of 3C147.}\vspace{-0.2cm}
 \tablenotetext{c}{Measured from multi-frequency synthesis images generated using aggregated continuum bandwidths and Natural (i.e., Briggs Robust$=$2) weighting.}
\end{deluxetable*}

Growing modern laboratory experimental results conversely suggested that water-ice free dust grains are stickier or at least as sticky as the water-ice coated ones (\citealt{Gundlach2018MNRAS,Steinpilz2019ApJ,Musiolik2019ApJ,Pillich2021A&A...652A.106P}).
\citet{Gundlach2018MNRAS} pointed out that the inconsistency between the modern and the earlier experimental results (e.g., \citealt{Gundlach2011Icar}) may be because the ice-coated dust samples in the earlier experiments were thermally processed due to the imperfect low-temperature operational techniques.
If this is indeed the case, then grown dust may be prone to form inward rather than outward of the water snow line in  protoplanetary disks (for some related discussion see \citealt{Pinilla2021A&A...645A..70P,Molyarova2021ApJ...910..153M,Musiolik2021arXiv210703188M} and references therein).
Since the laboratory dust samples do not necessarily have the same composition and morphology (e.g., porosity) as  interstellar dust, it remains essential to constrain the spatial distribution of $a_{\mbox{\scriptsize max}}$ in  protoplanetary disks by observations to fully address the issue of dust grain growth and terrestrial planet-formation.

Intriguingly, many recent Atacama Large Millimeter/Submillimeter Array (ALMA) multi-frequency or polarization observations towards the spatially extended, lower-temperature (e.g., $\lesssim$100 K) regions of  protoplanetary disks reported $a_{\mbox{\scriptsize max}}\lesssim$100 $\mu$m
(\citealt{Kataoka2016ApJ,Stephens2017ApJ,Bacciotti2018ApJ,Hull2018ApJ,Ohashi2018ApJ,Liu2019,Ohashi2019ApJ,Tazaki2019ApJ,Lin2020MNRAS,Ueda2020ApJ,Ohashi2020ApJ,Mori2021ApJ...908..153M}).
As pointed out by \cite{Sierra2020ApJ}, in the regions covered by these previous ALMA observations, millimeter-sized dust grains are more likely to be absent rather than hidden due to vertical grain-size segregation.
These recent observational results are in contrast to the earlier observational studies which claimed the detection of $>$1 mm sized dust grains (for a review see \citealt{Testi2014prpl.conf..339T}).
This discrepancy was because the earlier observational studies did not consider the effects of dust scattering opacity self-consistently (\citealt{Kataoka2015ApJ,Liu2019,Zhu2019ApJ}).
The latest observational measurements of $a_{\mbox{\scriptsize max}}$ based on self-consistent consideration of dust scattering opacity might indicate that water-ice coated dust grains are not as sticky as previously considered, supporting the latest laboratory experimental results. 

Constraining $a_{\mbox{\scriptsize max}}$ in regions inside the water snowlines is the natural next step  to test whether or not the water-ice free dust grains are indeed sticky enough to grow or to survive in such regions.
However, these regions are very small in general (e.g., $\lesssim$1--2 au scales; see the discussion in \citealt{Mori2021snowline} and references therein).
Limited by the sensitivity and angular resolution of the present date facilities, it is rather difficult to make such measurements.

The accretion outburst type young stellar objects (YSOs), namely the FUors (see \citealt{Audard2014} for a review), present elevated dust temperature thanks to their likely enhanced accretion rate (by several orders of magnitude relative to the quiescent Class II YSOs).
For example, the archetypal FUor, FU\,Ori ($d\sim$407.5 pc; \citealt{gaia2016,gaia2020}), has maintained a 10$^{-5}$--10$^{-4}$ $M_\odot$ yr$^{-1}$ accretion rate (e.g., \citealt{Perez2020ApJ} and references therein) over the last $\sim$80 yrs.
In their protoplanetary disks, water snowlines may expand to become considerably larger than those of typical protoplanetary disks (e.g., see the analysis of dust temperature in \citealt{Liu2017A&A,Liu2019ApJ...884...97L}).
This can also be supported by the detection of abundant gas-phase water line emission (e.g., \citealt{Fuente2020}).
When observed with sufficient spatial resolution and signal-to-noise ratio, these objects open a unique window to examine grain growth inward of  water snowlines  (see also the discussion in the proceeding by \citealt{Okuzumi2021}).

Concurrent with the aforementioned laboratory frameworks (\citealt{Gundlach2018MNRAS,Steinpilz2019ApJ,Musiolik2019ApJ,Pillich2021A&A...652A.106P}), previous spectral energy distribution (SED) analyses of FU\,Ori reported tentative detection of  $\sim$2--3 mm sized dust grains in its 300--400 K hot inner disk, which is characterized by a flux density {\it bump} at the $\sim$30 GHz observing frequency (\citealt{Liu2019ApJ...884...97L}; for some introductory background on the SED features, see Appendix \ref{appendix:dustspid}).
The SED analysis by \citet{Liu2019ApJ...884...97L} also found that \amax in the lower temperature (e.g., $<150$ K) regions of the \fuori disk is $\lesssim$200 $\mu$m, which may indicate that the  millimeter-sized dust formed in-situ in the hot inner disk.

The interpretation of \citet{Liu2019ApJ...884...97L} was uncertain because their observations (1) only sparsely sampled the 8--50 GHz frequency range (c.f. \citealt{Liu2017A&A}), (2) were impacted by the atmospheric delay bug that  might not have been fully corrected\footnote{Details of this issue has been given in  \url{https://science.nrao.edu/facilities/vla/data-processing/vla-atmospheric-delay-problem}. }, and (3) could potentially be confused by the time variability of the radio or (sub)millimeter emission (e.g., \citealt{Liu2018A&A,Johnstone2018ApJ,Francis2019ApJ,Wedeborn2020}).

To confirm the $\sim$30 GHz bump in the SED of FU\,Ori, and to tighten the constraints on the dust column density and $a_{\mbox{\scriptsize max}}$, we have observed the radio continuum using the NRAO Karl G. Jansky Very Large Array (JVLA) at X band (8-12 GHz), Ku band (12-18 GHz), K band (18-26 GHz), Ka band (29-37 GHz), and Q band (40-48 GHz).
These observations also simultaneously covered its close companion, FU\,Ori\,S.

\begin{figure*}
   \hspace{-1.3cm}
   \begin{tabular}{ p{4.2cm} p{4.2cm} p{4.2cm} p{4.2cm} }
        \includegraphics[height=5cm]{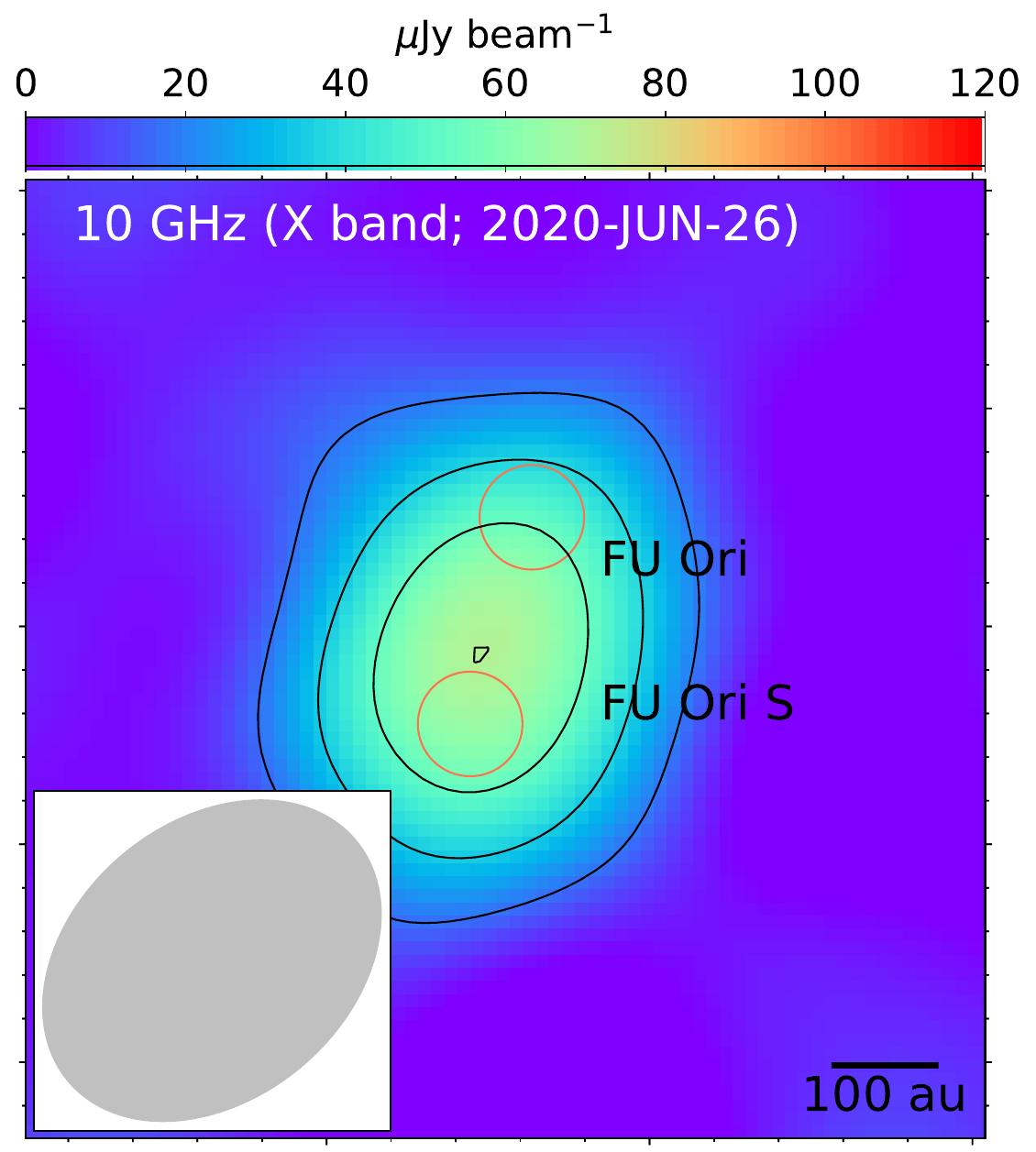}  &
        \includegraphics[height=5cm]{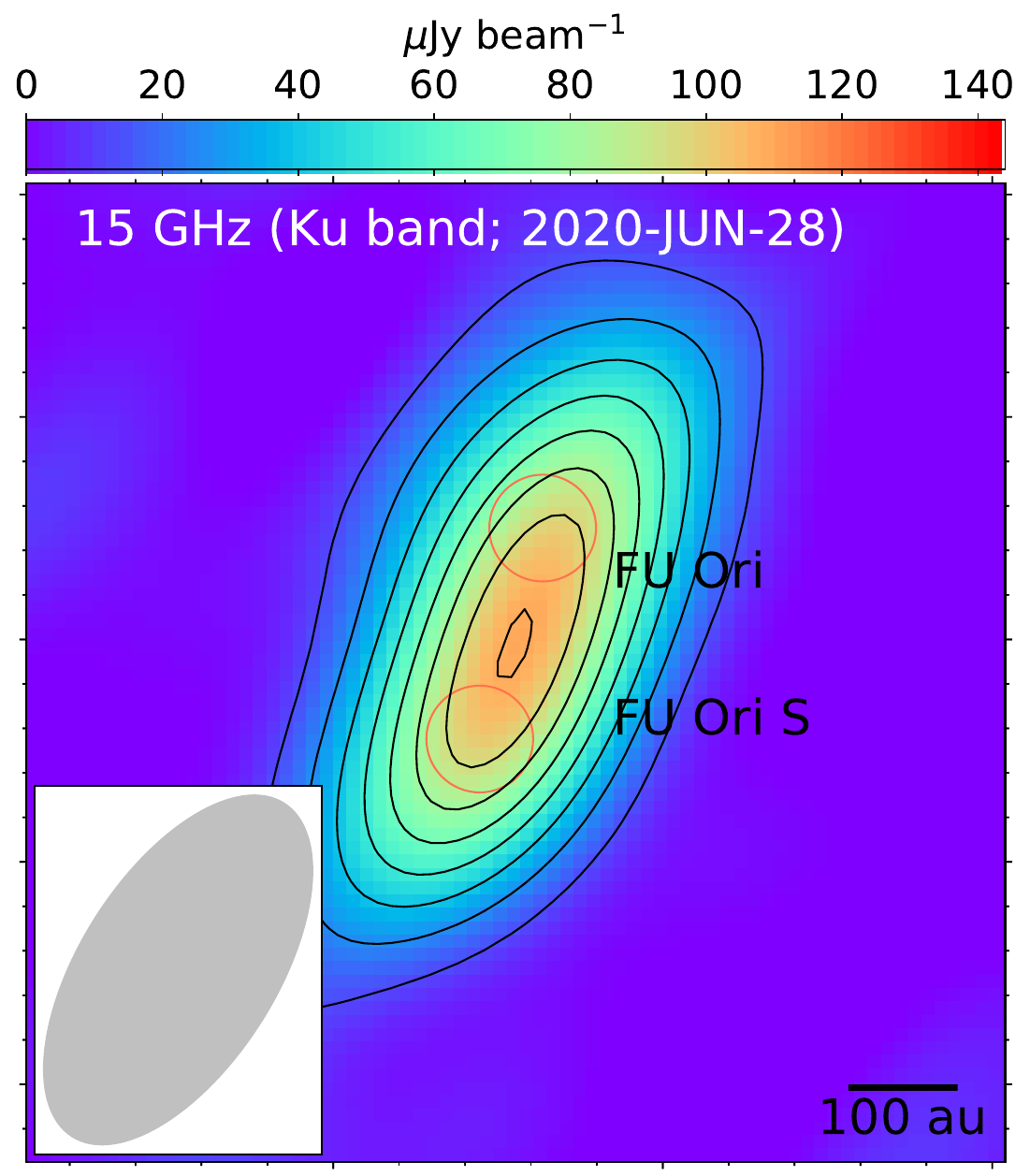} &
        \includegraphics[height=5cm]{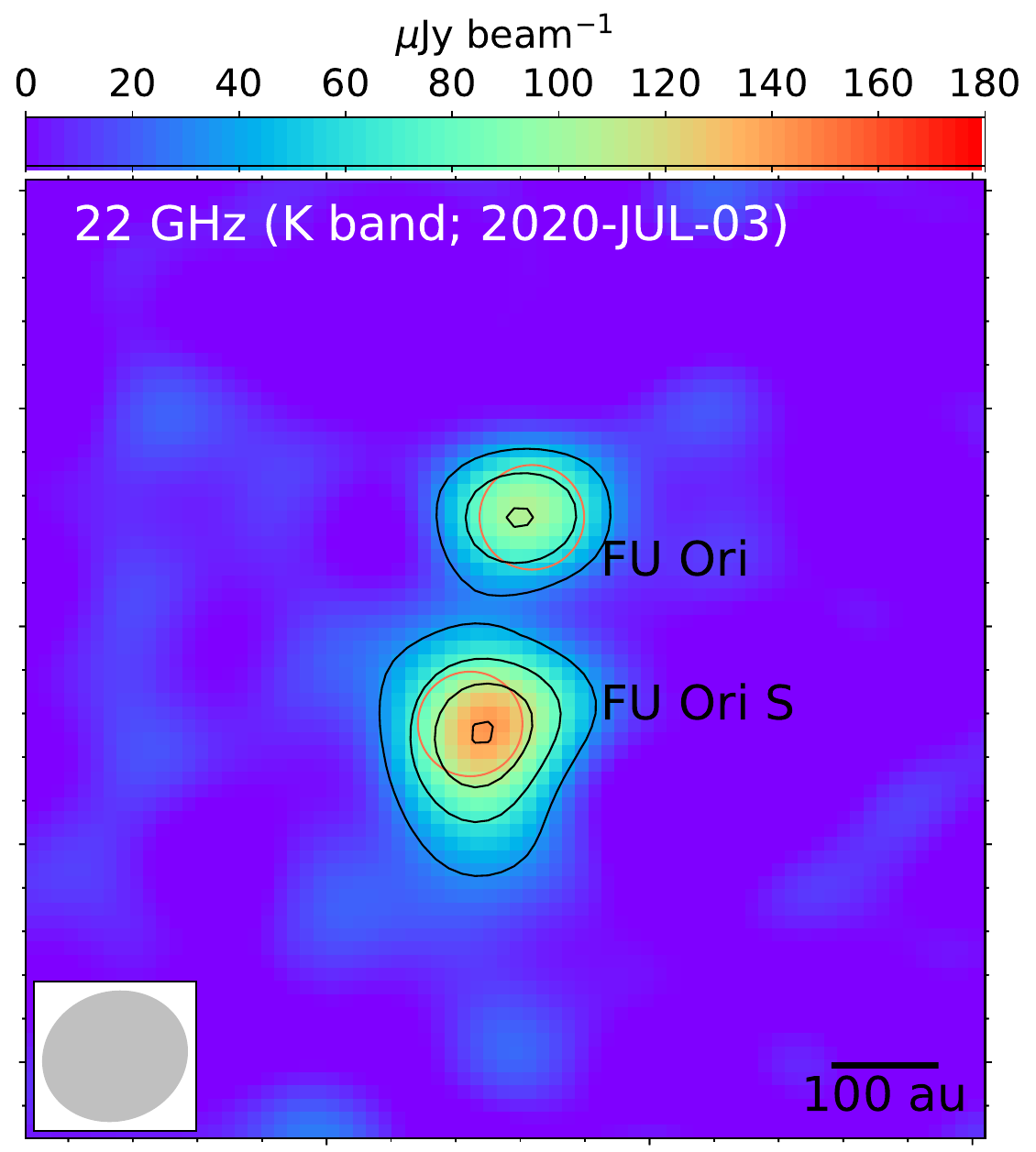}  &
        \includegraphics[height=5cm]{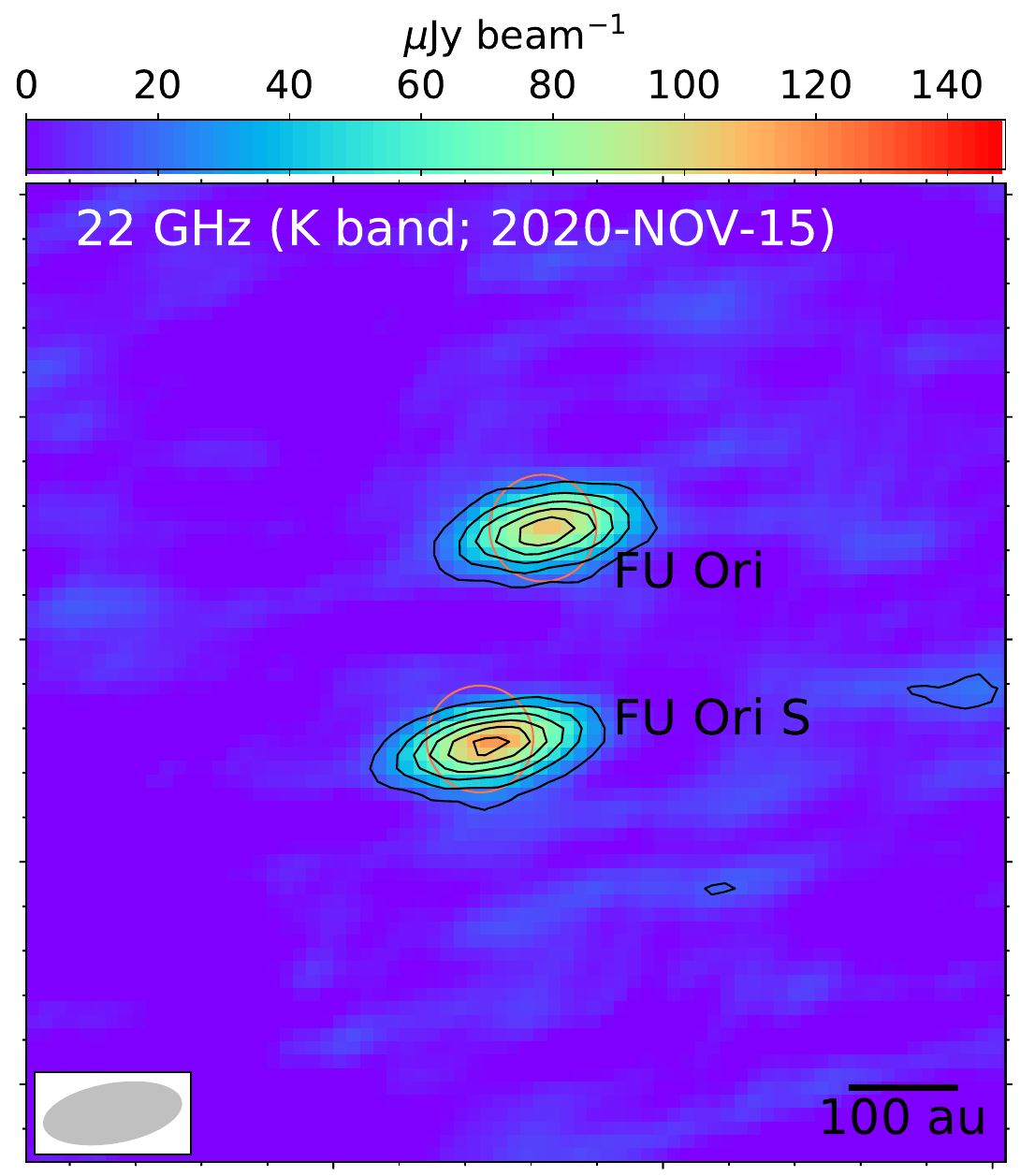}  \\
        \includegraphics[height=5cm]{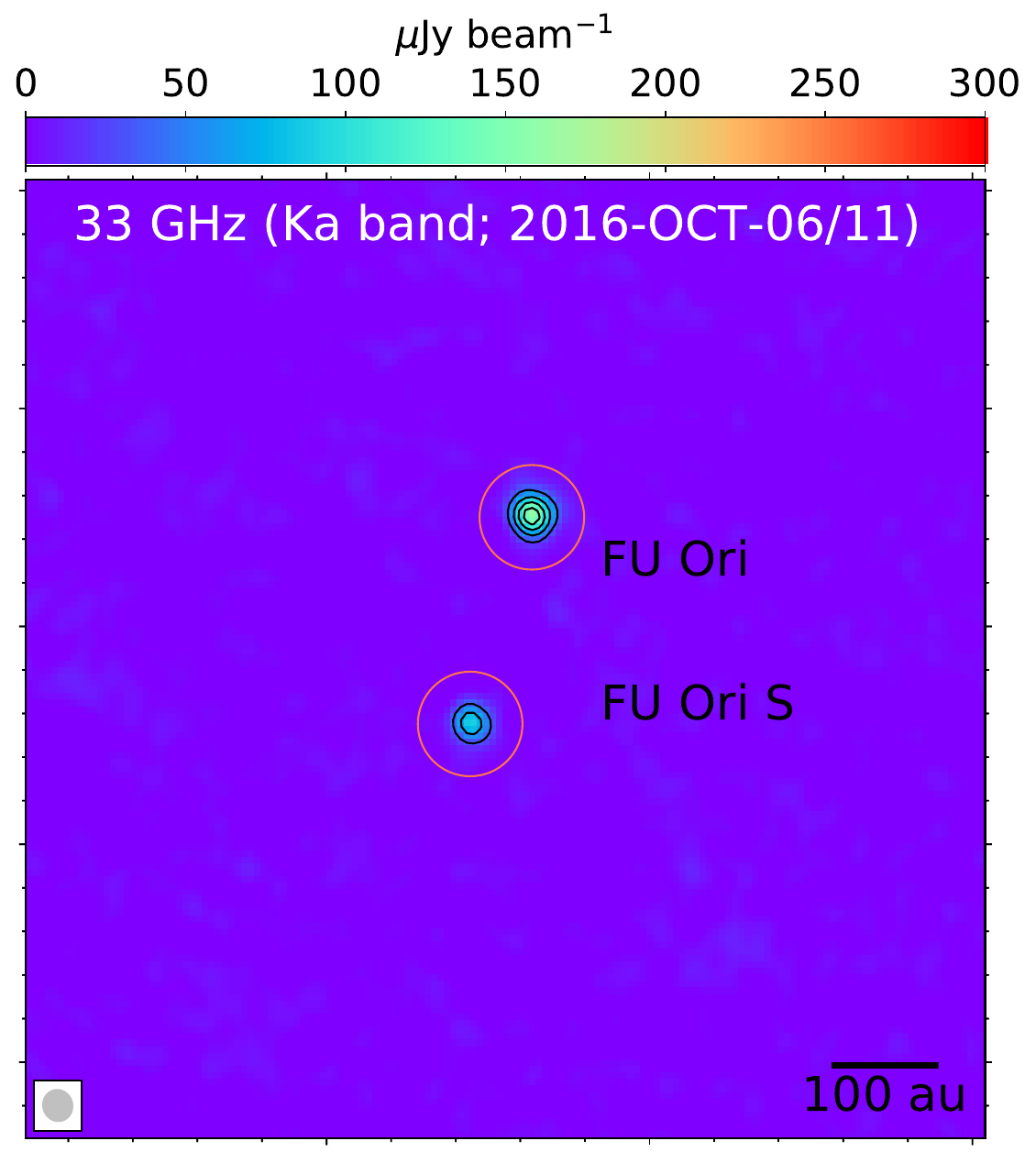} &
        \includegraphics[height=5cm]{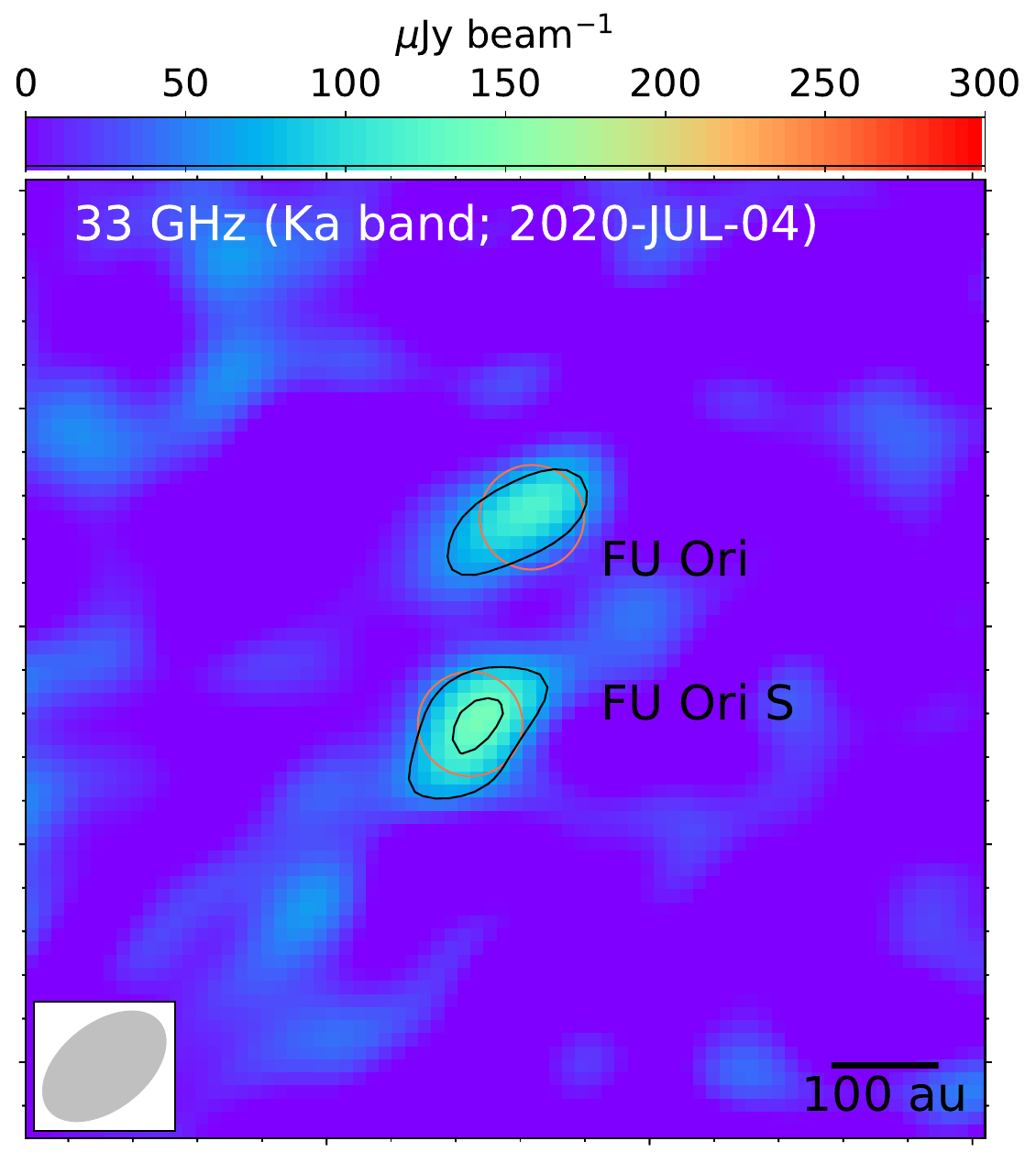} &
        \includegraphics[height=5cm]{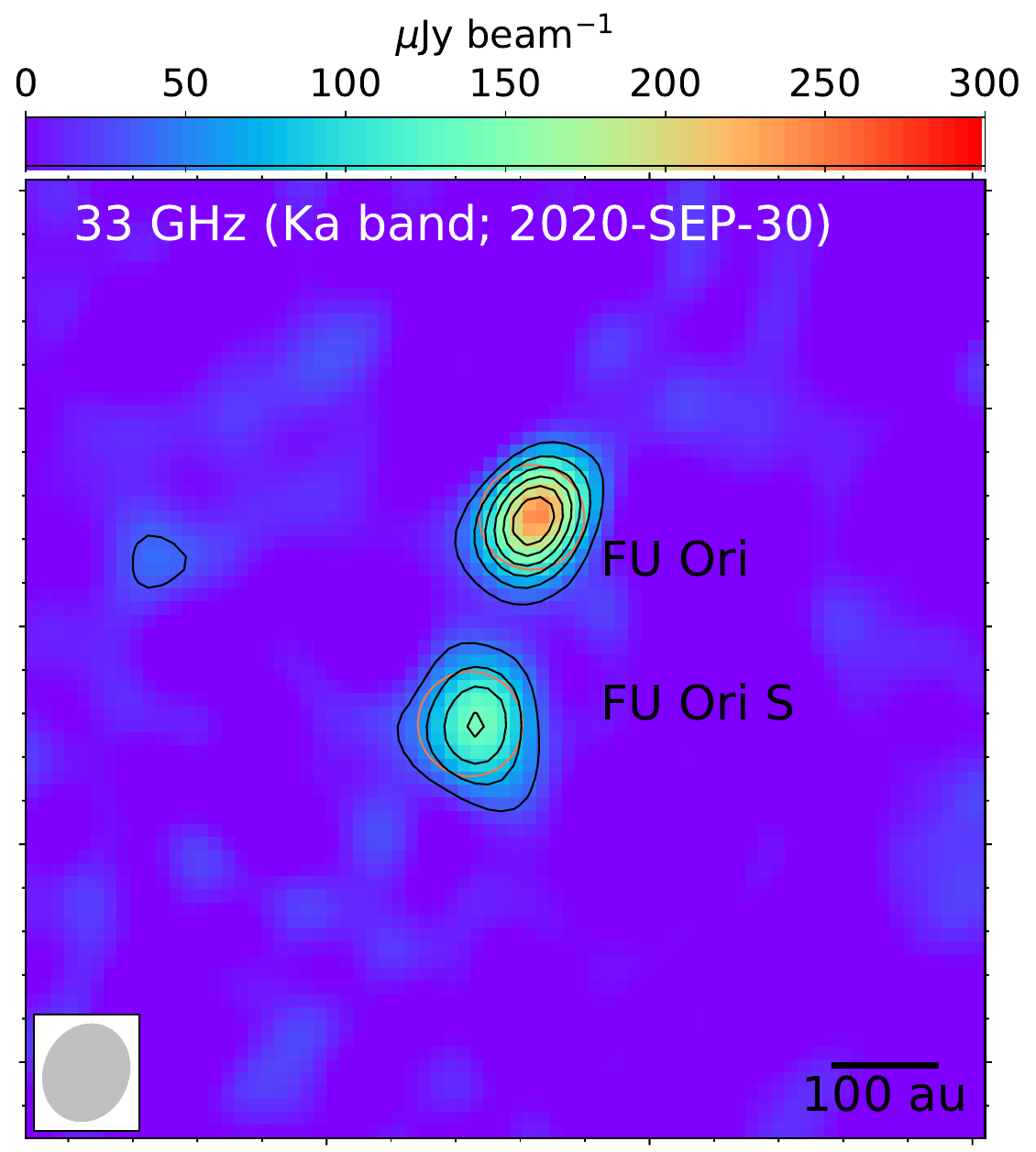} &
        \includegraphics[height=5cm]{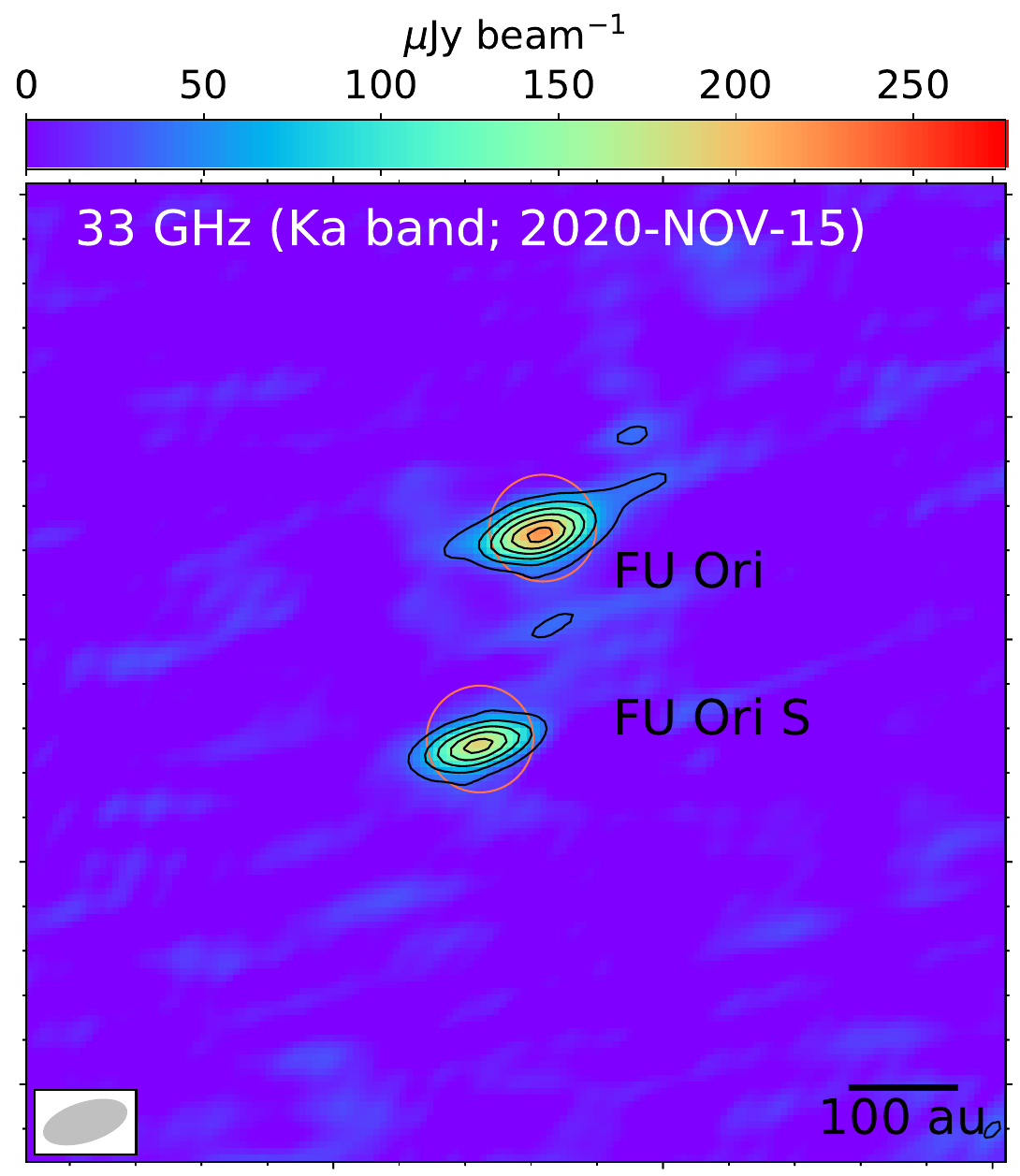} \\
        \includegraphics[height=5cm]{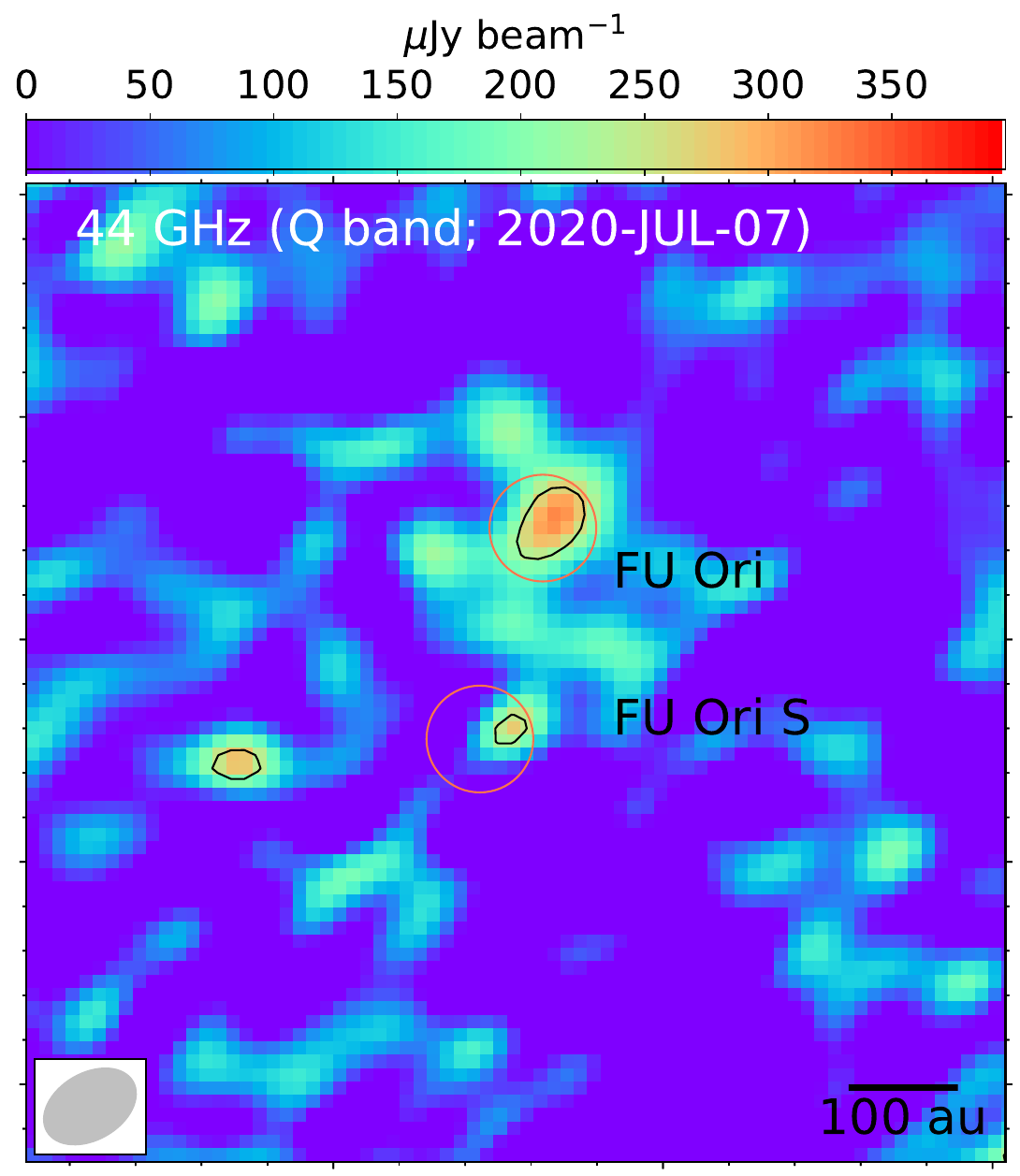}  &&& \\
   \end{tabular}
   \vspace{-0.3cm}
    \caption{JVLA continuum images towards FU Ori and its companion, FU Ori S (color and contours). The frequency bands and the date of the observations are labeled in individual panels. The red circles indicate the locations of FU Ori and FU Ori S, respectively. Contours in all panels start from 3$\sigma$ and have 3$\sigma$ intervals (c.f. Table \ref{tab:jvla} for the $\sigma$ values). }
    \label{fig:maps}
\end{figure*}

\begin{figure*}[h]
    \hspace{-1.5cm}
    \vspace{-0cm}
    \begin{tabular}{ p{7.5cm} p{8cm} p{3cm}  }
        \includegraphics[height=5.15cm]{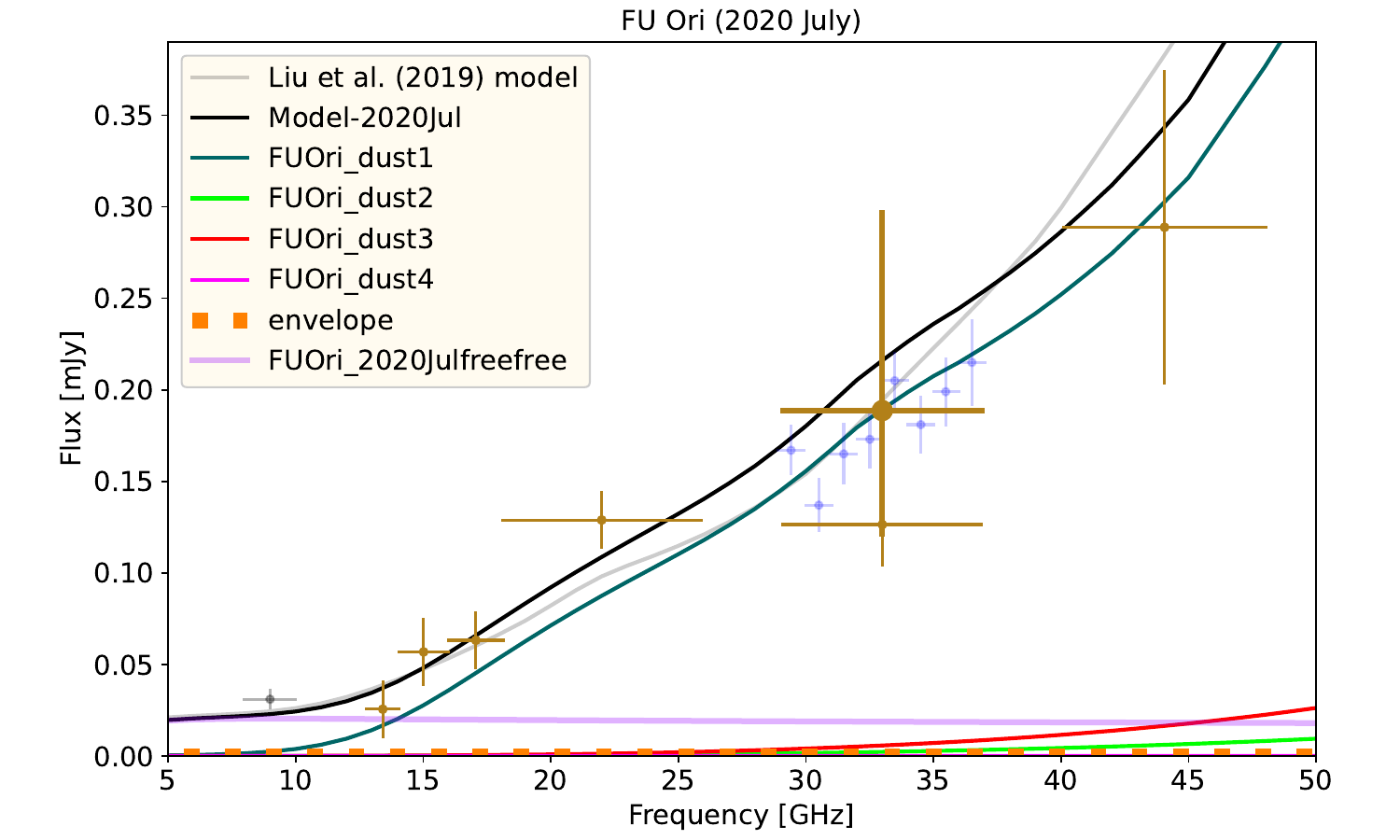} &  
        \includegraphics[height=5.15cm]{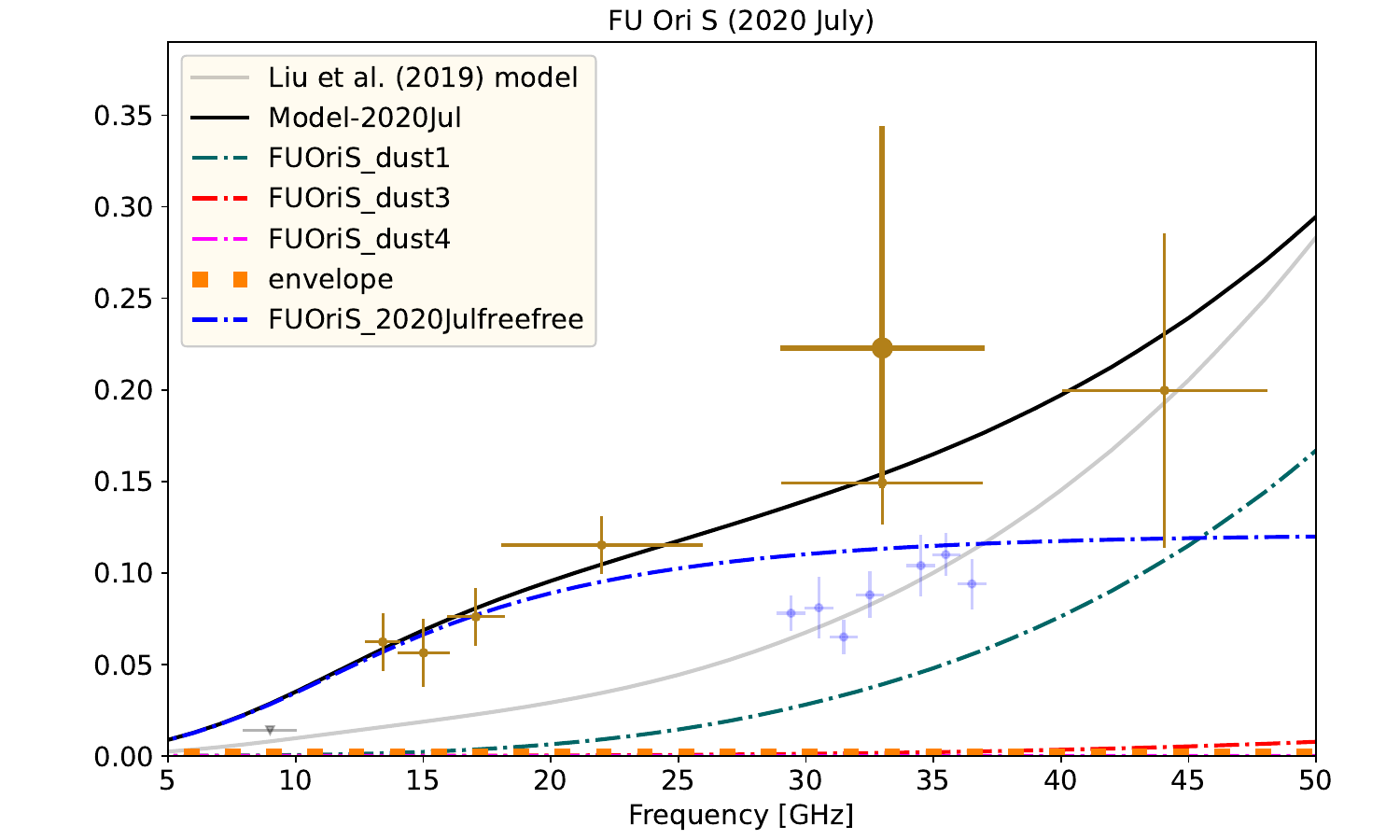} &
        \includegraphics[height=5.15cm]{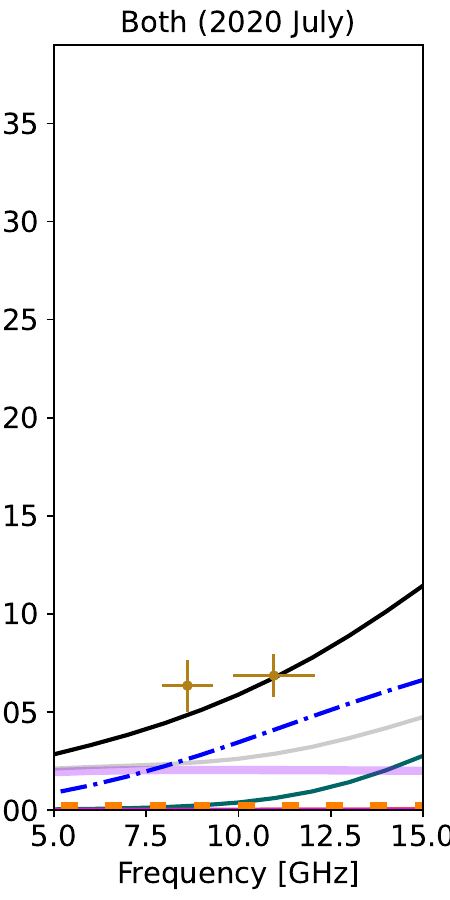} \\
    \end{tabular}
    
    \hspace{-1.5cm}
    \vspace{-0.4cm}
    \begin{tabular}{ p{7.5cm} p{8cm} p{3cm}  }
        \includegraphics[height=5.15cm]{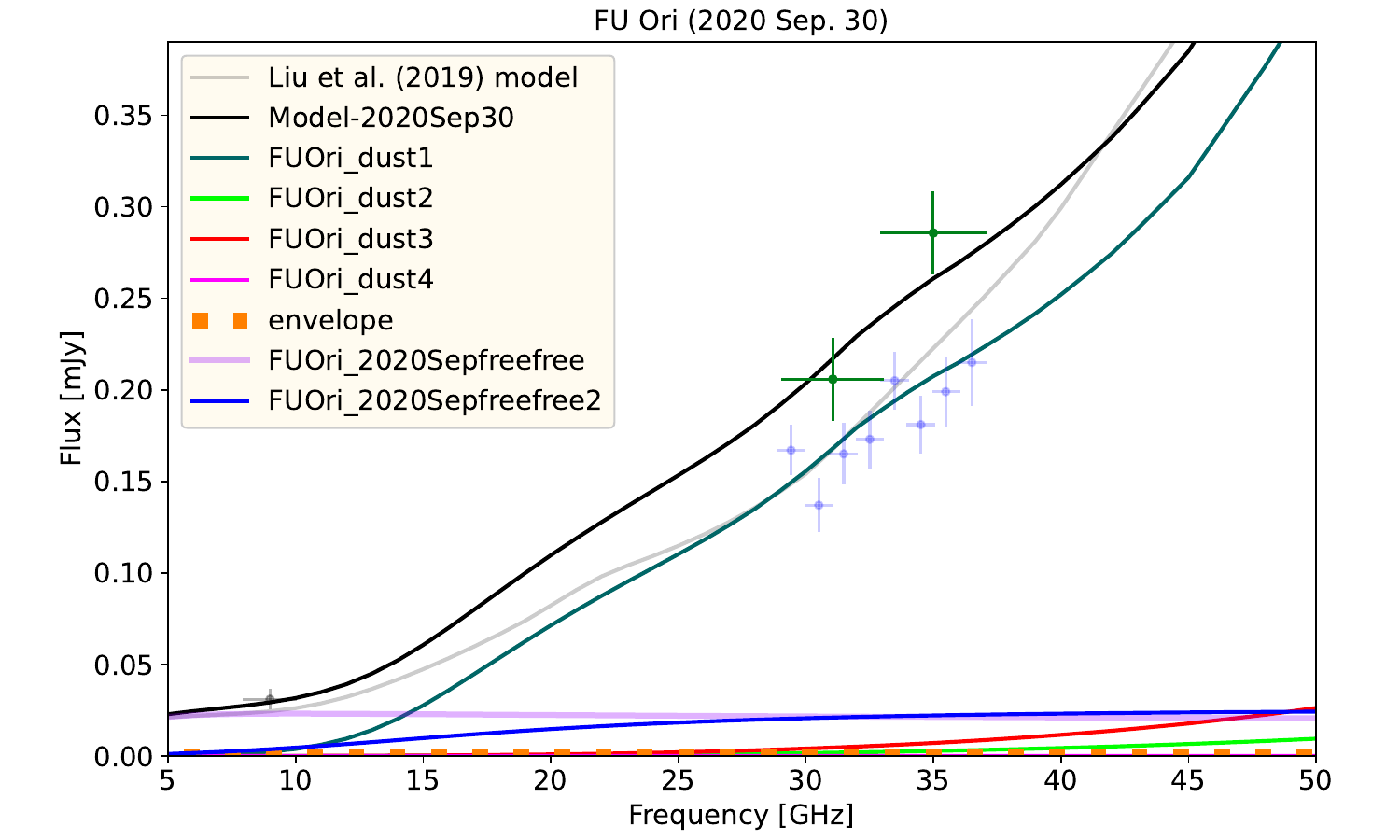} &  
        \includegraphics[height=5.15cm]{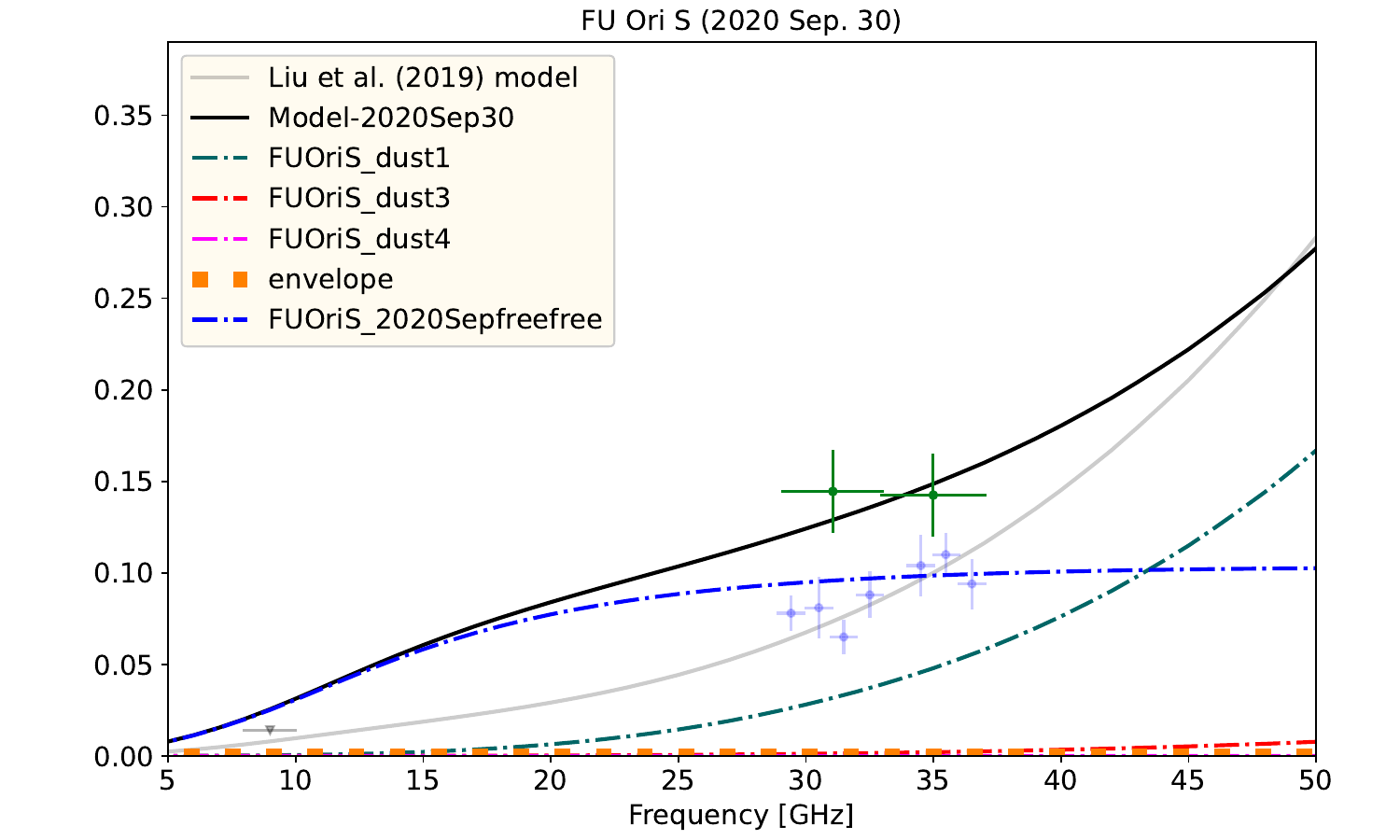} &
         \\
    \end{tabular}
    
    \hspace{-1.5cm}
    \vspace{-0cm}
    \begin{tabular}{ p{7.5cm} p{8cm} p{3cm}  }
        \includegraphics[height=5.15cm]{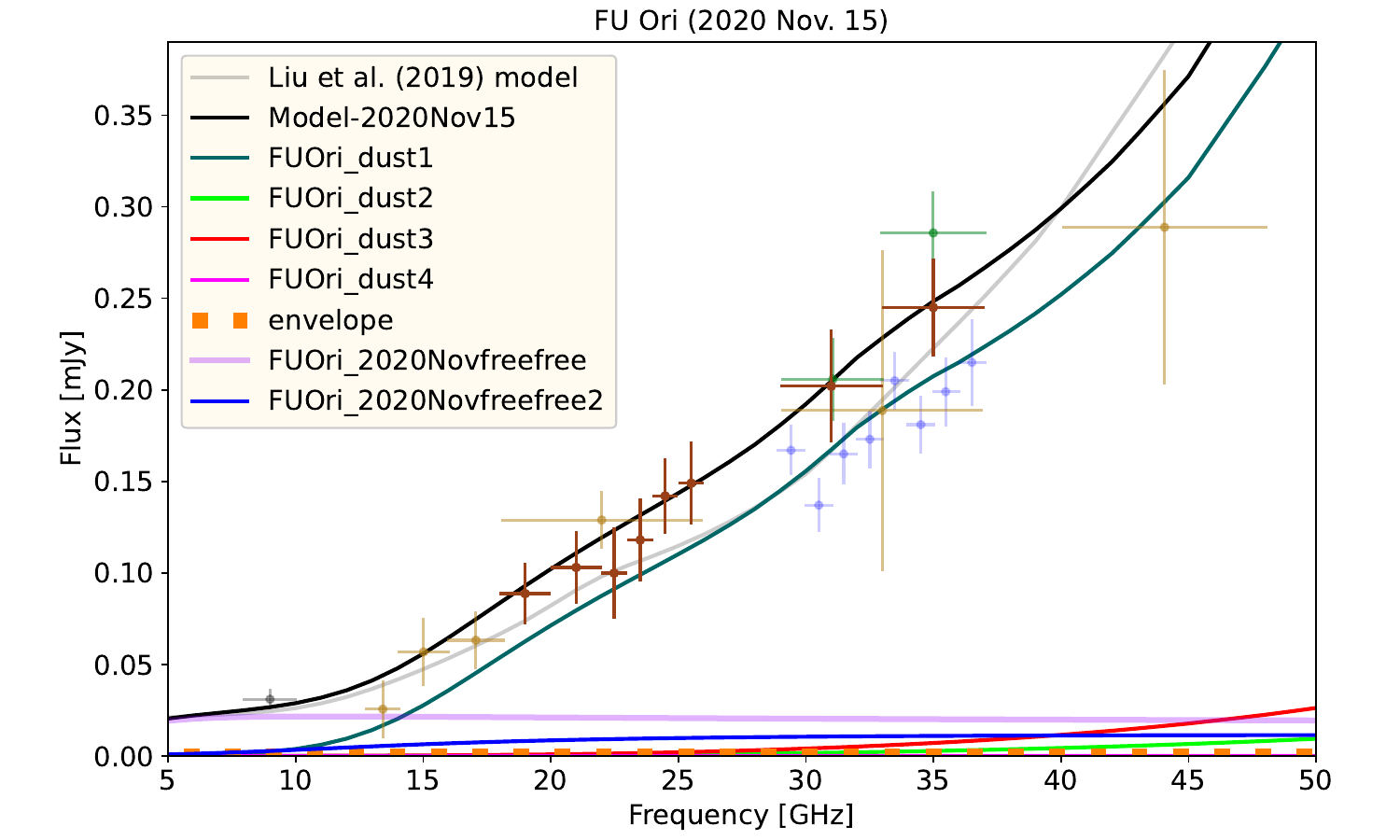} &  
        \includegraphics[height=5.15cm]{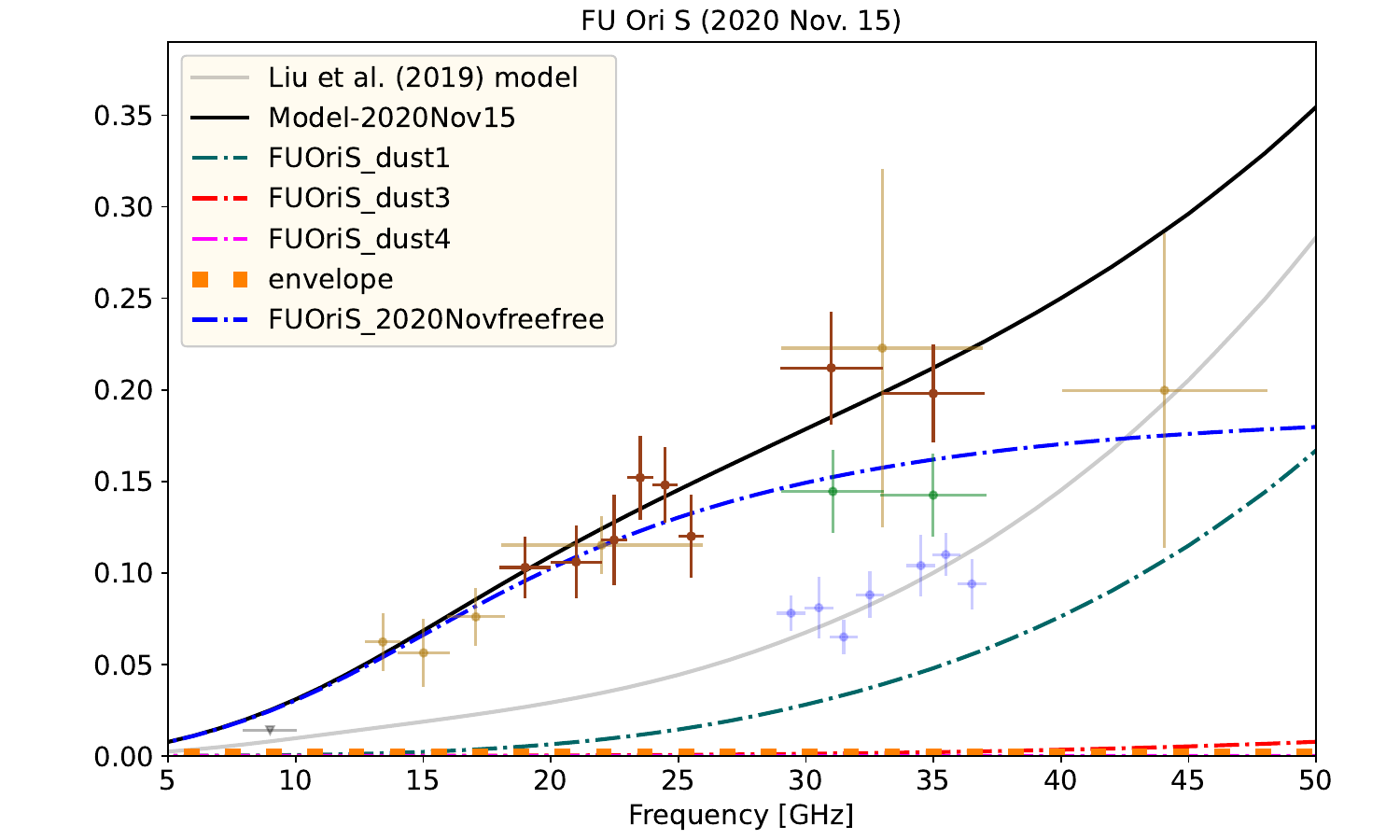} &
         \\
    \end{tabular}
    
    \vspace{-0.3cm}
    \caption{The resolved flux densities of \fuori ({\it left column}), \fuoris ({\it middle column}) or the combined flux densities in the case that the two sources cannot be separated ({\it right column}). Light brown, green and dark brown symbols show the measurement made in 2020 July, September 30, and November 15, respectively. 
    In the bottom row the  measurement made in 2020 July (after correcting the effects of phase errors, see Section \ref{subsub:fluxdensity}) and September 30 were also presented in lighter colors.
    In the top row we additionally use the thicker symbols to present the corrected Ka band (29--37 GHz) flux densities of the 2020 July observations, taking the large phase rms into consideration (Section \ref{sub:jvlaobs}).
    Then X band (8--10 GHz) measurement (3-$\sigma$ upper limit in the case of FU\,Ori\,S) taken on 2016 September 30 and the Ka band measurement taken on 2016 October 06 and 11 are presented in light gray and blue symbols, respectively (quoted from \citealt{Liu2017A&A,Liu2019ApJ...884...97L}).
    The vertical error bars for the Ku band data are $+/-$1.5$\sigma$ and take into consideration the larger uncertainty in flux density measurements (see Section \ref{subsub:fluxdensity}); the other vertical error bars are $+/-$1$\sigma$.
    The horizontal error bars present the frequency widths used for the imaging.
    Gray lines in individual panels present the SED models to describe the observations made before 2019, which were quoted from \citet{Liu2019ApJ...884...97L}.
    Black lines show the flux densities of the updated models combining all individual model components which are presented by lines in various colors (Section \ref{sub:sed}).
    These model components are labeled in the top row.
    We present the bottom panels in a log-log scale in Figure \ref{fig:fig2loglog}, which additionally include the ALMA Bands 3--7 data reported by \citet{Hales2015ApJ,Liu2019ApJ...884...97L,Perez2020ApJ}.
    }
    \label{fig:fuoriFlux}
\end{figure*}

During these observations, we serendipitously detected radio variability from both FU\,Ori and FU\,Ori\,S, which we immediately followed up with complementary optical observations using the Nanshan One-meter Wide-field Telescope (NOWT) and the Lulin Super Light Telescope (SLT).
This work also utilizes the ASAS-SN Variable Stars Database (\citealt{asassnOverall1,asassnOverall2,asassn1,asassn2}).
The details of our radio and optical observations are provided in Section \ref{sec:observation}.
The observational results are presented in Section \ref{sec:result}.
We provide a qualitative interpretation of our observational results in Section \ref{sub:interpretation}.
Section \ref{sub:sed} and \ref{sub:sedresult} introduces our strategy to quantitatively realize our interpretation with SED modeling, as well as the model results.
We discuss the physical implication of our results in Section \ref{sub:implication}.
In Section \ref{sub:future}, we briefly discuss how this experiment can be improved in the near future.
Our conclusion is given in Section \ref{sec:conclusion}.
Some introduction of the dust SED features related to the millimeter sized grains is provided in Appendix \ref{appendix:dustspid}.
Appendix \ref{appendix:loglog} summarizes our new JVLA observations together with the previous ALMA observations towards FU\,Ori and FU\,Ori\,S.
Based on the dust temperature profiles derived from the observations, we provide our hypothesis about the turbulence strength in the FU\,Ori hot inner disk in Appendix \ref{appendix:turbulent}.

\section{Observations}\label{sec:observation}

\subsection{JVLA}\label{sub:jvlaobs}
\subsubsection{Operation}\label{subsub:jvlaoperation}
We performed JVLA standard continuum mode observations toward FU Ori in the B and BnA array configurations in 2020 (project codes: 20A-106, 20B-423).
The pointing and phase referencing centers for our target source is  R.A. $=$ 05$^{\mbox{\scriptsize{h}}}$45$^{\mbox{\scriptsize{m}}}$22$^{\mbox{\scriptsize{s}}}$.357 (J2000), decl. $=$ +09$^{\circ}$04$'$12$\farcs$4 (J2000). 
We employed the 3 bit sampler to take the full RR, RL, LR, and LL correlator products.
The dates, projected baseline ranges, frequency coverages at the two intermediate frequency (IFs), the adopted flux, passband, and complex gain calibrators of our observations are summarized in Table \ref{tab:jvla}.

The original design of these observations was to observe at all selected frequency bands close in time.
To maximize the chance of being dynamically scheduled in suitable weather conditions, the high frequency bands (K, Ka, and Q) were observed for relatively short durations, tolerating the high thermal noise.
Because our target source is not ideal for the observations in the summer semester, the noise levels we actually achieved were still considerably higher than requested.
Although some individual measurements are rather uncertain, jointly using all data points still provides very indicative constraints on the emission mechanisms (Section \ref{sub:interpretation}, \ref{sub:sed}).
There is room to improve the observing strategy, which will be discussed in Section \ref{sub:future}.

The initial observations were made from June 26 to July 07 (hereafter the 2020 July observations).
The Atmospheric Phase Interferometer (API) rms for the Ka band observations during the 2020 July observations was considerably higher than the default limit.
Therefore, the Ka band observations were repeated on September 30.
The Ka band data taken in the 2020 July observations are usable although with some caution (more discussion in Section \ref{sec:result}).

When comparing these new Ka band observations with the Ka band data published in \citet{Liu2017A&A}, we found that both \fuori and \fuoris could be undergoing radio emission flares (more in Section \ref{sec:result}).
Therefore, we requested to follow up this event at K and Ka bands with Director's Discretionary Time (DDT), which was executed on November 15, 2020.
These DDT observations also attempted to constrain polarization properties.
In addition, we launched the complementary optical monitoring observations (Section \ref{sub:opticalobs}).

\subsubsection{Calibration}\label{subsub:jvlacal}
We manually followed the standard data calibration strategy using the Common Astronomy Software Applications (CASA; \citealt{McMullin2007}) package, release 5.6.2.
The brightest source in our observations was our absolute flux calibrator, 3C147.
The gain calibrator J0532+0732 has $\sim$1 Jy flux densities at all observing wavelengths and was not spatially resolved.
Given that 3C147 was spatially resolved in our observations at K, Ka, and Q bands, we instead took J0532+0732 as our delay and passband calibrators for all observations and used only the inner 400 $k\lambda$ data of 3C147 when referencing the absolute fluxes.
We also utilized the built-in image models for 3C147 during the calibrations.
After implementing the antenna position corrections, weather information, gain-elevation curve, and opacity model, we bootstrapped delay fitting and passband calibrations, and then performed complex gain calibration. 
We applied the absolute flux reference to our complex gain solutions, and then applied all derived solution tables to the target source. 

For the K and Ka band observations taken in November of 2020, we based our observations on 3C147 to solve the cross-hand delay and absolute polarization position angles using only the inner 400 $k\lambda$ data, and took 3C84 as a low polarization percentage calibrator when solving the leakage term (i.e., the D-term).

\begin{figure*}[ht]
    \hspace{-1.3cm}
    \begin{tabular}{ p{9cm} p{9cm} }
         \includegraphics[width=9.5cm]{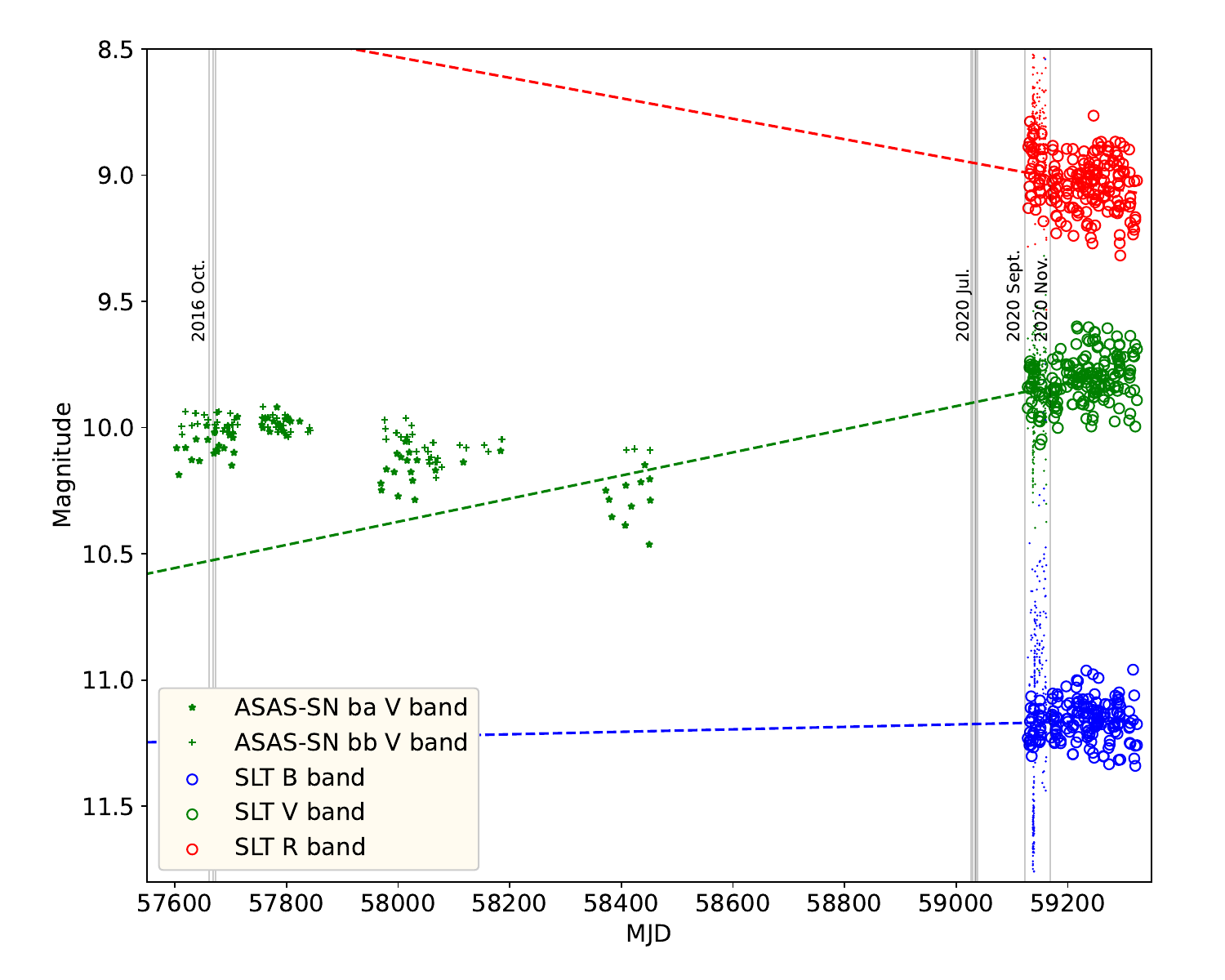} &
         \includegraphics[width=9.5cm]{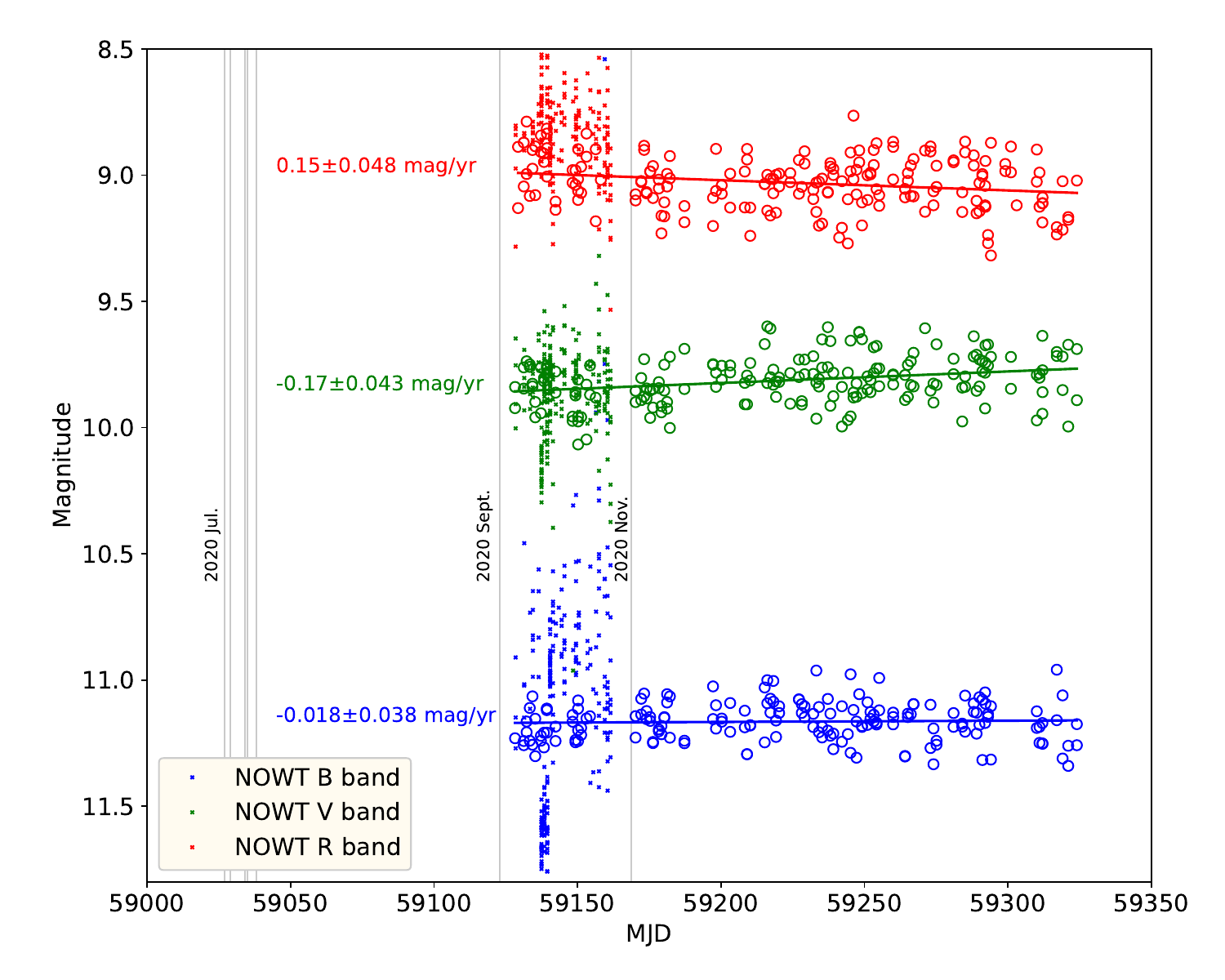} \\
    \end{tabular}
    \caption{Optical photometric monitoring observations toward FU\,Ori and FU\,Ori\,S.
    The observational data are shown by symbols. 
    Left panel shows the observations from 2016 to 2021, while the right panel shows the observations from mid 2020 to mid 2021.
    The linear regressions for the SLT data are shown by the solid lines in the right panel. In the left panel, we extrapolated these linear regression lines to 2016 which are shown by dashed lines.
    }
    \vspace{0.3cm}
    \label{fig:oirmonitor}
\end{figure*}

\subsubsection{Imaging}\label{subsub:jvlaimaging}

We performed the zeroth order (i.e., nterm$=$1) multi-frequency synthesis imaging \citep{Cornwell2008MultiscaleClean,Rau2011MultiscaleClean} using the CASA task {\tt tclean}.
From the previous ALMA and JVLA observations (e.g., \citealt{Hales2015ApJ,Liu2017A&A,Liu2019ApJ...884...97L,Perez2020ApJ}), the deconvolved sizes (diameters) of individual of FU\,Ori and FU\,Ori\,S are $\lesssim$0\farcs05 at the target frequencies.
Therefore, FU\,Ori and FU\,Ori\,S are approximately point sources in our new JVLA observations given that there was not much sampling at  $>$3000 $k\lambda$ baselines (Table \ref{tab:jvla}; c.f. the right panel of Figure 1 of \citealt{Perez2020ApJ}).
Therefore, we utilized all visibilities to maximize  sensitivity.

These two point sources are separable as long as there is adequate sensitivity at baselines longer than  $\sim$150 $k\lambda$.
For our observations (Table \ref{tab:jvla}), they can be clearly separated in the observations at K (18--26 GHz), Ka (29--37 GHz), and Q (40--44 GHz) bands; at Ku band (12--18 GHz) they can still be separated although we have to image with Briggs Robust$=$0 or uniform weighting; they cannot be separated in the X (8--12 GHz) band observations.

To assess the data quality, we first produced the naturally weighted (i.e., Briggs Robust$=$2.0) broad-band continuum images for each epoch of observations and at each band using the overall aggregated bandwidths.
For each epoch of observations, we also tried producing images with narrower frequency widths for a better constraint on spectral indices.
We chose the frequency widths such that both FU\,Ori and FU\,Ori\,S can be detected at $\sim$5--10$\sigma$, as far as it is possible.
Yet we did not try a complicated strategy to optimize the frequency widths which may make it less straightforward for the future observations to compare with the present one.

We had to image the K, Ka, and Q band data taken during the 2020 July observations using the aggregated frequency widths due to the relatively poor weather condition; the Ku and X bands data were imaged with every $\sim$2 GHz frequency width\footnote{These frequency widths are approximated values. Sometimes a few consecutive spectral windows may be flagged due to very serious radio frequency interference (RFI). Nevertheless, a small error in the frequency width is negligible since the thermal noise is inversely scaled with the square root of frequency width. This has been taken in to consideration when evaluating the centroid frequencies of the images.}.
We separately imaged the two IFs of the Ka band data taken on September 30 and November 15 (i.e., each 4 GHz bandwidths).
At K band the flux densities of FU\,Ori and FU\,Ori,S both increase with frequency (i.e., the signal was stronger at higher frequency).
Therefore, for the K band data taken on November 15, the lower 4 GHz IF was imaged with each 2 GHz frequency width. while the upper 4 GHz IF was imaged with each 1 GHz frequency width.

\subsubsection{Measurements of flux densities}\label{subsub:fluxdensity}

The flux densities of FU\,Ori and FU\,Ori\,S were measured from the images produced with narrower frequency widths.
We tried to measure them using the three methods: (1) directly quoting the peak intensities which is appropriate for point sources as long as the atmospheric phase errors were sufficiently calibrated, (2) performing 2D Gaussian fits, and (3) summing flux densities from the {\tt clean} components.
Visibility fitting is advantageous when the image is limited by dynamic range. 
However, it has less immunity to confusion and artifacts in general.
Therefore, we did not adopt visibility fitting.

We found that for observations taken in good weather conditions, the measurements based on methods (1) and (2) are consistent within 1$\sigma$ with no identified systematic bias. 
For noisy observations, method (2) sometimes yielded ambiguous fitting results if we do not fix the values of source positions and angular scales; on the other hand, fixing source positions and angular scales will make method (2) essentially not different from method (1).
Method (3) in general systematically underestimates the flux density by $\sim$1$\sigma$, which is expected since the {\tt clean} algorithm cannot distinguish weak signal from thermal noise.

In the end, we adopted method (1) for the Q, Ka, K, and X (unresolved) bands observations.
We refer to Section 5.4 of \citet{Howell2006hca..book.....H} which discussed why the approach of using the information within a small aperture may yield better accuracy when the signal-to-noise ratio is not high.

The images of FU\,Ori and FU\,Ori\,S are confused in the Robust$=$0 weighted Ku band images, which can be approximated by neither a single point source nor two separated point sources.
Therefore, for the individual FU\,Ori and FU\,Ori\,S, we took the flux densities summed from the clean components and added 0.5$\sigma$ as their flux densities.
We confirmed that the summed flux density of FU\,Ori and FU\,Ori\,S is consistent with the flux density integrated above the 2$\sigma$ contour in the restored {\tt clean} image.
When analyzing the Ku band results, we adopted the 1.5$\sigma$ error bars rather than 1$\sigma$ to take into account the larger measurement uncertainties.

We consider that the strategy of measuring flux density we applied is optimal for our specific observations and science purpose.
We argue that there is not yet a general ideal strategy for all observations.

\begin{figure}
    \hspace{-1.3cm}
    \begin{tabular}{ p{9cm} }
        \includegraphics[width=9.5cm]{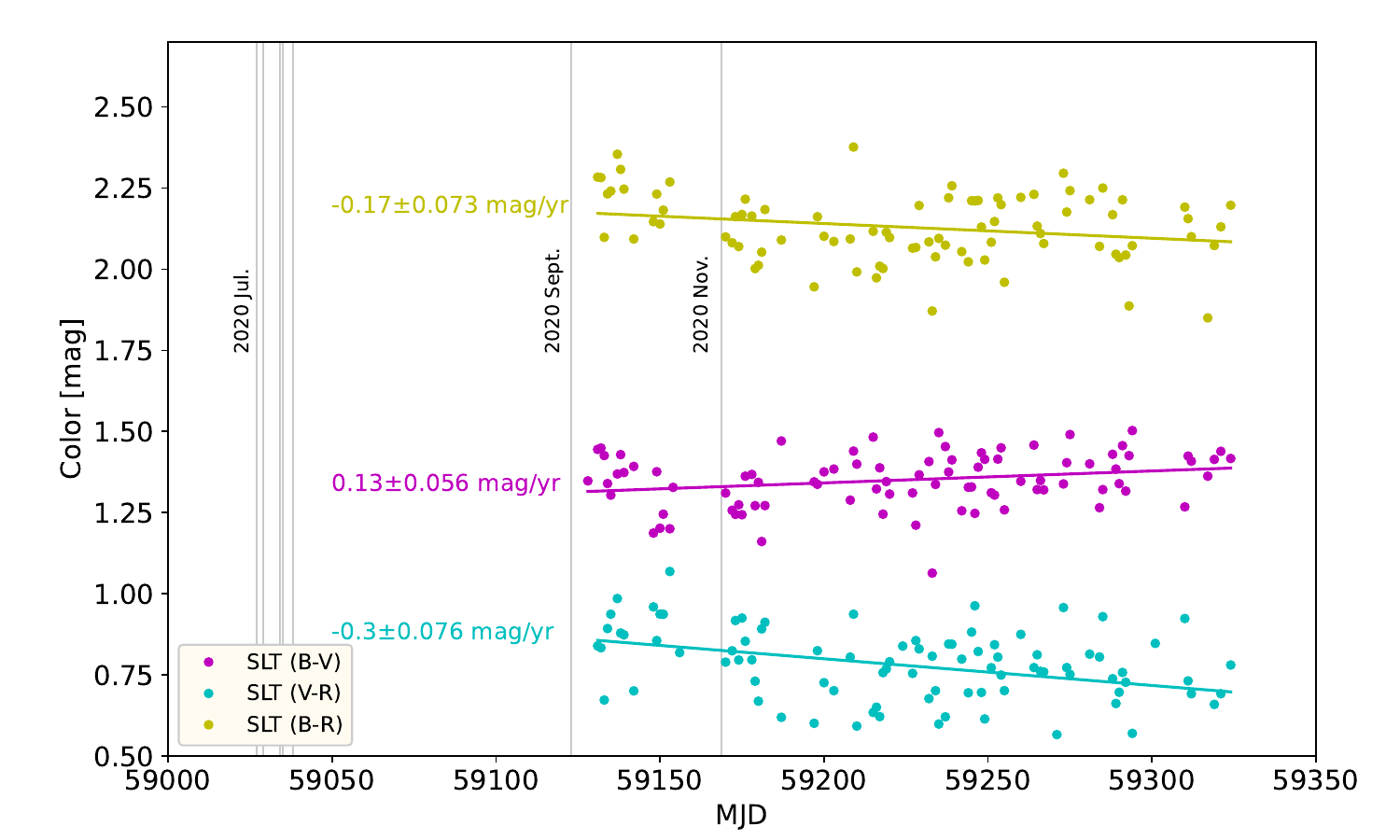} \\
        \includegraphics[width=9.5cm]{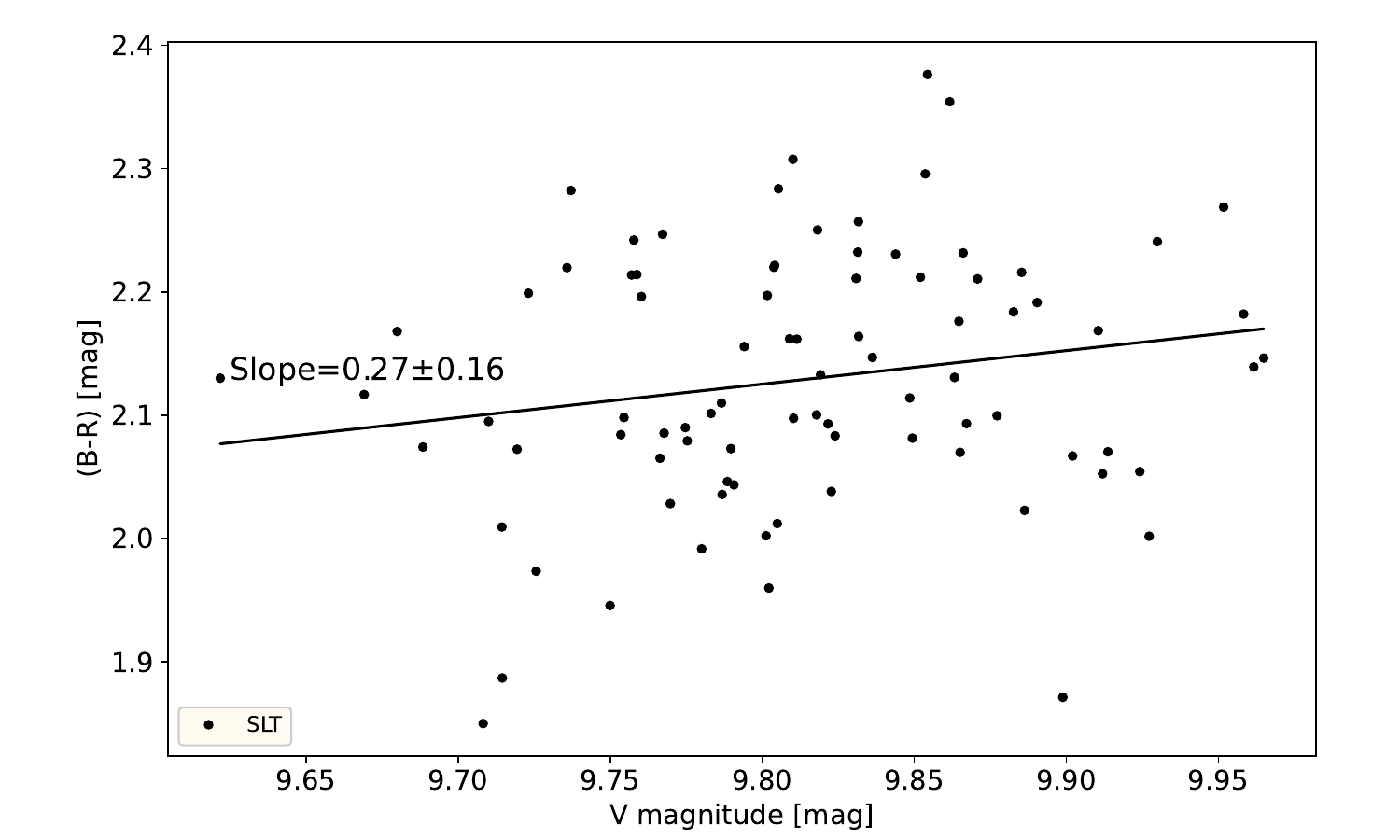} \\
    \end{tabular}
    \caption{Optical color of the FU\,Ori and FU\,Ori\,S binary system (spatially unresolved). {\it Top :--} The B-V, V-R, and B-R colors versus modified Julian date. {\it Bottom :--} The V-R color versus B band magnitude. Symbols in both panels show the observational data. The linear regressions for the data are shown in solid lines. The slopes of these regression lines are labeled. }
    \label{fig:color}
\end{figure}

The directly measured 29--37 GHz flux density on July 04 should be regarded as a lower limit of the true flux density owing to the high API (and phase) rms (see Table \ref{tab:jvla}; Section \ref{sub:jvlaobs}).
For a rough correction, we nominally assumed that the phase rms at the 11.7 GHz observing frequency and the $\ell=$300 m baseline length of the API is in the range of (API$^{\mbox{\scriptsize min}}_{\mbox{\scriptsize ka}}-$API$^{\mbox{\scriptsize lim}}_{\mbox{\scriptsize ka}}$)--(API$^{\mbox{\scriptsize max}}_{\mbox{\scriptsize ka}}-$API$^{\mbox{\scriptsize lim}}_{\mbox{\scriptsize ka}}$), where API$^{\mbox{\scriptsize min}}_{\mbox{\scriptsize ka}}=$9.8$^{\circ}$, API$^{\mbox{\scriptsize max}}_{\mbox{\scriptsize ka}}=$12.0$^{\circ}$, and API$^{\mbox{\scriptsize lim}}_{\mbox{\scriptsize ka}}=$7.0$^{\circ}$ are the observed minimum, maximum, and the default limiting API rms for the Ka band observations (Table \ref{tab:jvla}).
In addition, we assumed that the amplitude of phase fluctuation scales as $\nu^{1}$ \citep{Asaki1998RaSc} and $\ell^{0.5}$ (e.g., \citealt{Carilli1999RaSc}\footnote{The assumed power-law index is in between that of the theoretical 2D and 3D Kolmogorov turbulence.}), respectively, where $\nu$ and $\ell$ are observing frequency and baseline length.
Thereby, at the 33 GHz median observing frequency, the expected residual rms phase errors at the $\sim$5000 m median projected baseline length of our Ka band observations on July 04 (Table \ref{tab:jvla}) is $\sim$30$^{\circ}$--60$^{\circ}$.
Based on Equation 7.34 of \citet{Thompson}, we expect that the observed 33 GHz flux density had been reduced to 50\%--86\% of the true values due to the phase errors.
In the following analysis, we will adopt the corrected 29--37 GHz flux density for July 04 if not specifically mentioned.

\subsection{Optical monitoring observations}\label{sub:opticalobs}

We have performed time-series photometric observations at Johnson {\it BVR} bands using the Super Light 0.4 m Telescope (SLT) located in Taiwan\footnote{For technical details see \url{http://www.lulin.ncu.edu.tw/english/index.htm}.} and the Nanshan One-meter Wide-field Telescope (NOWT) located in Xinjiang \citep{NOWT2020RAA....20..211B}. 

The Johnson {\it R} band data were calibrated by referencing to the {\it VRI} photometry of $\lambda$ Ori star-forming region reported by  \citet{Dolan2002AJ....123..387D}.
The Johnson {\it B} and {\it V} bands data were calibrated by referencing to the AAVSO Photometric All Sky Survey (APASS) DR9 (\citealt{Henden2016JAVSO..44...84H}) which is consistent with the calibration of the ASAS-SN survey (\citealt{asassnOverall1,asassnOverall2}), except that when we evaluated the {\it V$-$R} color, both the Johnson {\it V} and {\it R} bands data were calibrated by referencing to  {\it VRI} photometry of the $\lambda$ Ori star-forming region.
The calibrations based on APASS and the {\it VRI} photometry of $\lambda$ Ori star-forming region yielded a  difference of $\lesssim$0.01-0.02 magnitudes, which is negligible for our science case.

\section{Results} \label{sec:result}
\subsection{Radio observations} \label{sub:radioresult}
Figure \ref{fig:maps} shows the naturally weighted new JVLA Stokes I continuum images produced using the aggregated bandwidths (Section \ref{sub:jvlaobs}) as well as the JVLA Ka band image presented in \citet{Liu2017A&A}.
The achieved synthesized beams and rms noises in the new JVLA images are summarized in Table \ref{tab:jvla}.
From the observations on 2020 November 15, we did not find $>$3$\sigma$ detection of Stokes Q, U, and V.
Figure \ref{fig:fuoriFlux} shows the radio flux densities measured from FU\,Ori, FU\,Ori\,S, and from both sources when they cannot be spatially separated by our JVLA observations.

In Figure \ref{fig:maps}, the locations of the 8--12 GHz and 12--18 GHz peaks in our new observations are closer to FU\,Ori\,S, meaning that FU\,Ori\,S was brighter than FU\,Ori at these two frequency bands. 
However, in the 8--10 GHz observations in 2016 September, FU\,Ori was marginally detected while FU\,Ori\,S was not (\citealt{Liu2019ApJ...884...97L}).
Moreover, the 8.4 GHz nondetection in the data taken in 1988 October constrained the 3$\sigma$ upper limit on the individual of FU\,Ori and FU\,Ori\,S to be $<$50 $\mu$Jy (\citealt{Rodriguez1990PASP}).
The 8--12 GHz flux density we detected on 2020 June 26 is higher than the upper limits reported by \citet{Rodriguez1990PASP} and is also higher than the flux density of FU\,Ori in 2016 September (Figure \ref{fig:fuoriFlux}). 
These observations indicate that FU\,Ori\,S has become brighter at low frequency.

The flux densities of these two sources are comparable in the two epochs of 18--26 GHz observations although the significant frequency variations at this band complicated the comparison (Figures \ref{fig:maps}, \ref{fig:fuoriFlux}).
FU\,Ori is brighter than FU\,Ori\,S at 40--48 GHz band and at higher frequencies (\citealt{Hales2015ApJ,Liu2019ApJ...884...97L,Perez2020ApJ}).
In summary, FU\,Ori presents a steeper spectral slope than FU\,Ori\,S at 8--48 GHz frequency.

In the 2020 July observations (Figure \ref{fig:fuoriFlux}), the 18--26 GHz and 40--48 GHz flux densities of FU\,Ori are comparable with the modeled spectral energy distribution published by \citet{Liu2019ApJ...884...97L}. 
Due to the phase errors, the directly measured 29--37 GHz flux density on July 04 was slightly lower than the model prediction and also lower than the measurements taken in 2016 October (\citealt{Liu2017A&A, Liu2019ApJ...884...97L}).
As presented in the top row of Figure \ref{fig:fuoriFlux}, the corrected (c.f., Section \ref{subsub:fluxdensity}) 29--37 GHz flux density of FU\,Ori on July 04 is indeed more consistent with the model prediction and the measurement taken in 2016 October despite the enlarged error bars.
The September 30 and November 15 observations consistently showed that at 29--37 GHz, FU\,Ori had become slightly brighter than how it was in 2016 October.

The 29--37 GHz flux densities of FU\,Ori\,S on July 04 both before and after correcting for the effect of phase errors were above the model prediction of \citet{Liu2019ApJ...884...97L} and the measurements taken in 2016 October.
In the 2020 July observations, we also detected high 12--26 GHz flux densities from FU\,Ori\,S.
From the point of view of SED modeling (Section \ref{sub:sed}), it is  hard to reconcile these new observations with the 8--10 GHz band upper limit in 2016 September unless we consider time variability.
The 29--37 GHz flux density of FU\,Ori\,S from 2020 July to November was persistently higher than the measurement of 2016 October and may also present some variability within this time period.

The variability of FU\,Ori\,S and FU\,Ori has a consequence on their relative brightness at 29--37 GHz: FU\,Ori appeared fainter than FU\,Ori\,S in 2020 July (Figure \ref{fig:maps}) but was brighter than FU\,Ori\,S on September 30 and November 15.
Our images of two point sources are unlikely to be limited by intensity dynamic range.
Therefore, the observed variation in relative brightness can hardly be attributed to calibration errors or imaging artifacts.

\subsection{Optical observations} \label{sub:opticalresult}

Figure \ref{fig:oirmonitor} shows the results of the optical photometric monitoring observations.
The Johnson {\it BVR} bands observations taken with the SLT and NOWT are reasonably consistent with each other.
We will derive the optical colors based on the SLT data which have lower noise.

The ASAS-SN data show that the Johnson {\it V} band magnitude of the FU\,Ori and FU\,Ori\,S binary system had an increasing trend from late 2016 to late 2018 or early 2019 (i.e., the Johnson {\it V} band flux has a decreasing trend).
The Johnson {\it V} band magnitudes detected in 2020-2021 appeared lower than the values in late 2018.
The linear regressions for the SLT data indicate that both the Johnson {\it B} and {\it V} band magnitudes had a decreasing trend in 2020-2021, while the Johnson {\it R} band magnitudes had an increasing trend.
The extrapolation of the Johnson {\it V} band regression line shows good consistency with the latest ASAS-SN Johnson {\it V} band measurements (Figure \ref{fig:oirmonitor}), indicating that the decreasing trend of the Johnson {\it V} band magnitude has not much changed since late 2018 or early 2019.

Figure \ref{fig:color} shows the Johnson {\it B-V}, {\it V-R}, and {\it B-R} colors observed by the SLT in 2020-2021, and compares the Johnson {\it B-R} color with Johnson {\it V} band magnitude over this time period.
The trends appear complicated. 
The Johnson {\it B-V} color was becoming redder over time while the Johnson {\it V-R} and {\it B-R} colors were becoming bluer over time.
The Johnson {\it B-R} color appears positively correlated with the Johnson {\it V} band magnitude (i.e., the source appeared bluer when it was brighter at Johnson {\it V} band).
Our tentative interpretation for these optical colors will be provided in Section \ref{sec:discussion}.

%

\begin{deluxetable}{ llll }
\tablecaption{Model parameters for FUOri\_dust1 and FUOriS\_dust1.\label{tab:dustmodel}}
\tablewidth{700pt}
\tabletypesize{\scriptsize}
\tablehead{
\colhead{$T_{\mbox{\scriptsize dust}}$} &
\colhead{$\Sigma_{\mbox{\scriptsize dust}}$} &
\colhead{$\Omega_{\mbox{\scriptsize dust}}$\tablenotemark{a}} &  
\colhead{\amax} \\
\colhead{(K)} &
\colhead{(g\,cm$^{-2}$)} &
\colhead{($10^{-14}$ sr)} & 
\colhead{(mm)} \\
} 
\startdata
\multicolumn{4}{c}{FUOri\_dust1} \\
370$^{+60}_{-42}$ &   63$^{+20}_{-13}$  &   4.1$^{+0.6}_{-0.6}$   &   1.59$^{+0.18}_{-0.14}$   \\
\multicolumn{4}{c}{FUOriS\_dust1} \\
150$^{+15}_{-23}$ &   19$^{+5.3}_{-5.1}$  &   4.1$^{+0.8}_{-0.3}$   &   $\lesssim$0.19$^{+0.12}_{-0.12}$    \\
\enddata
\tablenotetext{a}{1 sr $\sim$4.25$\times$10$^{10}$ square arcsecond.}
\end{deluxetable}

\begin{deluxetable*}{l  lll  lll  lll}
\tablecaption{Model parameters for the free-free emission components\label{tab:freefreemodel}}
\tablewidth{700pt}
\tabletypesize{\scriptsize}
\tablehead{
& \multicolumn{3}{c}{FU\,Ori free-free} & \multicolumn{3}{c}{FU\,Ori free-free2} & \multicolumn{3}{c}{FU\,Ori\,S free-free} \\
\colhead{Time} &
\colhead{$T_{e}$} &
\colhead{EM} &  
\colhead{$\Omega_{\mbox{\scriptsize ff}}$} &
\colhead{$T_{e}$} &
\colhead{EM} &  
\colhead{$\Omega_{\mbox{\scriptsize ff}}$} &
\colhead{$T_{e}$} &
\colhead{EM} &  
\colhead{$\Omega_{\mbox{\scriptsize ff}}$} \\
\colhead{(UTC)} &
\colhead{(10$^{3}$ K)} &
\colhead{(cm$^{-6}$pc)} & 
\colhead{(10$^{-14}$sr)\tablenotemark{a} } &
\colhead{(10$^{3}$ K)} &
\colhead{(cm$^{-6}$pc)} & 
\colhead{(10$^{-14}$sr)\tablenotemark{a} } &
\colhead{(10$^{3}$ K)} &
\colhead{(cm$^{-6}$pc)} & 
\colhead{(10$^{-14}$sr)\tablenotemark{a} } \\
} 
\startdata
2016 Sept./Oct. &
6.8$^{+3.2}_{-2.2}$ &
1.7$^{+0.7}_{-0.5}\times10^{7}$ &
1.5$^{+0.5}_{-0.4}$   &
$\cdots$ & $\cdots$ & $\cdots$ &
12$^{+4.5}_{-3.0}$ &
3.1$^{+0.7}_{-0.6}\times10^{9}$ &
3.1$^{+0.4}_{-0.5}\times10^{-2}$   \\
2020 July &
7.1$^{+3.3}_{-2.5}$ &
1.9$^{+0.6}_{-0.5}\times10^{7}$ &
1.3$^{+0.4}_{-0.3}$   &
$\cdots$ & $\cdots$ & $\cdots$ &
15$^{+4.5}_{-3.4}$ &
2.8$^{+1.0}_{-0.8}\times10^{9}$ &
7.8$^{+1.9}_{-1.7}\times10^{-2}$   \\
2020 September 30 &
5.9$^{+2.3}_{-1.8}$ &
1.8$^{+0.9}_{-0.6}\times10^{7}$ &
1.4$^{+0.6}_{-0.5}$   &
7.6$^{+2.8}_{-2.3}$ &
1.9$^{+0.8}_{-0.6}\times10^{9}$ &
2.0$^{+1.1}_{-0.6}\times10^{-2}$   &
15$^{+5.9}_{-4.7}$ &
2.7$^{+1.1}_{-0.7}10^{9}$ &
6.9$^{+2.1}_{-1.7}\times10^{-2}$   \\
2020 November 15 &
6.7$^{+3.5}_{-2.3}$ &
2.6$^{+0.7}_{-0.7}\times10^{7}$ &
1.0$^{+0.3}_{-0.3}$   &
7.6$^{+3.1}_{-2.5}$ &
1.1$^{+0.5}_{-0.4}\times10^{9}$ &
1.5$^{+0.7}_{-0.5}\times10^{-2}$   &
19$^{+5.6}_{-4.7}$ &
7.3$^{+2.5}_{-1.7}10^{9}$ &
5.2$^{+1.0}_{-0.9}\times10^{-2}$   \\
\enddata
\tablenotetext{a}{1 sr $\sim$4.25$\times$10$^{10}$ square arcsecond.}
\end{deluxetable*}

\section{Discussion}\label{sec:discussion}
Due to the complexity of our new observations and the existing data, and for the sake of making this manuscript self-contained, we first qualitatively discuss our overall interpretation in Subsection \ref{sub:interpretation}.
We realized our interpretation by constructing SED models and employed the Markov chain Monte Carlo (MCMC) method to optimize the free parameters in the models, which is introduced in Subsection \ref{sub:sed}.
The MCMC results are introduced in Subsection \ref{sub:sedresult}.
Based on the results, we also briefly discuss the deficiency of our present experimental setup and make a suggestion of how to improve those in the near future in Subsection \ref{sub:future}.

\subsection{Qualitative interpretation}\label{sub:interpretation}

The absence of Stokes Q, U, and V detection (Section \ref{sub:radioresult}) is consistent with that our new JVLA observations detected predominantly thermal emission.
The fractional radio flux density variations over a few months timescales are not large (Figure \ref{fig:fuoriFlux}), which also make the dominant emission mechanisms appear more like the thermal ones rather than nonthermal (e.g., see \citealt{Liu2014} and references therein for more discussion).

The previous X-ray observations have detected time varying high energy activities from our target sources (\citealt{Skinner2010ApJ...722.1654S}).
The X-ray emission may also be accompanied by nonthermal radio emission.
Nevertheless, such X-ray and nonthermal radio emissions should be emanated from regions that are very close to the protostellar surfaces.
In actively accreting YSOs, such nonthermal radio emission is often fully obscured by free electrons (c.f., \citealt{Feigelson1999ARA&A..37..363F} and references therein).
Therefore, it is sensible to interpret the observed 8--48 GHz flux densities by combinations of dust thermal emission and free-free emission.

Given that dust emission is unlikely to be prominent at 8--12 GHz (c.f., \citealt{Liu2017A&A,Liu2019ApJ...884...97L}), the observed X band flux density at any time tightly constrains the total budget of free-free emission in the overall SED of FU\,Ori and FU\,Ori\,S at that time.
At higher frequencies, the portions of flux densities that cannot be interpreted with free-free emission have to be attributed to dust thermal emission.

The stationary part of the $\lesssim$40 GHz emission of FU\,Ori\,S can be explained with the dominating optically thick dust emission mixed with some optically thin free-free emission; otherwise, it can be dominated by free-free emission over a range of optical depths (see also the discussion in Section 4.1 of \citealt{Liu2017A&A}).
Previously, \citet{Liu2019ApJ...884...97L} considered the former to be more likely in the interpretation of their 29--37 GHz data taken in 2016.
However, it appears that the latter is more consistent with our new 8--48 GHz observations.
Such a difference may be the consequence that the free-free emission of FU\,Ori\,S has become brighter from 2016 to 2020 while its dust emission has not changed considerably.
The 8-48 GHz spectral indices of FU\,Ori\,S became lower when it became brighter in 2020. 
In other words, the variable part of the 8--48 GHz SED of FU\,Ori\,S appears to have a flat spectral index which is indeed more consistent with free-free emission (more in Subsections \ref{sub:sed}, \ref{subsub:fuoriSresult}).

We note that the 8--10 GHz upper limit in 2016 September (Figure \ref{fig:fuoriFlux}) indicates that some stationary dust emission may still be necessary for the interpretation of the 8--48 GHz SED of FU\,Ori\,S.
Our rather weak (although significant) 8--12 GHz detection on June 26 and the ALMA 86--160 GHz observations presented in \citet{Liu2019ApJ...884...97L} also disfavor interpreting the 8--48 GHz SED of FU\,Ori\,S solely by optically thick free-free emission (Section \ref{sub:sed}).

The SED of FU\,Ori is steeply rising (with frequency) at $\sim$10--20 GHz, meaning that dust emission dominates over free-free emission at this frequency range and at higher frequencies (Figure \ref{fig:fuoriFlux}).
Its spectral slope increases more slowly around $\sim$25--30 GHz and becomes steep again at $\gtrsim$40 GHz, which may be consistent with having a persistent dust SED bump at $\sim$30 GHz (see also Appendix \ref{appendix:loglog} the discussion in Section 4.4.1 of \citealt{Liu2019ApJ...884...97L}).
Interpreting the flatter spectral index of FU\,Ori at $\sim$30 GHz by a high contribution of free-free emission instead of a dust SED bump will easily make the extrapolated 8--12 GHz  flux density exceed what had been detected.
Being dominated by dust instead of free-free emission may also explain why FU\,Ori appears more stationary than FU\,Ori\,S at 40--48 GHz (Figure \ref{fig:fuoriFlux}).

The small radio variability of FU\,Ori, if not due to absolute flux calibration errors, can be explained by time varying free-free emission, or parametrically time varying dust temperature ($T_{\mbox{\scriptsize dust}}$), dust column density ($\Sigma_{\mbox{\scriptsize dust}}$), and \amax.
In reality, $\Sigma_{\mbox{\scriptsize dust}}$ and \amax are unlikely to vary significantly on timescales of only a few years unless $T_{\mbox{\scriptsize dust}}$ is varied dramatically enough to sublimate dust or to change dust properties (e.g., \citealt{Stammler2017A&A...600A.140S,Molyarova2021ApJ...910..153M}).
Physically, the time variations of free-free emission and $T_{\mbox{\scriptsize dust}}$ are not mutually exclusive.

For the FU\,Ori and FU\,Ori\,S binary system, optical and infrared emission is dominated by the disk of FU\,Ori instead of the protostellar photosphere of either of these two sources \citep{Hartmann1996ARA&A,Turner1997ApJ...480..754T,Wang2004ApJ...601L..83W,Liu2016SciA,Takami2018ApJ,Laws2020ApJ...888....7L}.
When $T_{\mbox{\scriptsize dust}}$ is varying with time, we may expect the flux densities at optical/infrared and radio bands to vary in the same sense.
In addition, the variability should appear considerably more prominent at the optical and infrared bands since the radio observations are well in the Rayleigh-Jeans limit.
At optical, infrared, and (sub)millimeter bands, the observed variation on decadal timescales is small (Section \ref{sub:opticalobs}, \ref{sec:result}; \citealt{Herbig1977ApJ...217..693H,Kolotilov1985SvAL...11..358K,Green2016ApJ,Liu2018A&A}).
From Figure \ref{fig:oirmonitor} it can also be seen that the optical {\it V} band magnitude has not varied significantly until mid-April of 2021\footnote{FU\,Ori became a day-time source after then, which cannot be observed from ground based optical observatories.}.
The small variability at optical and/or infrared bands can be attributed to very small variations of the disk temperature of FU\,Ori; otherwise, it may also be due to the more significant variability of the less luminous source of the binary system, FU\,Ori\,S.
It is sensible to consider that the dust-mass weighted averages of $T_{\mbox{\scriptsize dust}}$, $\Sigma_{\mbox{\scriptsize dust}}$, and \amax in the FU\,Ori disk are approximately invariant at least during the time period of our new JVLA observations and attribute its variability at 8--48 GHz to free-free emission.

A caveat is that when $T_{\mbox{\scriptsize dust}}$ is changed due to the time varying non-passive heating mechanisms such as viscous heating, adiabatic compression, or shocks (e.g., \citealt{Evans2015MNRAS.453.1147E,Vorobyov2018A&A,Vorobyov2020A&A...644A..74V}), the effects can be initially more prominent in the deeply obscured, higher density disk mid-plane, during which only the (sub)millimeter or radio emission is brightened.
Eventually, the heat will still be transported to the disk surface.
However, in this case, the brightening at optical or infrared bands may lag.
With the present observational data, there is still the possibility that $T_{\mbox{\scriptsize dust}}$ in the mid-plane but not at the surface of the FU\,Ori disk was significantly changed over the time period of our JVLA observations.
Our optical photometric monitoring observations have covered an extended time period after the latest epoch of the JVLA observations (Figure \ref{fig:oirmonitor}), which have made this possibility relatively unlikely.
To strictly rule out this possibility, it might be necessary to extend the optical and infrared photometric monitoring observations for another year or two.

\subsection{SED modeling}\label{sub:sed}
The noise levels of our new measurements at 8--48 GHz do not permit deriving spectral indices using every two adjacent data points in the frequency domain.
Because the spectral indices of dust and free-free emission are not rapidly changing with frequency, the SED models presented in this section utilize the constraints given by all data points to view overall trends of spectral index variations over broad frequency ranges.
The central goal of our SED modeling is to constrain the \amax values in the dusty disks around FU\,Ori and FU\,Ori\,S.

\begin{figure*}[ht]
    \centering
    \includegraphics[width=16cm]{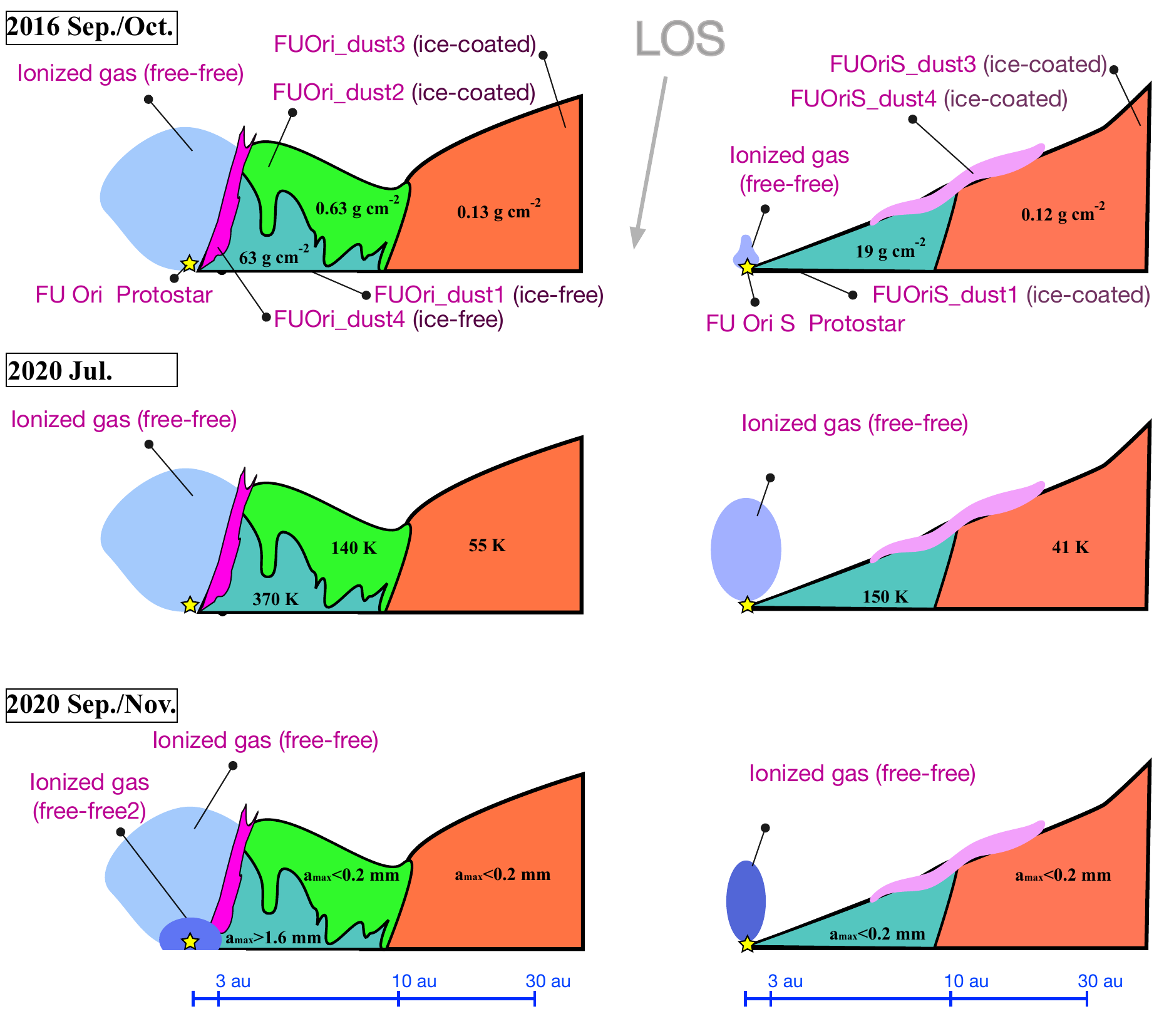}
    \caption{Schematic pictures to describe our spectral energy distribution modeling. Left and right column show the models for FU\,Ori and FU\,Ori\,S, respectively. They are approximately viewed from top. From top to bottom rows show the situation in 2016 September/October, 2020 July, and 2020 September/November. The invariant dust emission components FUOri\_dust1--4 and FUOriS\_dust1,3,4 are color-coded differently and are labeled (see the introduction in Section \ref{subsub:components}). The color-codings are in the same sense with the SED models presented in Figure \ref{fig:fuoriFlux} and \ref{fig:fuoriFluxPre2019}. The best fits of the dust column density, temperature, and maximum grain size ($a_{\mbox{\scriptsize max}}$) for FUOri\_dust1--3 and FUOriS\_dust1,3, which dominate the overall dust mass budget, are labeled in the top, middle, and bottom panels, respectively.}
    \vspace{0.4cm}
    \label{fig:schematic}
\end{figure*}

We produced SED models for the new 8--48 GHz observations and for the JVLA and higher frequency (ALMA, SMA, {\it Herschel}, {\it Spitzer}, VLTI/GRAVITY) observations presented in \citet{Liu2019ApJ...884...97L}.
To avoid overfitting the infrared spectra, we followed the approach of \citet{Liu2019ApJ...884...97L} to artificially reduce the weight by assigning the flux errors of the {\it Herschel}/SPIRE, {\it Herschel}/PACS, and {\it Spitzer}/IRS data to be 1000, 10, and 1 times the detected fluxes.
This adjustment is needed since fitting the details of these infrared spectra requires models of fine 3-dimensional temperature profiles (e.g., \citealt{Yang2018ApJ...860..174Y,Zhang2021A&A...646A..18Z}), which are very degenerated when the target sources are not spatially resolved and are not dominated by passive heating.
Another reason that we should avoid overfitting the {\it Herschel} spectra is that {\it Herschel} spectra are inevitably confused by the foreground/background dust emission.
Since the main focus of the present work is on interpreting the interferometric data in the Rayleigh-Jeans limit, it is sufficient to approximate with averaged temperatures without deriving the detailed temperature profiles based on the infrared spectra.

We assume that the $>$50 GHz emission is dominated by dust thermal emission and thus is approximately stationary.
The 8--48 GHz measurements are treated as four independent epochs of data: the 2016 September/October observations reported in \citet{Liu2017A&A} and \citet{Liu2019ApJ...884...97L} are treated as one single epoch; the 2020 July observations are treated as the second epoch; the 2020 September 30 and November 15 observations are treated as two independent epochs. 
We attempted to fit these data with stationary dust emission components and time varying free-free emission components.

\subsubsection{Emission mechanisms}\label{subsub:emission}

\begin{figure*}
    \hspace{2cm}
    \includegraphics[width=13.5cm]{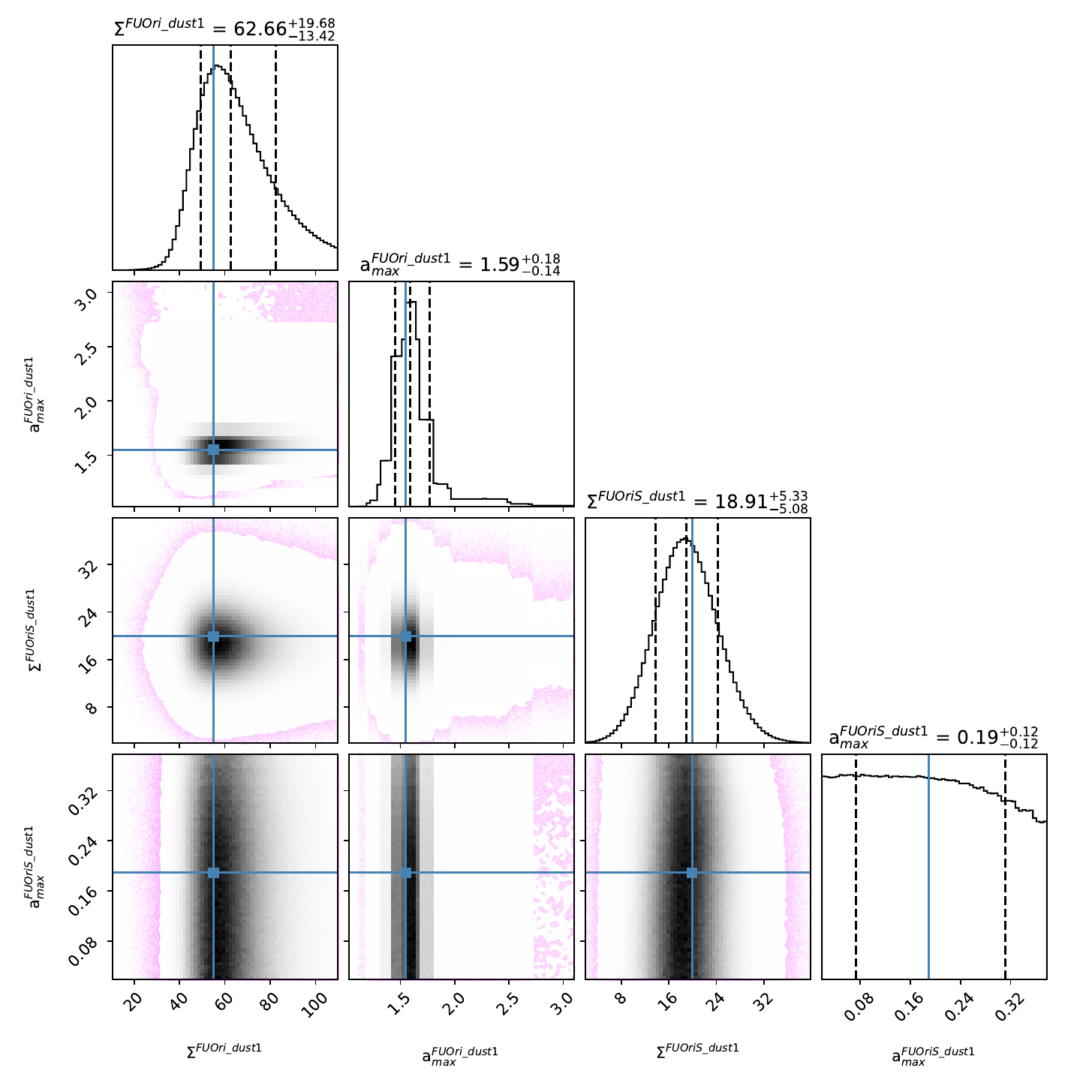}
    \caption{MCMC corner plot and the posterior distributions (Section \ref{subsub:optimization}) for the $\Sigma_{\mbox{\scriptsize dust}}$ and \amax parameters of the FUOri\_dust1 and FUOriS\_dust1 components (Table \ref{tab:dustmodel}; Section \ref{subsub:components}). The solid blue lines show the mean initial positions of the MCMC samplers in the second round of our MCMC fits (Section \ref{subsub:optimization}).  The 1D histograms are marginal posterior distributions of these parameters. The vertical dashed lines indicated the 16th, 50th, and 84th percentiles of the samples, respectively, which are labeled on the top of each histogram. In the 2D histograms, the well-sampled areas are represented with grayscales; individual samplers are plotted in light purple colors in the regions that have relatively poor sampling rates. }
    \vspace{0.3cm}
    \label{fig:corner}
\end{figure*}

Due to the complicated physics involved in the accretion outburst of FU\,Ori, 
it is not feasible to describe the radial $T_{\mbox{\scriptsize dust}}$ and $\Sigma_{\mbox{\scriptsize dust}}$ profiles with analytical forms.
To facilitate SED modeling, some simplification is necessary.
We built on the strategy outlined in Section 4.2 of \citet{Liu2019ApJ...884...97L} to model the radio and higher frequency SEDs of FU\,Ori and FU\,Ori\,S with components (described in Section \ref{subsub:components}) of dust and free-free emission, taking into consideration their mutual obscuration.
The physical properties within each component were assumed to be uniform.
Specifically, we quoted their Equation 3 to describe the overall flux densities ($F_{\nu}$) of FU\,Ori and FU\,Ori\,S as:

\begin{equation}\label{eqn:multicomponent}
    F_{\nu} = \sum\limits_{i} F_{\nu}^{i} e^{-\sum\limits_{j}\tau^{i,j}_{\nu}}, 
\end{equation}
where $F_{\nu}^{i}$ is the flux density of the dust or free-free emission component $i$, and $\tau^{i,j}_{\nu}$ is the optical depth of the emission component $j$ to obscure the emission component $i$.
Physically, each dust emission component can represent a parcel of dust at a certain (radial, azimuthal, and vertical) location in the FU\,Ori or FU\,Ori\,S disk.
The abstracted geometric information is provided by $\tau^{i,j}_{\nu}$.

We approximated the spectral profile and optical depth of free-free emission according to the formulation outlined in \citet{Mezger1967} and \citet{Keto2003}.
In this way, the spectral profile depends on three free parameters: electron temperature ($T_{\mbox{\scriptsize e}}$), emission measure (EM)\footnote{The emission measure EM$=$ $\int n_{\mbox{\scriptsize e}}^{2}d\ell$, with $n_{\mbox{\scriptsize e}}$ being the electron number density, and $\ell$ is the linear size scale of the free-free emission component along the line of sight.}, and solid angle ($\Omega_{\mbox{\scriptsize ff}}$).

We quoted the DSHARP size-dependent dust opacity tables which were constructed for the purpose of modeling circumstellar dust emission.
These tables assumed that dust grains are morphologically compact, which may be supported by the recent ALMA observations (e.g., \citealt{Tazaki2019ApJ}).
In addition, the dust grains with and without water ice \citep{Warren1984ApOpt..23.1206W} are composed of astronomical silicates \citep{Draine2003ARA&A..41..241D}, troilite, refractory organics (\citealt{Henning1996A&A...311..291H}; see Tabe 1 of \citealt{Birnstiel2018}).
The default (i.e., water-ice coated) DSHARP opacity was quoted for dust components with $<$170 K temperatures; otherwise, the ice-free opacity table was quoted.

We assumed that the dust grain size ($a$) distribution $n(a)$ follows $a^{-q}$ in between the minimum and maximum grain sizes ($a_{\mbox{\scriptsize min}}$, $a_{\mbox{\scriptsize max}}$) and is 0 beyond this range.
We fixed $a_{\mbox{\scriptsize min}}$ to $10^{-4}$ mm nominally since the spectral profile of dust is not sensitive to this parameter.
The values of $a_{\mbox{\scriptsize max}}$ was optimized together with dust temperature ($T_{\mbox{\scriptsize dust}}$), column density ($\Sigma_{\mbox{\scriptsize dust}}$), and projected solid angle ($\Omega_{\mbox{\scriptsize dust}}$) to fit the observations.
We followed Equation 6 of \citet{Birnstiel2018} to evaluate the dust mass opacities, and followed Equations 10--20 of \citet{Birnstiel2018} to approximate the dust spectral profile, taking anisotropic dust self-scattering into consideration.

We assumed that all dust components are approximately face-on given that we do not have enough measurements to constrain the inclinations as free parameters, which will not qualitatively change our analysis.
We optimize $T_{\mbox{\scriptsize dust}}$ as a free parameter instead of iteratively solving it using 3-dimensional radiative transfer since the innermost $\sim$10 au regions of FUors or any other actively accreting YSOs may be dominated by viscous heating (\citealt{Calvet1991ApJ,Liu2019ApJ...884...97L, Takami2019ApJ,Liu2021,Labdon2021A&A}).

\subsubsection{Emission components}\label{subsub:components}

\begin{figure}
    \hspace{-1.0cm}
    \vspace{-0cm}
    \begin{tabular}{ p{7.5cm} }
        \includegraphics[height=5.15cm]{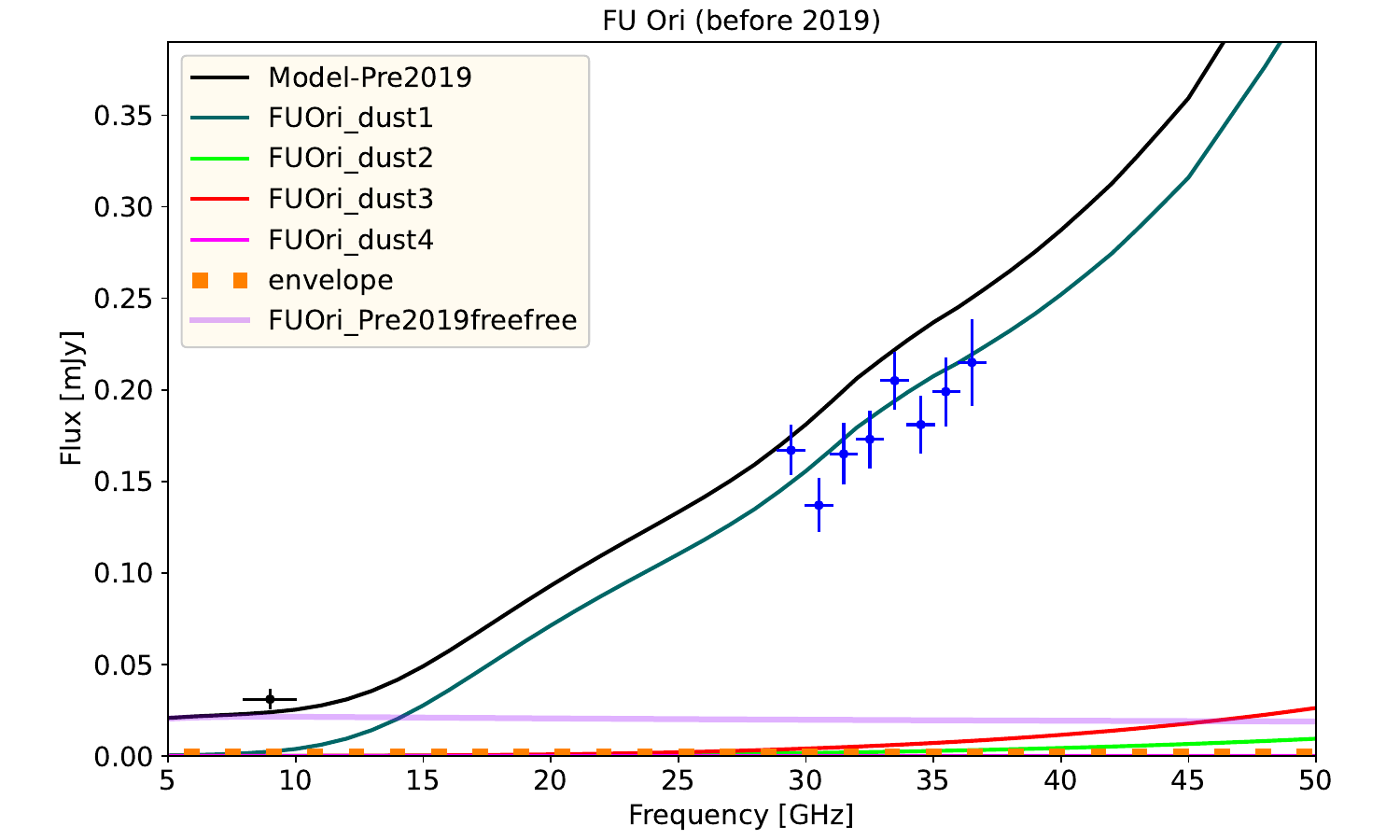} \\  
        \includegraphics[height=5.15cm]{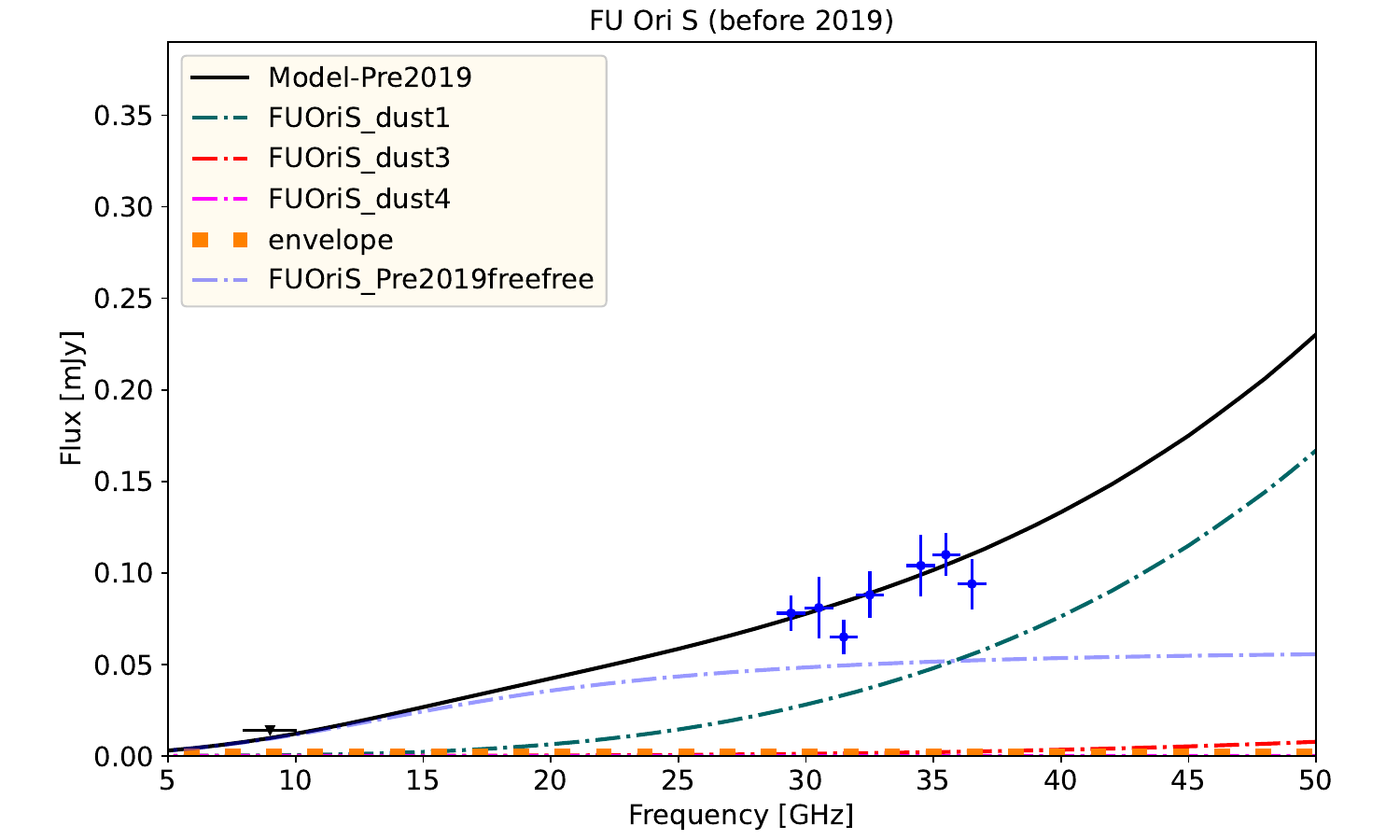} 
    \end{tabular}
    
    \vspace{0cm}
    \caption{The JVLA X and Ka band data presented in \citet{Liu2017A&A,Liu2019ApJ...884...97L}, and our updated models to fit these data (labeled in each panel; see Section \ref{sub:sed} for the explanations).
    }
    \label{fig:fuoriFluxPre2019}
\end{figure}

As described by Equation \ref{eqn:multicomponent}, our approach is to approximate the dust and free-free emission sources as a finite number of emission components.
The total number of emission components is limited by the independent observational measurements we presently possess.

For FU\,Ori, \citet{Liu2019ApJ...884...97L} found that their JVLA 8--10 GHz and 29--37 GHz observations and ALMA 86--160 GHz observations, the General Relativity Analysis via VLT InTerferometrY (GRAVITY) instrument on the Very Large Telescope Interferometer (VLTI) 2--2.45 $\mu$m observations, and the ALMA 225 and 345 GHz observations reported in \citet{Hales2015ApJ} and \citet{Perez2020ApJ} can be described by a free-free emission component (FUOri\_freefree) and only four dust components namely FUOri\_dust1--4: FUOri\_dust4 represents the very high temperature surface of the inner 3 au disk; FUOri\_dust1 represents the mid-plane of the $\lesssim$10 au disk which is obscured by the lower temperature surface FUOri\_dust2; and FUOri\_dust3 is the lower temperature, lower column density outer disk. 
Similarly, their presented observations on FU\,Ori\,S can be described by a free-free emission component (FUOriS\_freefree) and three dust components (FUOriS\_dust1,3 and 4): FUOriS\_dust1 and 4 represent the mid-plane and surface of the inner $\lesssim$10 au disk while FUOriS\_dust3 is the lower temperature, lower column density outer disk.
These dusty disk components are illustrated in Figure \ref{fig:schematic}.
After including a spatially extended common envelope component that is obscuring all the aforementioned emission components but was resolved out by JVLA and ALMA, the {\it Herschel} and {\it Spitzer} spectra reported in \citet{Green2006ApJ,Green2013ApJ,Green2016AJ}, and Submillimeter Array (SMA)  observations published in \citet{Liu2018A&A} can also be reasonably well described.

After some trial adjustment of free parameters, we confirmed that those previous observations of FU\,Ori and FU\,Ori\,S, the JVLA observations taken in 2020 July, September 30, and November 15 can be approximated by the stationary dust emission components that are very similar to those in the models of \citet{Liu2019ApJ...884...97L}, and the time variable free-free emission components.

In contrast to \citet{Liu2019ApJ...884...97L}, we included one additional optically thick free-free emission component (FUOri\_freefree2) in the model for FU\,Ori (i.e., using two free-free emission components in total) to artificially avoid interpreting its 8--48 GHz variability by the time varying $T_{\mbox{\scriptsize dust}}$ (more in Subsection \ref{sub:sedresult}).
The projected solid angle of FUOri\_freefree2 corresponds to a $\sim$1 au diameter, which is 2 orders of magnitude smaller than that of optically thin free-free emission component, FUOri\_freefree.
The emission of FUOri\_freefree2 was negligible in 2016 September/October and in 2020 July.
Physically, FUOri\_freefree2 may represent some radiatively ionized gas located close to the host protostar, or one or multiple thermally ionized gas knots associated with some hot spots in the FU\,Ori disk (illustrated in the bottom row of Figure \ref{fig:schematic}).

In the MCMC optimization (Section \ref{subsub:optimization}), we found that the samplers tended to converge to a seemingly unreasonably high dust column density ($\sim$200 g\,cm$^{-2}$) and the relatively high \amax values (e.g., 3-5 mm) in the inner $\sim$10 au FU\,Ori disk if FUOri\_freefree2 is not included when $T_{\mbox{\scriptsize dust}}$ is forced to be a constant of time.
The high $\Sigma_{\mbox{\scriptsize dust}}$ and \amax values help enhance the dust emission at low frequency bands.
Presently, we cannot rule out the possibility that the FUOri\_freefree2 component was unnecessary.
The \amax of FU\,Ori derived from our MCMC models with two free-free emission components may (or may not) be underestimated by a factor of up to $\sim$3 (Subsection \ref{sub:sedresult}).

While we presently do not include more dust emission components, imperfections in the models may still be attributed to oversimplification, or measurement errors (e.g., absolute flux errors).
In particular, $\Sigma_{\mbox{\scriptsize dust}}$ likely has significant spatial dependence in the innermost few au.
Some related discussion for the case of FU\,Ori\,S will be provided in Subsection \ref{subsub:fuoriSresult}.
The values of \amax may also have spatial dependence in regions where dust grain growth/fragmentation/migration/trapping is efficient (e.g., \citealt{Vorobyov2018A&A}).
The values of \amax and $\Sigma_{\mbox{\scriptsize dust}}$ derived based on the MCMC fits for the integrated SEDs represent the intensity-weighted averaged values.

\subsubsection{Optimization}\label{subsub:optimization}
We used the MCMC fitting routine {\texttt emcee} to optimize the model free parameters and to help assess the uncertainties of those parameters.
Both FU\,Ori and FU\,Ori S were fitted together because they share the same envelope.
We adopted the model parameters of FUOri\_dust2,3,4,  FUOriS\_dust3,4, and the common envelope quoted from \citet{Liu2019ApJ...884...97L} without advancing them, as these dust components have little contribution over the 8--48 GHz frequency range of our new JVLA observations.
They are primarily constrained by the previous $>$50 GHz observations that can be reasonably well fit by the modeling parameters of \citet{Liu2019ApJ...884...97L}.

The components FUOri\_dust1 and FUOriS\_dust1 may significantly contribute to the flux densities detected at 8--48 GHz (see Figure 5 of \citealt{Liu2019ApJ...884...97L}).
Using MCMC and assuming flat priors, we seek (1) the solutions of FUOri\_dust1 and FUOriS\_dust1 that are optimal for all epochs of observations (i.e., these two components do not vary from epoch to epoch), and (2) the parameters to describe the free-free emission in individual epochs of observations.

It turned out that the additional free-free component (i.e., FUOri\_freefree2) was only necessary for the data from 2020 September and 2020 November.
The parameters (e.g., electron temperature, solid angle) for the free-free emission components are degenerate since our JVLA observations (Section \ref{sec:observation}) were not dedicated to constraining them.
Nevertheless, as long as the observed SED profiles can be reproduced by the models, such degeneracy does not seriously impact our discussion about dust properties.

The $T_{\mbox{\scriptsize dust}}$ and $\Omega_{\mbox{\scriptsize dust}}$ parameters of FUOri\_dust1 and FUOriS\_dust1 are also degenerated to some extent.
For example, the fits can converge to a rather large $\Omega_{\mbox{\scriptsize dust}}$ value (e.g., as far as permitted by the {\it Herschel} spectra) and small $T_{\mbox{\scriptsize dust}}$, although this degeneracy turned out to be not particularly serious in our MCMC fits.
In fact, the upper limit of $\Omega_{\mbox{\scriptsize dust}}$ has been constrained by the high angular resolution JVLA and ALMA observations (\citealt{Liu2017A&A,Perez2020ApJ}).
Therefore, the $\Omega_{\mbox{\scriptsize dust}}$ and $T_{\mbox{\scriptsize dust}}$ given by our fits should be regarded as upper and lower limits, respectively.
The actual values of $\Omega_{\mbox{\scriptsize dust}}$ can be smaller if dust is concentrated in spatially unresolved substructures (e.g., rings, crescents, etc).
On the other hand, $T_{\mbox{\scriptsize dust}}$ should not exceed the $\sim$1500 K sublimation temperature in any case.

We performed two rounds of MCMC fits: in the first round all free parameters were iteratively advanced; in the second round we marginalized the $T_{\mbox{\scriptsize dust}}$ and $\Omega_{\mbox{\scriptsize dust}}$ of FUOri\_dust1,2 and all the parameters for the free-free emission by adopting the best fits of the first round, and only advanced the $\Sigma_{\mbox{\scriptsize dust}}$ and \amax of FUOri\_dust1 and FUOriS\_dust1 iteratively.
This strategy helps yield better and more comprehensive samplings for $\Sigma_{\mbox{\scriptsize dust}}$ and \amax.
In addition, in the second round, we only compared the models with the JVLA data taken in 2020 to alleviate the potential biases due to neglecting the possibility of the long-term time variability of $T_{\mbox{\scriptsize dust}}$.

In the first round, we used 150 walkers with 12000 iterative steps each; the results from the first 2000 steps were discarded.
In the second round, we used 300 walkers with 80000 iterative steps each; again, the results from the first 2000 steps were discarded.
For each walker, we rejected the steps which yielded  9 GHz flux densities of FU\,Ori\,S that are higher than the 3$\sigma$ upper limit in 2016 September (Figure \ref{fig:fuoriFlux}).
The results are introduced in the following subsection.

\begin{figure}
    \hspace{-2.1cm}
    \begin{tabular}{ p{9cm} }
       \includegraphics[height=6cm]{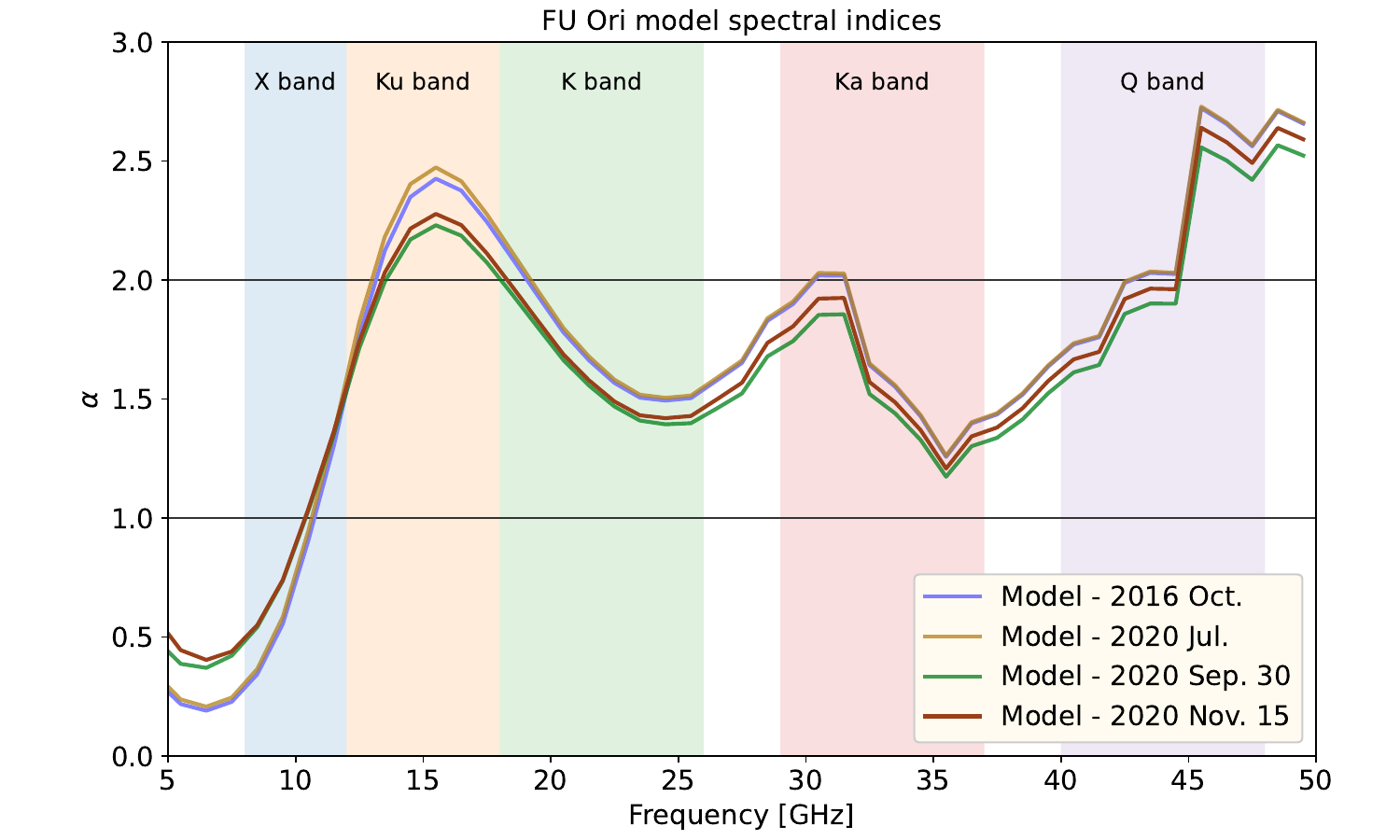}   \\
       \includegraphics[height=6cm]{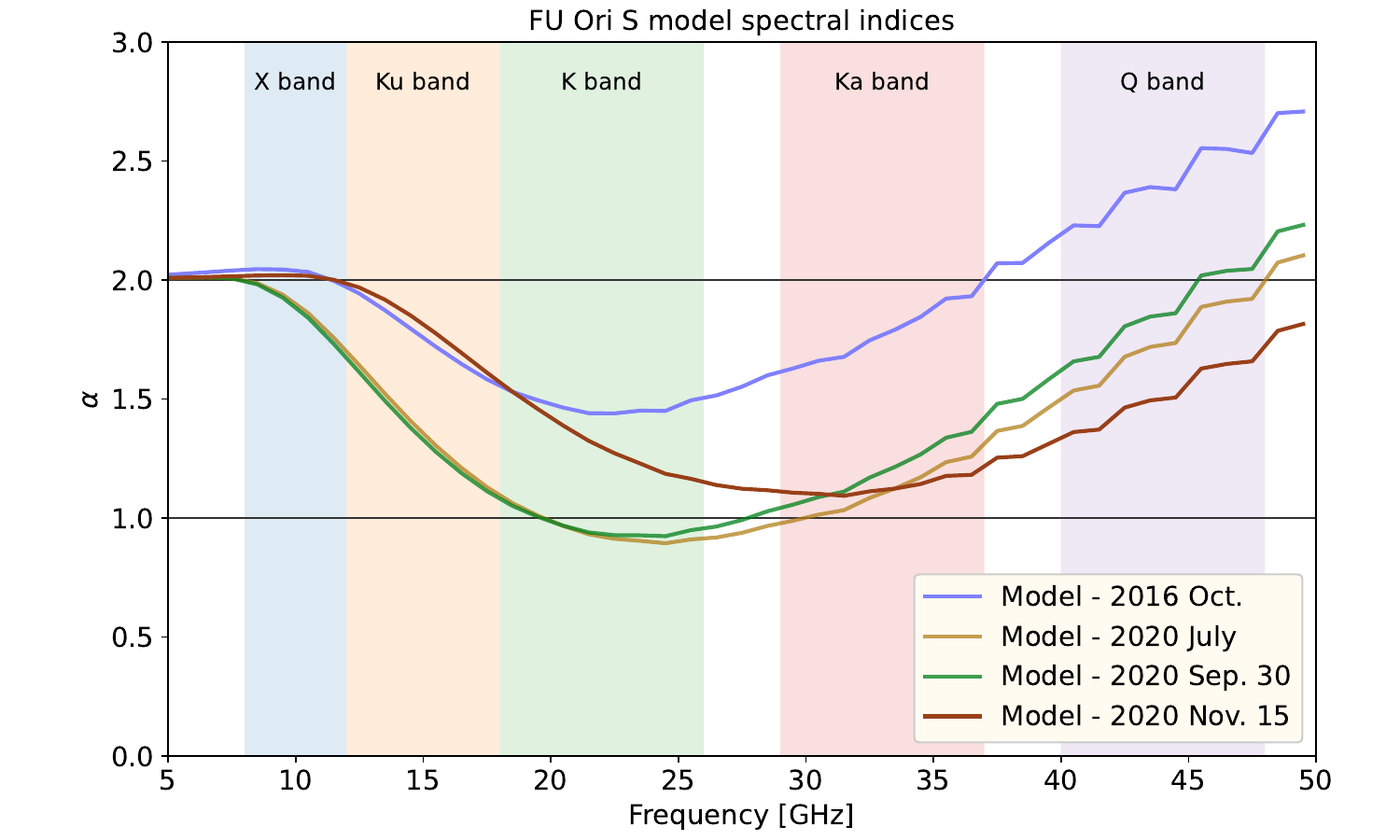}  \\
    \end{tabular}
    \caption{The spectral indices derived from the models for FU\,Ori and FU\,Ori\,S presented in Figure \ref{fig:fuoriFlux}.}
    \label{fig:modelapha}
\end{figure}

\subsection{Modeling results}\label{sub:sedresult}
Tables \ref{tab:dustmodel} and \ref{tab:freefreemodel} summarize the results of our MCMC fits, which include the 50th percentile of the samples as the best fit for each parameter and define the errors based on the 16th and 84th percentiles.
Figure \ref{fig:corner} shows the corner plot and the MCMC posteriors (Section \ref{subsub:optimization}) for the $\Sigma_{\mbox{\scriptsize dust}}$ and \amax parameters of the FUOri\_dust1 and FUOriS\_dust1 components (Table \ref{tab:dustmodel}; c.f., Figure \ref{fig:schematic}).
The flux densities of the individual dust and free-free emission model components to fit the 2020 observations are presented in Figure \ref{fig:fuoriFlux}; those to fit the earlier observations (i.e., reported by \citealt{Liu2017A&A}, \citealt{Liu2019}) are presented in Figure \ref{fig:fuoriFluxPre2019}.
The 8--48 GHz spectral indices of the models for each epoch of observations are presented in Figure \ref{fig:modelapha}.
The results of individual models of FU\,Ori and FU\,Ori\,S are discussed in the following.

\subsubsection{FU\,Ori} \label{subsub:fuoriresult}
The new model (with 2 free-free emission components and four dust emission components) reproduced reasonably well  the 8--48 GHz flux densities of FU\,Ori observed in 2020 (Figure \ref{fig:fuoriFlux}).
The newly derived $T_{\mbox{\scriptsize dust}}$, $\Sigma_{\mbox{\scriptsize dust}}$ and $\Omega_{\mbox{\scriptsize dust}}$ of FUOri\_dust1 are consistent with the previous derivation of \citet{Liu2019ApJ...884...97L}. 
This is expected because the newly measured flux densities at 8--48 GHz are comparable to the previous model prediction.
Our new measurements at 15--26 GHz better constrained the $\Sigma_{\mbox{\scriptsize dust}}$ of FUOri\_dust1, making its value $\sim$15\% higher than the derivation of \citet{Liu2019ApJ...884...97L}.
From Figure \ref{fig:corner} it can be seen that some solutions with further higher $\Sigma_{\mbox{\scriptsize dust}}$ also can be consistent with the JVLA observations.
Our update of $\Sigma_{\mbox{\scriptsize dust}}$ over the results of \citet{Liu2019ApJ...884...97L} does not significantly impact the fits for the $>$50 GHz data (see also Appendix \ref{appendix:loglog}) since FUOri\_dust1 becomes very optically thick and is obscured by the disk surface component FUOri\_dust2 at high frequencies.

Our updated \amax value for FUOri\_dust1 is $\sim$1.59$^{+0.18}_{-0.14}$ mm (see Figure \ref{fig:corner}), which is slightly lower than the 2.4$^{+0.40}_{-0.32}$ mm value reported by \citet{Liu2019ApJ...884...97L}.
The uncertainty of \amax is mainly resulted from the uncertainty of the models for free-free emission.
The value of \amax becomes smaller when stronger free-free emission is assumed in the fits of the 8--48 GHz SED.

When we only use one free-free emission component in the model for FU\,Ori instead of two, the marginal distribution of the MCMC posteriors for the \amax of FUOri\_dust1 peaks at 2.7 mm and is skewed towards higher values (e.g., the values up to $\sim$5 mm are also reasonably probable).
In general, assuming strong free-free emission will yield over-intensities at X band (8--12 GHz) and Ku band (12--18 GHz) as compared with the observations.
Therefore, the MCMC fits (with a flat prior) favor the larger \amax value and suppressing the free-free emission component(s).
The marginal distribution of the MCMC posteriors for the \amax values is skewed towards larger sizes even when we adopted two free-free emission components (Figure \ref{fig:corner}) although in this case the high \amax tail is only detectable with a large number of MCMC samples.
The best fit of \amax given in Table \ref{tab:dustmodel} thus may be regarded as a conservative lower limit.

\begin{figure*}
    \hspace{-1cm}
    \begin{tabular}{ p{8.5cm} p{8.5cm} }
        \includegraphics[height=6cm]{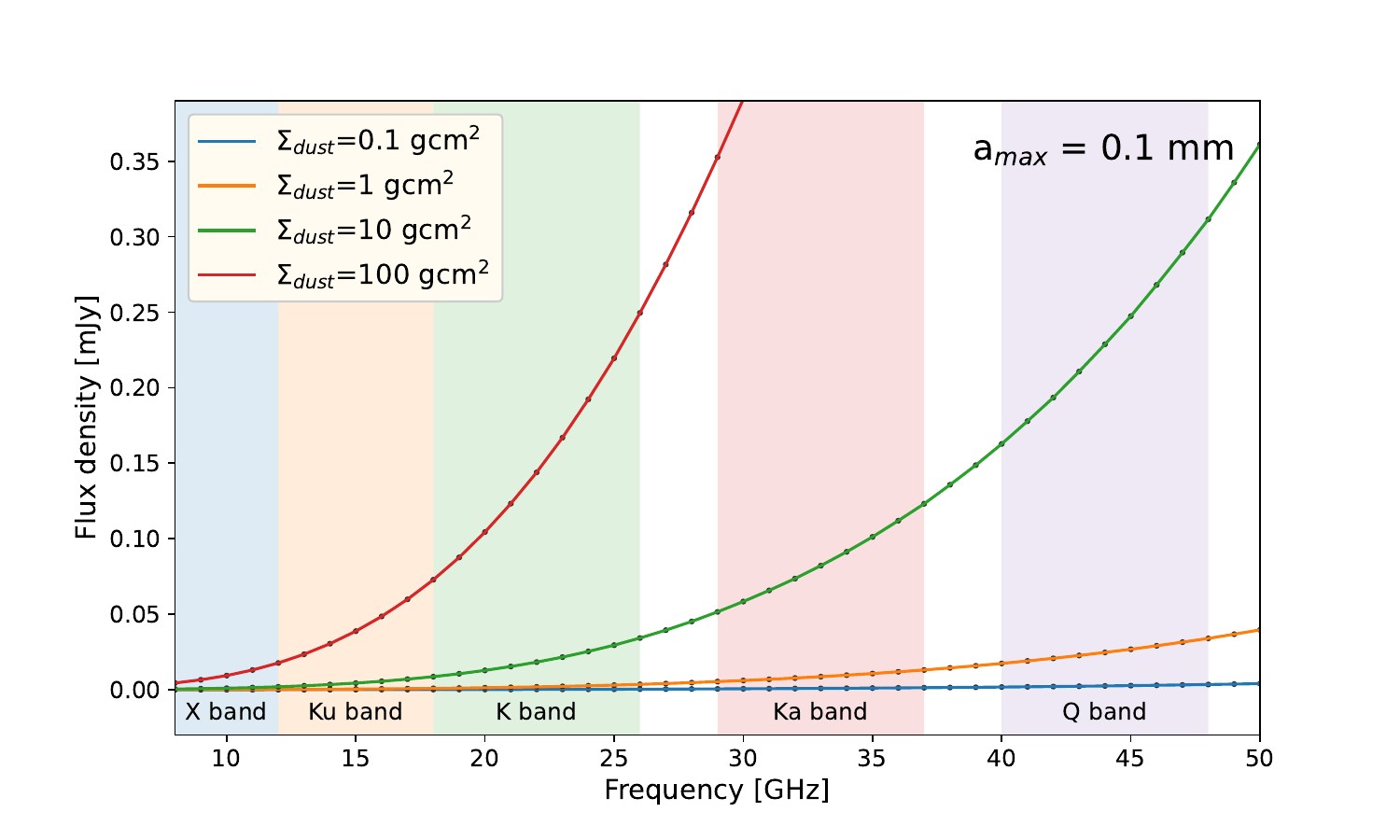} &  
        \includegraphics[height=6cm]{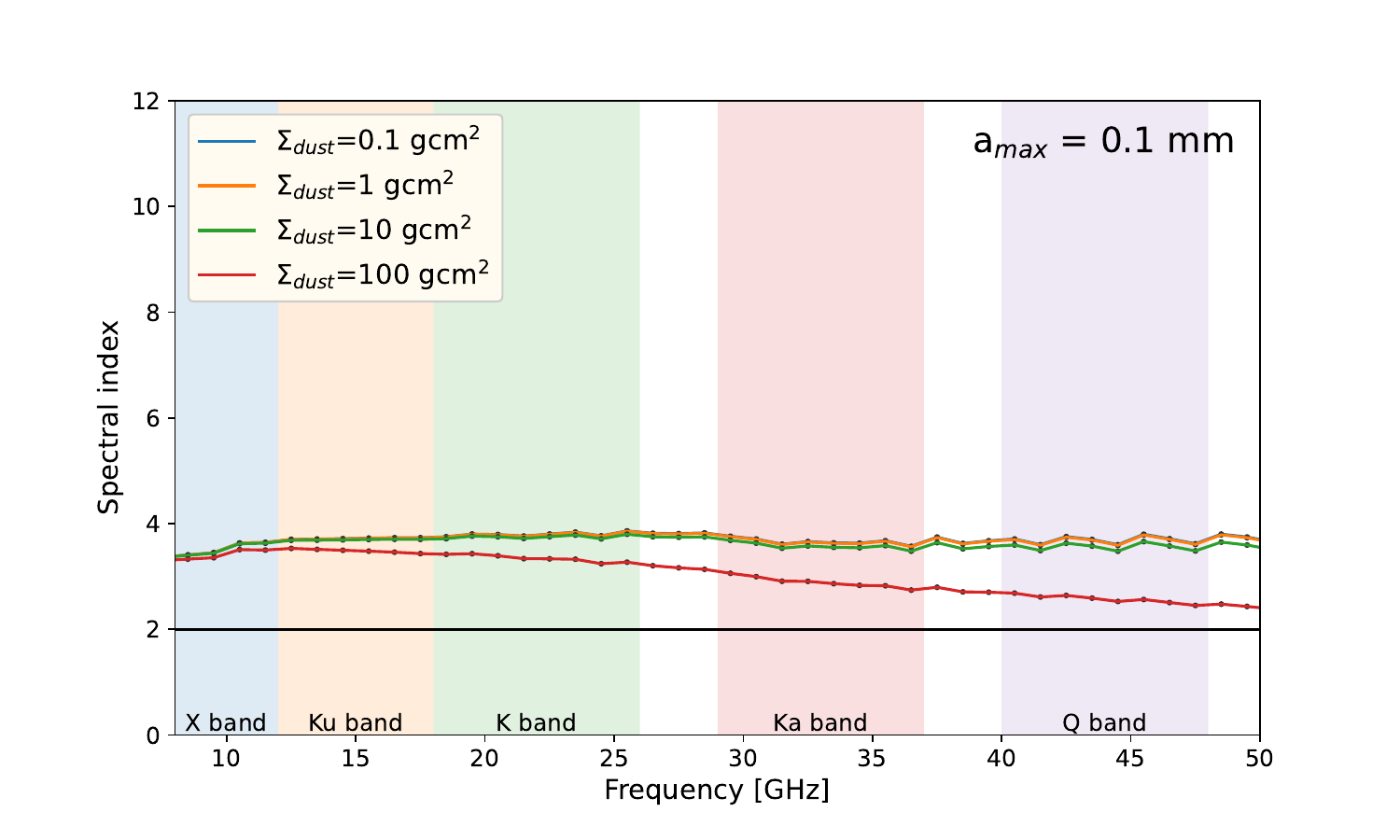} \\
    \end{tabular}
    
    \hspace{-1cm}
    \begin{tabular}{ p{8.5cm} p{8.5cm} }
        \vspace{-1.2cm}\includegraphics[height=6cm]{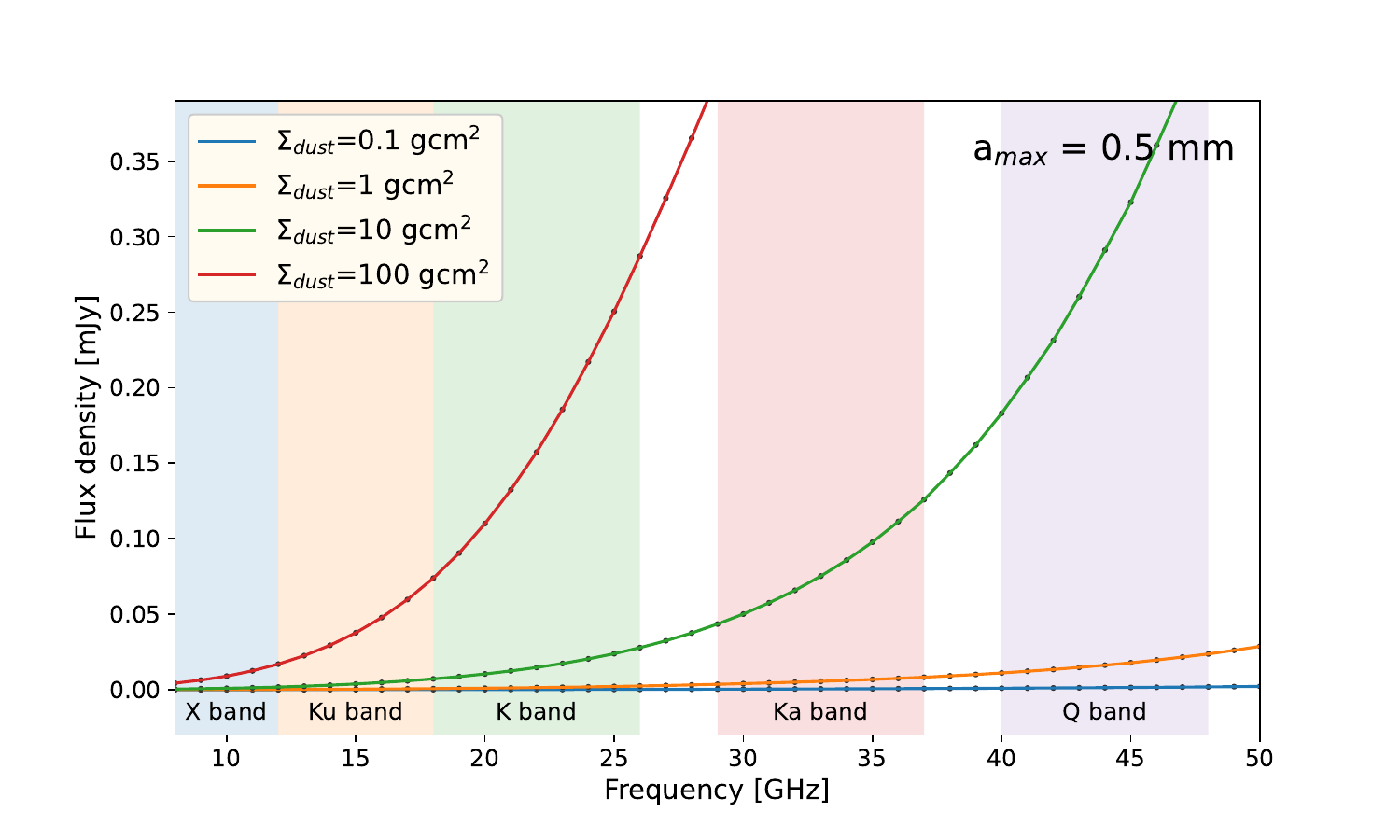} &  
        \vspace{-1.2cm}\includegraphics[height=6cm]{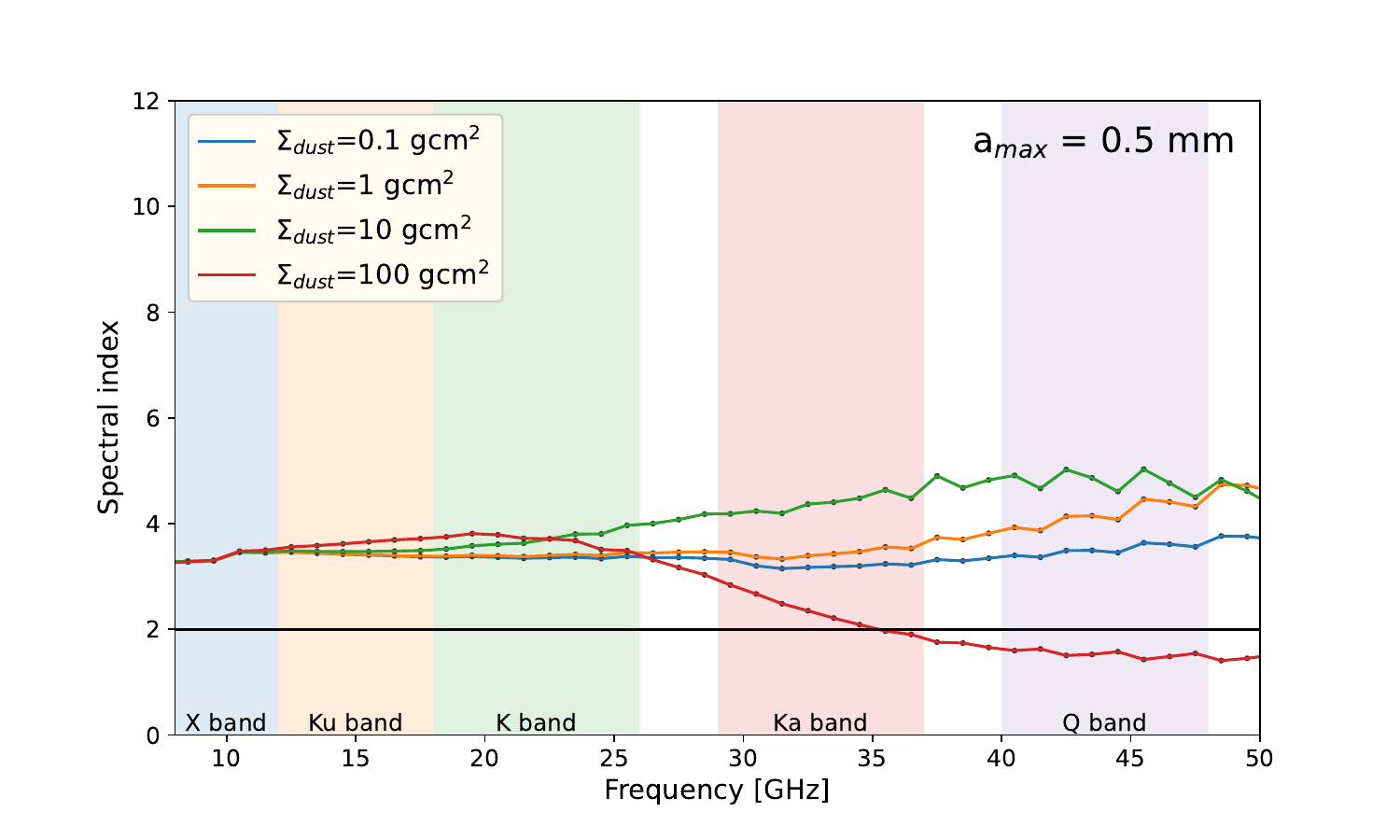} \\
    \end{tabular}
    
    \hspace{-1cm}
    \begin{tabular}{ p{8.5cm} p{8.5cm} }
        \vspace{-1.2cm}\includegraphics[height=6cm]{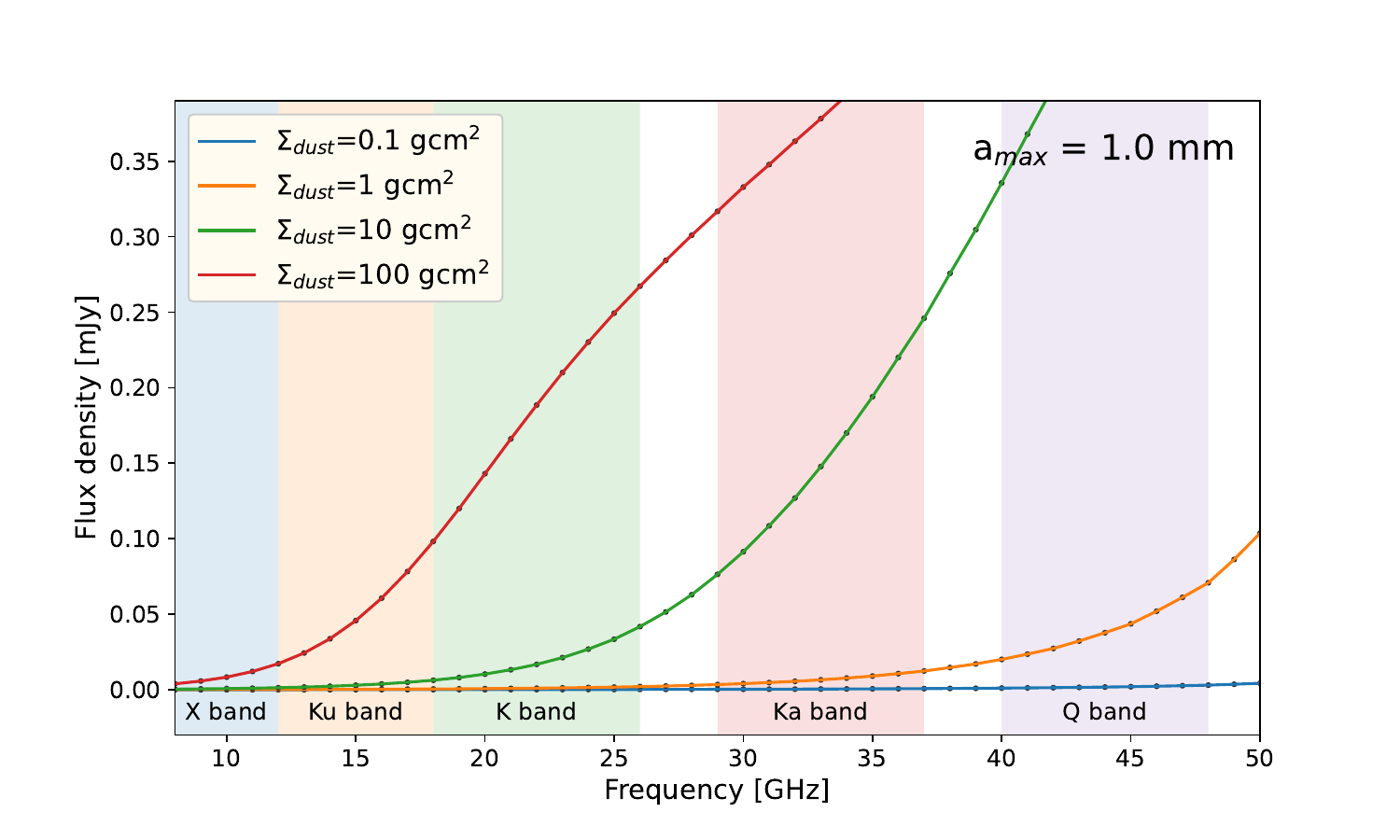} &  
        \vspace{-1.2cm}\includegraphics[height=6cm]{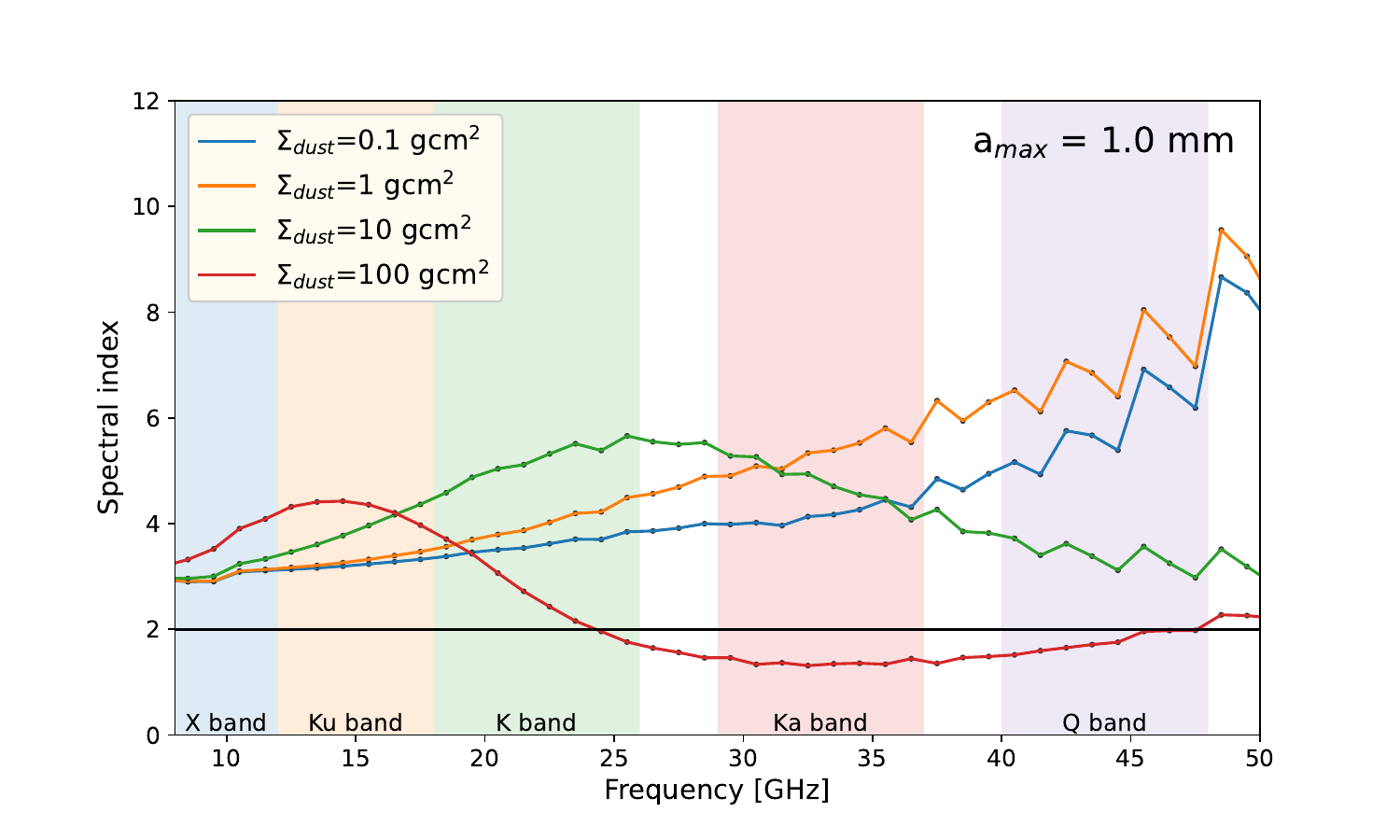} \\
    \end{tabular}
    
    \hspace{-1cm}
    \begin{tabular}{ p{8.5cm} p{8.5cm} }
        \vspace{-1.2cm}\includegraphics[height=6cm]{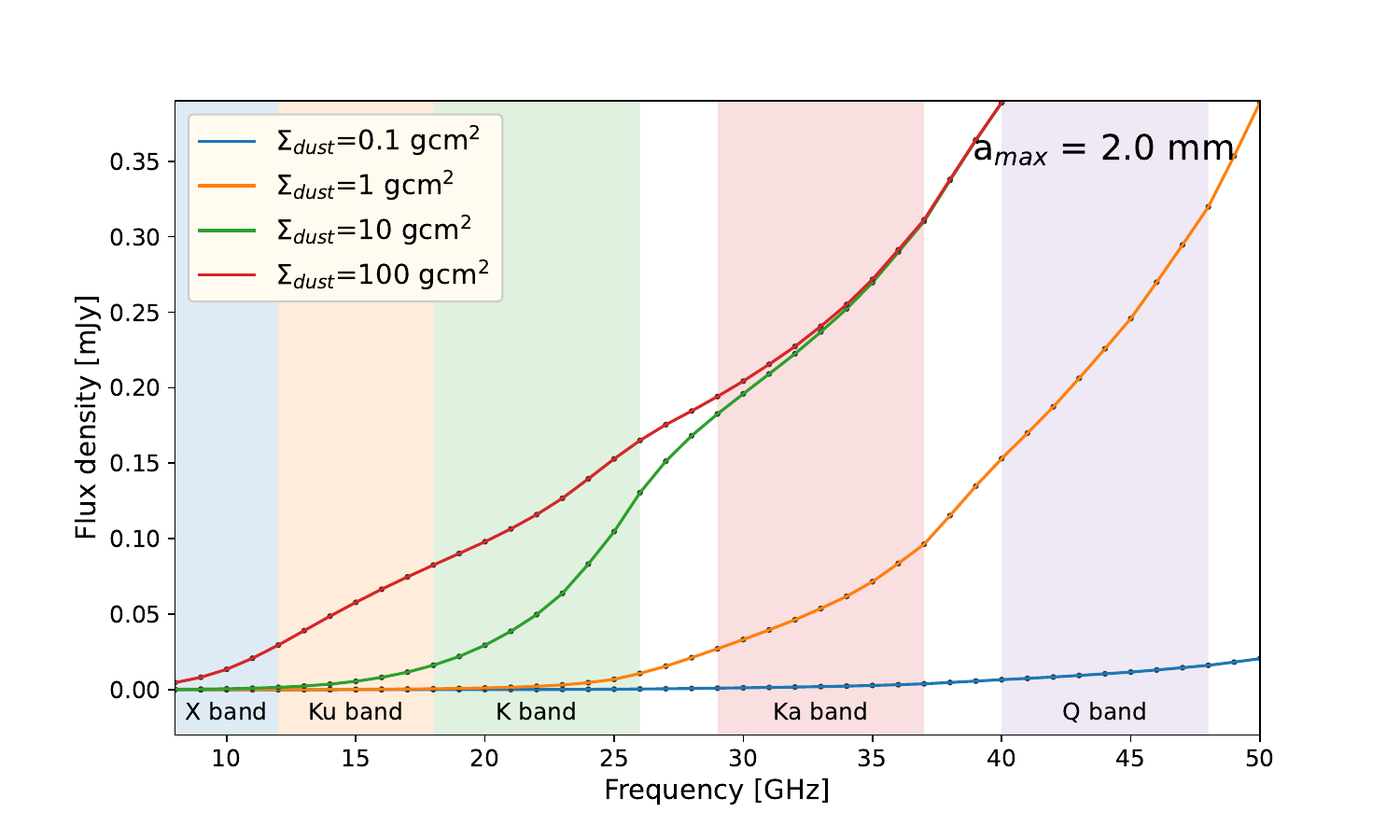} &  
        \vspace{-1.2cm}\includegraphics[height=6cm]{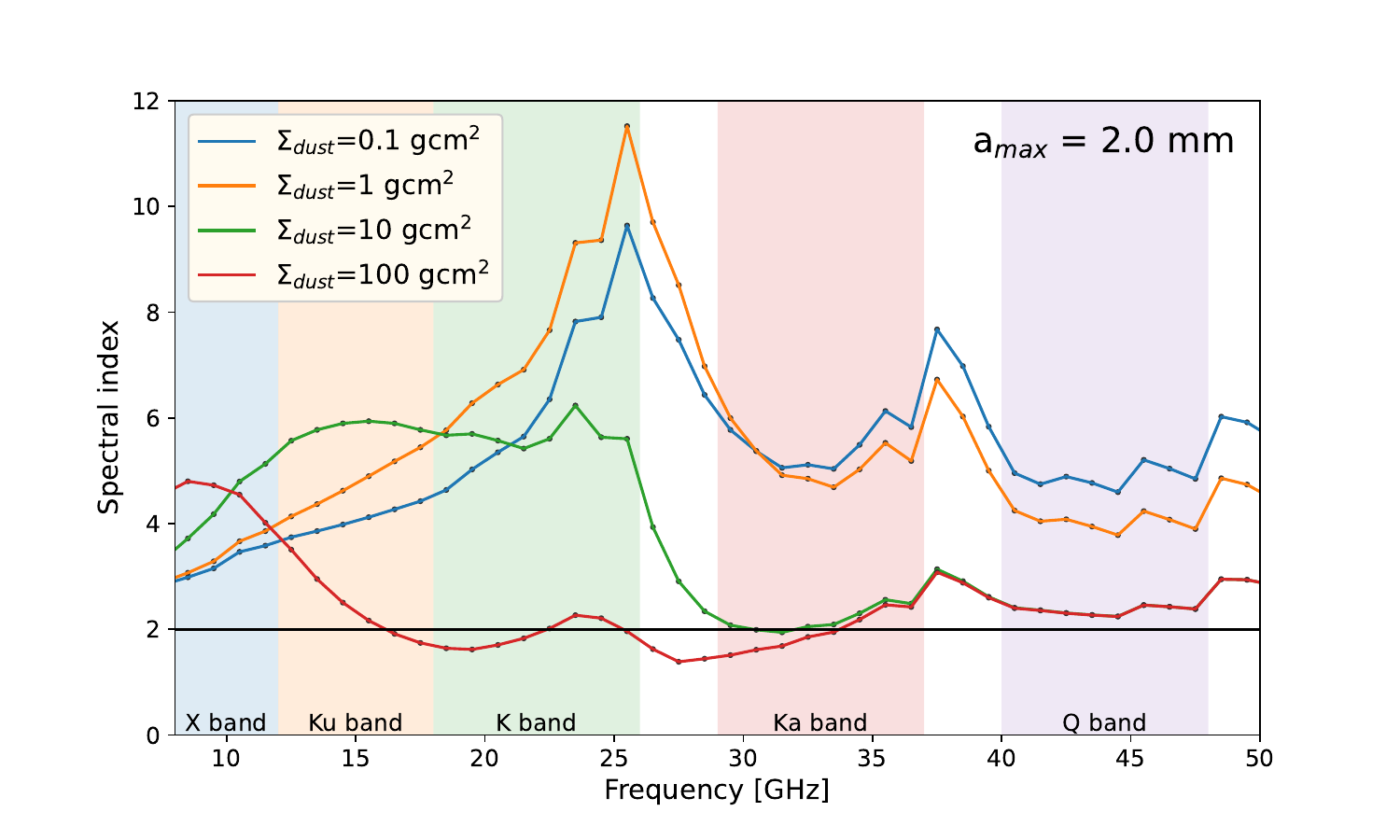} \\
    \end{tabular}
    
    \caption{Flux densities ({\it left}) and spectral indices ({\it right}) of a circular, 350 K, 10 au radius face-on dust slab which has a \amax$=$0.1, 0.5, 1, 2 mm maximum grain size (\amax) and the 0.1, 1, 10, and 100 g\,cm$^{-1}$ dust column density ($\Sigma_{\mbox{\scriptsize dust}}$). The assumed distance is 407.5 pc. These were evaluated based on the assumption of the water-ice free DSHARP opacity table (\citealt{Birnstiel2018}). The evaluations based on the default (ice-coated) DSHARP opacity table are qualitatively similar to this figure. }
    \label{fig:hypothesis}
\end{figure*}

Comparing the upper panel of Figure \ref{fig:modelapha} with Figure \ref{fig:hypothesis} also shows that it is very difficult to fit our JVLA observations with \amax$\lesssim$1 mm in FUOri\_dust1, which will yield too low flux densities and too high spectral indices at 8--48 GHz unless the assumed $\Sigma_{\mbox{\scriptsize dust}}$ is as high as a few times 10$^{2}$ g\,cm$^{-2}$.
Such a high $\Sigma_{\mbox{\scriptsize dust}}$ value implies that the overall dust mass within the $\sim$10 au radius of FUOri\_dust1 is a few times 10$^{3}$ Earth-mass (M$_{\oplus}$).
With a conventionally assumed gas-to-dust mass ratio of 100, the corresponding overall disk mass is higher than 0.3 $M_{\sun}$.
This seems unrealistic given that the mass of the host protostar itself may be only 0.3--0.5 $M_{\sun}$ (\citealt{Zhu2007ApJ...669..483Z}).

We conclude that the mid-plane of the inner $\sim$10 au FU\,Ori disk is populated with grown dust with \amax $\gtrsim$1--3 mm.
We do not rule out the presence of components with larger grain sizes (e.g., $>$1 cm).
To robustly test grain growth at the regime of \amax$\gtrsim$1 cm, the observations need to better sample the $\gtrsim$6 cm wavelengths (i.e., $\lesssim$5 GHz frequencies) which will require the sensitivity and angular resolution of the next generation Very Large Array (ngVLA; \citealt{Murphy2018ASPC}) or the Square Kilometer Array (SKA; e.g., \citealt{Ilee2020MNRAS.498.5116I}).

Our new model still has tension with the 29--37 GHz observations taken in 2016 (Figure \ref{fig:fuoriFluxPre2019}) because it yields a slight over-intensity of dust emission in that frequency range.
The discrepancy between the new model and the data is small enough to be attributed to absolute flux calibration errors of the JVLA observations or the attenuation of the observed flux density due to the delay bug in those previous JVLA observations (Section \ref{sec:intro}).
There is no strong evidence that observational/calibration effects significantly impacted our analyses because such tension did not occur in the fits for FU\,Ori\,S (Figure \ref{fig:fuoriFluxPre2019}). 
Otherwise, such a discrepancy may be explained by that the $T_{\mbox{\scriptsize dust}}$ of the FUOri\_dust1 component (i.e., averaging over the disk mid-plane in the inner 10 au) is up to $\sim$10\% higher in 2020 than in 2016.
Incorporating this possibility in our MCMC fits (Section \ref{subsub:components}, \ref{subsub:optimization}) by lowering the $T_{\mbox{\scriptsize dust}}$ of the FUOri\_dust1 component in 2016 by 10\% will not yield any non-trivial implication, and thus was not implemented.

The possibility that at least some parts in the inner $
\sim$10 au region of the FU\,Ori disk were warming up may be supported by the optical monitoring observations which show lower {\it V} band magnitude in 2020 than in 2016 September/October.
The positive correlation between the Johnson {\it B-R} color and the {\it V} band magnitude (Figure \ref{fig:color}; Section \ref{sub:opticalresult}) also supports that the FU\,Ori disk, which has been the dominating optical emission source, was warming up during 2020-2021.
So how we can interpret the increasing trend of the Johnson {\it R} band magnitude (Figure \ref{fig:oirmonitor}) and the complicated trends of the Johnson {\it B-V}, {\it V-R}, and {\it B-R} colors in 2020-2021?
A tentative hypothesis is that the outer, lower-temperature regions of the disk that were dominating the Johnson {\it R} band magnitude was cooling down.
Such cooling reduced the thermal pressure, which subsequently led to inflows of gas that induced heating (by compression, viscous heat dissipation, or shocks) in an inner region that was dominating the Johnson {\it V} band magnitude.
The gas inflows may be thermally ionized and may enhance the accretion and ionizing irradiation of the host protostellar, leading to enhanced free-free emission.
The regions which are dominating the optical emission were on $\lesssim$1 au spatial scales (\citealt{Labdon2021A&A}) where the dynamic timescales are comparable with the timescales of the photometric monitoring presented in Figure \ref{fig:oirmonitor}.
Alternatively, the brightening in the optical {\it BV} bands may also partly be attributed to the reduction of dust extinction. 
This can be due to the depletion of small dust grains, which can either be because of the ongoing dust grain growth or enhanced sublimation.

Modeling the optical light curves requires self-consistently solving the vertical and radial thermal and small-dust density structures, which is beyond the scope of the present paper.
Uncertainty in our interpretation is that there might be a systematic bias between the ASAS-SN, SLT, and NOWT data calibration although there is no evidence to demonstrate that this is indeed the case.
The lower Johnson {\it V} band magnitude in 2020 could also be due to that FU\,Ori\,S instead of FU\,Ori was brightened (Subsection \ref{subsub:fuoriSresult}).
To discern these possibilities, it is necessary to conduct new multi-frequency monitoring observations.

\subsubsection{FU\,Ori\,S}\label{subsub:fuoriSresult}
The newly derived $T_{\mbox{\scriptsize dust}}$ and $\Omega_{\mbox{\scriptsize dust}}$ for FUOriS\_dust1 are also consistent with the previous derivation of \citet{Liu2019ApJ...884...97L}.
The JVLA observations taken in 2016 and 2020 did not provide a strong constraint on the \amax of FUOriS\_dust1 due to the very significant contribution of free-free emission relative to dust emission at 8--48 GHz.
As a consequence, the marginal posterior distribution of \amax does not present an obvious peak (e.g., as the case of FUOri\_dust1; see Figure \ref{fig:corner}).
Instead, the distribution is skewed towards $<$0.2 mm grain sizes such that the 50th percentile of our samplers (i.e., given in Table \ref{tab:dustmodel}) should be regarded as a loose upper limit.

The upper limit of \amax is (weakly) constrained due to that FUOriS\_dust1 dominates the emission at 90--230 GHz where the observed spectral indices were close to 2.0 (\citealt{Liu2019ApJ...884...97L}).
For a \amax value of $\sim$0.2 mm, the effect of dust scattering helps suppress the spectral index at 90--230 GHz, making FUOriS\_dust1 fit the previous ALMA observations more easily.
However, if \amax of FUOriS\_dust1 is $\gtrsim$0.3 mm, the dust self-scattering will yield the too significant anomalously high/low spectral indices at 90--230 GHz (c.f. discussion in Appendix Section \ref{appendix:dustspid}), leading to tension with the previous ALMA observations.

Due to the high noise of our new JVLA observations at 29--48 GHz, the \amax value of FUOriS\_dust1 is marginally consistent with $<$0.1 mm. 
However, when \amax is smaller than 0.1 mm, the effect of dust scattering is very week at 90--230 GHz such that it requires a very high dust column density to yield a $\sim$2.0 spectral index at 90--230 GHz.
The high column density will in turn lead to a high contribution of optically thin dust emission at 29--48 GHz, making the spectral index of our model approach the high-end allowed by the present observations.
We expect the parameter space (for FUOriS\_dust1) with \amax$<$0.1 mm to be ruled out once the S/N at 29--48 GHz is improved.
Therefore, in our second round of MCMC fits, we artificially forbade such possibilities in order to better resolve the more probable parameter space.
Tightening the lower limit of \amax can also be achieved by obtaining additional $\lesssim$5 au resolution observations at 345--700 GHz (e.g., using ALMA).
Finally, we cannot rule out the possibility that there is $\gtrsim$1 mm sized dust in the mid-plane of the FU\,Ori\,S disk, which may only be robustly diagnosed when the free-free emission become a few times weaker.

As compared with the fits of \citet{Liu2019A&A...624A.114L}, the  $\Sigma_{\mbox{\scriptsize dust}}$ of FUOriS\_dust1 is reduced from $\sim$32 g\,cm$^{-2}$ to $\sim$19 g\,cm$^{-2}$ (Table \ref{tab:dustmodel}).
The update of $\Sigma_{\mbox{\scriptsize dust}}$ does not affect the fits at $>$50 GHz due to that FUOriS\_dust1 is optically thick at high frequency (see also Appendix \ref{appendix:loglog}).
The effect of lowering the value of $\Sigma_{\mbox{\scriptsize dust}}$ is significantly reducing the contribution of dust emission at 8--48 GHz which is compensated by the brighter free-free emission component in our new fits.

Our present best fit may yield a too high spectral index at 29--48 GHz as compared with the observations taken in 2020 July and November 15.
This is not a fundamental failure of our modeling strategy.
A better fit to the data can be achieved by dividing FUOriS\_dust1 into two internal components with different $\Sigma_{\mbox{\scriptsize dust}}$ and \amax values (but without changing the overall projected solid angle): the spatially more compact one with a higher \amax value ($\sim$0.5--1 mm) efficiently obscures the free-free emission at $\sim$30--50 GHz while the spatially more extended one with \amax$\lesssim$0.2 mm dominates the dust emission at $\sim$90--250 GHz.
In fact, there has been some hint that this should be the case, which was why the index 2 of the FUOriS\_dust components has been reserved (in \citealt{Liu2019ApJ...884...97L} and the present work).
Physically, this can be illustrated either by a significant radial dependence of either $\Sigma_{\mbox{\scriptsize dust}}$ and/or \amax within FUOriS\_dust1 (e.g., due to the inward migration of grown dust).
Since the averaged $T_{\mbox{\scriptsize dust}}$ of FUOriS\_dust1 (Table \ref{tab:dustmodel}) is close to the 150--170 K sublimation temperature of water-ice, another appealing conjecture would be that $\gtrsim$1 mm sized dust grains are forming inward of the snowline of FU\,Ori\,S.
In our baseline experimental design, we expected these two dust components to be separable in the SED fits if the achieved noise is low enough such that we can image the 29--37 GHz and 40--48 GHz bands in each 2 GHz baseband (e.g., the $<$29 GHz data were too dominated by free-free emission to help constrain dust properties).
With the noise level we actually achieved, sub-dividing FUOriS\_dust1 into two internal components will yield a larger number (8) of free parameters than the (5) constraints (JVLA Ka, Q bands and ALMA Band 3, 4, 6), making the MCMC fits too degenerate to be comprehensive.

In our best fits, the free-free emission component is optically thick at $\lesssim$20-30 GHz and becomes optically thin at the higher frequency where dust emission becomes dominant.
This leads to a minimized spectral indices at 20--30 GHz (see the right panel of Figure \ref{fig:modelapha}).
The spectral indices observed in 2020 are lower than those observed in 2016 September/October due to that the brightened free-free emission in 2020.
This might be related to a temporarily enhanced accretion rate of FU\,Ori\,S that resulted in enhanced ionizing radiation.
This also partly or fully explains the enhanced emission of this binary system at  Johnson {\it V} band (Figure \ref{fig:oirmonitor}; Section \ref{sub:opticalresult}).
The values of $\Omega_{\mbox{\scriptsize ff}}$ and EM of FUOriS\_freefree may both be changing with time.
This may also be the case for FU\,Ori, although the degeneracy in determining the parameters of free-free emission is higher in the source which is dominated by dust emission.

\subsection{Implication}\label{sub:implication}
The origin of the $>$1 mm sized dust grains in the hot inner disk of FU\,Ori remains uncertain.
These millimeter-sized dust grains survived the high temperature in the inner few au FU\,Ori disk without fragmenting back down to smaller sizes (e.g., $\lesssim$100 $\mu$m), supporting that pebbles, planetesimals, and/or the Earth-like terrestrial planets can form in-situ (e.g., for a thorough discussion see \citealt{Liu2020RAA....20..164L} and references therein).

If water-ice free dust grains are fragile (e.g.,  \citealt{Gundlach2011Icar,Gundlach2015ApJ...798...34G}), then the coagulation or survival of the millimeter-sized grains may require the inner few au FU\,Ori disk to be not turbulent (e.g., maintaining a magnetorotational instability (MRI) dead zone such that dust grains do not collide with each other at high velocities; for an example see model 2 of \citealt{Molyarova2021ApJ...910..153M}; c.f., \citealt{Vorobyov2020A&A...644A..74V}).
This might be contradictory with the observations of the $T\propto r^{-0.75}$ radial temperature profile (\citealt{Liu2019ApJ...884...97L,Labdon2021A&A}) which can be consistent with that the FU\,Ori disk is  turbulent inward of the $\lesssim$10 au radius ($r$).
The comparison of the $<$1 year rise time of FU\,Ori with numerical simulations
(e.g., Figure 2 in \citealt{Kadam2020ApJ...895...41K} and Figure 7 \citealt{Vorobyov2020A&A...644A..74V}) also indicate that at least the innermost part of the FU\,Ori disk is very viscous (which maybe due to that it is turbulent).

Otherwise, water-ice free dust grains might be sticky. The stickiness of dust grains can be characterized with the fragmentation velocity $v_{\mbox{\tiny frag}}$ that is higher for stickier grains (for a review see \citealt{Birnstiel2016SSRv..205...41B}).
Assuming that the maximum grain size is limited by the fragmentation barrier, the material density of dust grains $\rho_{\mbox{\scriptsize s}}=$3.0 g\,cm$^{-3}$, dust and gas are perfectly thermalized, then following Equation 34 of \citet{Birnstiel2016SSRv..205...41B}, we can derive the lower limits of $v_{\mbox{\tiny frag}}$ by assuming that the gas-to-dust mass ratio $\zeta_{\mbox{\scriptsize g2d}}$ is 100.
We note that in protoplanetary disks, in general, the values of $\zeta_{\mbox{\scriptsize g2d}}$ can potentially be considerably lower than 100 although the existing observational constraints are very uncertain (c.f., \citealt{Miotello2017A&A...599A.113M} and references therein).
Adopting the dust column density, temperature, and $a_{\mbox{\scriptsize max}}$ obtained from our SED fitting for FUOri\_dust1 (Table \ref{tab:dustmodel}), the assumptions of the Sunyaev \& Shakura viscous $\alpha_{\mbox{\scriptsize t}}$ values 0.01, 0.1, and 1.0 will correspond to the $v_{\mbox{\tiny frag}}$ lower limits of 2.3, 7.4, and 23 m\,s$^{-1}$, respectively.
These $v_{\mbox{\tiny frag}}$ values are already higher than that of the previously considered poorly sticky rocky grains ($\sim$1 m\,s$^{-1}$; \citealt{Blum2000Icar..143..138B}).
If in reality the effective $\alpha_{\mbox{\scriptsize t}}$ is $\gtrsim$0.1 (see Appendix C for some hypothesis about the values of $\alpha_{\mbox{\scriptsize t}}$), our lower limit of \amax will support the results of the latest analytical calculations (\citealt{Kimura2015ApJ}) and laboratory experiments (\citealt{Gundlach2018MNRAS,Steinpilz2019ApJ,Musiolik2019ApJ}).
Conversely, if we adopt the $v_{\mbox{\tiny frag}}$ indicated by the latest laboratory experiments (\citealt{Gundlach2018MNRAS,Steinpilz2019ApJ,Musiolik2019ApJ,Pillich2021A&A...652A.106P}), then we might consider the inner $\sim$10 au region of FU\,Ori to be turbulent with an effective $\alpha_{\mbox{\scriptsize t}}$ in the range of 0.1--1.
Finally, we note that \citet{Hubbard2017ApJ...840L...5H} pointed out that the molten dust grains in the specific environment where the temperature is modestly above 1000 K may be sticky enough to bypass the bouncing and fragmentation barriers.

\subsection{How to improve this experiment?}\label{sub:future}

Growing evidence has shown that the radio and/or (sub)millimeter flux densities of FUors can vary with time (e.g., \citealt{Liu2018A&A,Johnstone2018ApJ,Francis2019ApJ,Wedeborn2020}).
In addition, our present work and the measurements published in \citet{Liu2019ApJ...884...97L} indicate that the spectral profiles of FUors can present complicated features at 8--350 GHz and higher frequencies, although the interpretation for the higher frequency measurement may be ambiguous to some extent in any case due to the high optical depths.

To robustly diagnose \amax and dust column density within the snowline of FUors, we suggest that it is important to obtain sensitive and simultaneous flux density measurements at 8--50 GHz, accompanied by some SED measurements at 90--350 GHz.
Considering the sensitivities and the site (weather) condition of the present and upcoming facilities, a feasible approach in the near future is to use ALMA Band 1 (\citealt{Huang2012SPIE}) to observe at 30--50 GHz on a target source for a few hours, coordinating with triggered JVLA observations to take lower-frequency data at the same time.
The JVLA observations over a few hours' duration can interleave with X, Ku, and K band observations. 
The ALMA observations can be complemented by snapshots at Band 3 or 4 and snapshots at Band 6 or 7\footnote{For this science case, snapshots are sensitive enough for the ALMA Band 3--10 receivers. The limitation is that the ALMA cryogenic system only permits activating 3 receivers at a time. One possibility to permit simultaneously observing at more frequency bands in the future may be developing the sub-arrays capability.}.

The development of wider bandwidth capabilities (e.g., ALMA Band 2+3, \citealt{Gonzalez2016} and the coordination with the next generation facilities including the ngVLA (e.g., \citealt{Andrews2018ASPC..517..137A}) and SKA (e.g., \citealt{Braun2015aska.confE.174B,Carilli2015aska}) will further improve the precision of these studies and  increase schedule flexibility.
Combining the observations of the ngVLA and SKA, one can achieve (sub)milli-arcsecond angular resolutions and $\lesssim$1 $\mu$Jy\,beam$^{-1}$ rms noise at centimeter bands for a large number of protoplanetary disks, which will be particularly elucidating for the physical processes of pebble formation.

\section{Conclusion} \label{sec:conclusion}
We observed the accretion outburst YSO, FU\,Ori, and its companion FU\,Ori\,S, using the JVLA at X (8--12 GHz), Ku (12--18 GHz), K (18--26 GHz), Ka (29--37 GHz), and Q (40--48 GHz) bands around 2020 July, at Ka band on 2020 September 30, and at K and Ka bands on 2020 November 15.
We have performed the complementary Johnson {\it BVR} bands optical photometric monitoring observations using the SLT and NOWT.
We found that:

\begin{enumerate}
    \item The new radio observations and the previous radio and (sub)millimeter observations towards FU\,Ori indicate that the maximum dust grain size \amax is $\gtrsim$1.6 mm in its inner $\sim$10 au, $\sim$400 K hot inner disk where water-ice likely has been sublimated. Grown dust either formed in-situ in the hot inner disk, otherwise migrated to the hot inner disk and survives the high temperature.
    We note that the latter case is not inconsistent with the theoretical timescale estimates of  \citet{Schoonenberg2017A&A...605L...2S} since it does not require grown dust to start piling up in the inner $\sim$10 au region only after the onset of the FU\,Ori outburst.
    \item The constrained lower limit of \amax in the FU\,Ori hot inner disk implies that either this region remains to harbor an MRI dead zone with low turbulence (i.e., corresponding to $\alpha_{\mbox{\scriptsize t}}<$0.01), or the fragmentation velocity $v_{\mbox{\tiny frag}}$ of the water-ice free dust grains in the protoplanetary disks is considerably higher than that of the previously considered poorly sticky rocky grains (1 m\,s$^{-1}$; \citealt{Blum2000Icar..143..138B}). The latter is true if the actual value of \amax is higher than the lower limit derived from our observations, the hot inner disk is turbulent (e.g., $\alpha_{\mbox{\scriptsize t}}\gtrsim$0.01), or if the gas-to-dust mass ratio is lower than 100 (which are not mutually exclusive).
    The recent analytical calculations (\citealt{Kimura2015ApJ}) and laboratory experiments (\citealt{Gundlach2018MNRAS,Steinpilz2019ApJ,Musiolik2019ApJ,Pillich2021A&A...652A.106P}) indicated that $v_{\mbox{\tiny frag}}$ of the water-ice free dust grains in the protoplanetary disks can be $\gtrsim$10 m\,s$^{-1}$.
    \item The observed \amax value in the inner $\sim$10 au FU\,Ori\,S disk ($T_{\mbox{\scriptsize}}\sim$150 K) is $\lesssim$100--200 $\mu$m. Due to the confusion of the relatively strong free-free emission in FU\,Ori\,S at low frequencies, we cannot rule out the possibility that there are millimeter-sized or larger dust grains in the disk mid-plane. We cannot verify this possibility either. 
    \item The free-free emission in FU\,Ori\,S is brighter in 2020 than in 2016 September, leading to the significantly varying 8--48 GHz flux densities. The optical Johnson {\it V} band magnitude of the binary system also presents small variations and might be slightly brighter in 2020 than in 2016. These radio and optical variabilities may be partly attributed to the variable accretion activities of FU\,Ori\,S although they appeared not as dramatic as the onset of the FU\,Ori outburst.
    \item The 8--48 GHz flux density of FU\,Ori might also be higher in 2020 than in 2016 October although we cannot rule out the possibility that this is due to some observational or calibration artifacts. Otherwise, the observed 8--48 GHz variability may be explained by very slight temperature variation of the FU\,Ori disk, or the formation of some optically thick free-free emission knots around the disk. These two possibilities are not mutually exclusive and may be both related to viscous heating, adiabatic compression, or shock in the FU\,Ori disk.
\end{enumerate}

Coagulating water-ice coated dust grains, although is taking place in the environment of molecular clouds (e.g., \citealt{Chen1993ApJ...409..319C}) is not necessarily the most efficient or the only mode of dust grain growth in protoplanetary disks.
The importance of our result is that it provides a {\it possibility} to understand why the terrestrial planets (e.g., Earth, Mars) and the asteroid-belt objects in the inner Solar System are deficient in  water.


\begin{acknowledgments}
The National Radio Astronomy Observatory is a facility of the National Science Foundation operated under cooperative agreement by Associated Universities, Inc.
This paper makes use of the following ALMA data: ADS/JAO.ALMA \#2011.0.00548.S,  \#2016.1.01228.S, and \#2017.1.00388.S. ALMA is a partnership of ESO (representing its member states), NSF (USA) and NINS (Japan), together with NRC (Canada), MOST and ASIAA (Taiwan), and KASI (Republic of Korea), in cooperation with the Republic of Chile. The Joint ALMA Observatory is operated by ESO, AUI/NRAO and NAOJ.
This work has made use of data from the European Space Agency (ESA) mission
{\it Gaia} (\url{https://www.cosmos.esa.int/gaia}), processed by the {\it Gaia}
Data Processing and Analysis Consortium (DPAC,
\url{https://www.cosmos.esa.int/web/gaia/dpac/consortium}). Funding for the DPAC
has been provided by national institutions, in particular the institutions
participating in the {\it Gaia} Multilateral Agreement.
This work is based [in part] on observations made with the Spitzer Space Telescope, which is operated by the Jet Propulsion Laboratory, California Institute of Technology under a contract with NASA.
Herschel is an ESA space observatory with science instruments provided by European-led Principal Investigator consortia and with important participation from NASA.
The Submillimeter Array is a joint project between the Smithsonian Astrophysical Observatory and the Academia Sinica Institute of Astronomy and Astrophysics, and is funded by the Smithsonian Institution and the Academia Sinica (\citealt{Ho2004}).
H.B.L. is supported by the Ministry of Science and
Technology (MoST) of Taiwan (Grant Nos. 108-2112-M-001-002-MY3 and 110-2112-M-001-069-).
M.T. is supported by the Ministry of Science and Technology (MoST) of Taiwan (grant No. 106-2119-M-001-026-MY3, 109-2112-M-001-019, 110-2112-M-001-044).
M.T., S.Y.L., and H.B.L. are supported by the Ministry of Science and Technology (MoST) of Taiwan (Grant Nos. 108-2923-M-001-006-MY3).
E. I. Vorobyov and V. Elbakyan were supported by the
Russian Fund for Fundamental Research, Russian-Taiwanese project
19-52-52011.
S.P. acknowledges support ANID/FONDECYT Regular grant 1191934.
Y.-L. Yang acknowledges the supports from the Virginia Initiative of Cosmic Origins Postdoctoral Fellowship.
\end{acknowledgments}

\facilities{JVLA, ALMA, LOT-1m, SLT-0.4m, TAOS/BEST, NOWT}


\software{
          astropy \citep{2013A&A...558A..33A},  
          Numpy \citep{VanDerWalt2011}, 
          CASA \citep[v5.6.2; ][]{McMullin2007},
          emcee \citep{Foreman-Mackey2013PASP},
          corner \citep{corner},
          PyAstronomy \citep{pya}
          }



\appendix

\section{Spectral index features related to millimeter sized dust} \label{appendix:dustspid}

When dust grains grow to millimeter sizes or beyond, the Rayleigh limit ($\lambda\sim2\pi$\amax) is shifted into the 0--50 GHz frequency coverage of the JVLA.
In this case, the effective dust scattering opacity (i.e., excluding forward scattering; $\kappa^{\mbox{\scriptsize sca,eff}}$) can become $\sim$1 order of magnitude higher than the absorption opacity $\kappa^{\mbox{\scriptsize abs}}$ (e.g., \citealt{Kataoka2015ApJ,Birnstiel2018,Liu2019,Zhu2019ApJ}).
The frequency variation of albedo $\omega$ (defined as $\kappa^{\mbox{\scriptsize sca,eff}}$/($\kappa^{\mbox{\scriptsize sca,eff}}$+$\kappa^{\mbox{\scriptsize abs}}$)) then starts playing a role in changing the spectral indices.
In this case, the shape of the dust SED in the frequency coverage of the JVLA is complicated because it is no longer always a monotonic function of frequency. 

As an example, assuming the DSHARP dust opacity tables appropriate for the studies of protoplanetary disks (\citealt{Birnstiel2018}), when \amax is 2--3 mm, $\omega$ increases with frequency at $\sim$30 GHz and decreases with frequency at higher frequencies (e.g., for some examples of dust opacities see Figure 2 of \citealt{Liu2019}).
As a result, the SED is anomalously flattened at $\sim$30 GHz and is anomalously steepened at $\sim$40--90 GHz.
At 10--30 GHz, the spectral indices may become very high if dust is not optically thick, which is related to a feature in $\kappa^{\mbox{\scriptsize abs}}$ (see also the discussion in \citealt{Pavlyuchenkov2019MNRAS}).
These features may lead to the presence of one or more spectral index {\it bump(s)} in the frequency domain.
For a broad range of dust column density, it is possible to constrain \amax by identifying these SED features from multi-frequency radio interferometric observations (Figure \ref{fig:hypothesis}).
We note that the qualitative spectral features induced by the high  $\kappa^{\mbox{\scriptsize sca,eff}}$ values are relatively fundamental, insensitive to dust composition as long as the dust grains are not carriers of free charges  (\citealt{Jackson1998clel.book.....J}).

On the other hand, with the broad 8--48 GHz frequency coverage of our experimental design (Section \ref{sub:jvlaobs}), it is appropriate to regard the spectral indices $\alpha(\nu)$ of the target sources as constants of frequency only if they fulfill the following two asymptotic conditions.
The first is the extremely optically thin, Rayleigh-Jeans, and small dust grain ($a_{\mbox{\scriptsize max}}<$1 mm) limit, where $\alpha(\nu)$ asymptotically approaches $\sim$3.8 (e.g., Figure \ref{fig:hypothesis}).
The second is the extremely optically thick and Rayleigh-Jeans limit with either very large or very small $a_{\mbox{\scriptsize max}}$ such that the Rayleigh criterion (in terms of frequency, c/($2\pi a_{\mbox{\scriptsize max}}$) ) is located well outside of this frequency coverage.  (e.g., \amax$\ll$1 mm or \amax$\gg$10 cm).
In this case, $\alpha(\nu)$ asymptotically approaches $\sim$2.0.
When the observed $\alpha(\nu)$ is apparently inconsistent with the aforementioned two limits, the value of $a_{\mbox{\scriptsize max}}$ is likely a few mm.
In this case, we can simultaneously constrain dust temperature, column density, and maximum grain size based on SED fits/models.

We note that the aforementioned two limits may not be particularly realistic for our present case study.
In the extremely optically thin limit, the expected flux density is very low throughout our frequency coverage which is unlikely to be detected.
In the extremely optically thick limit, in the case of either very small or very large \amax, throughout the 8--48 GHz frequency coverage, the dust absorption mass opacities are small. 
This is because very small dust grains cannot emit/absorb efficiently at long wavelengths while the surface area per unit mass is small in the case of very large dust grains.
Therefore, in such cases, a very high dust column density is required to achieve the optically thick constant $\alpha(\nu)=$2.0 limit over our frequency coverage (Figure \ref{fig:hypothesis}).
Moreover, in the case with \amax$\gg$10 cm the values of $\omega$ will be close to 1.0, which will lead to additional attenuation of the observed flux densities.
If we require the projected area of the disks to be large enough such that they are detectable, the implied overall dust mass by such a high dust column density may not be realistic.
Therefore, observing a constant $\alpha(\nu)=$2.0 over our frequency coverage may still favor that $a_{\mbox{\scriptsize max}}\gtrsim$1 mm although it may require mixing multiple dust or free-free emission components to interpret the SED.
In light of these concerns, there is no strong reason to assume that $\alpha(\nu)$ is a constant over the 8--48 GHz frequency range.

In case of misunderstanding, we note that when $\alpha(\nu)\neq$1.0 is constant in frequency, the flux densities presented in linear instead of log scales will trend up or down.
When the observed flux densities are consistent with a linear function with frequency, $\alpha(\nu)$ must vary with frequency unless that linear function passes through the origin with a slope consistent with 1.0.

\section{Radio--submillimeter SED and our best fit models} \label{appendix:loglog}
\begin{figure}
    \hspace{-1.5cm}
    \begin{tabular}{ p{9cm} p{9cm} }
        \includegraphics[width=9.8cm]{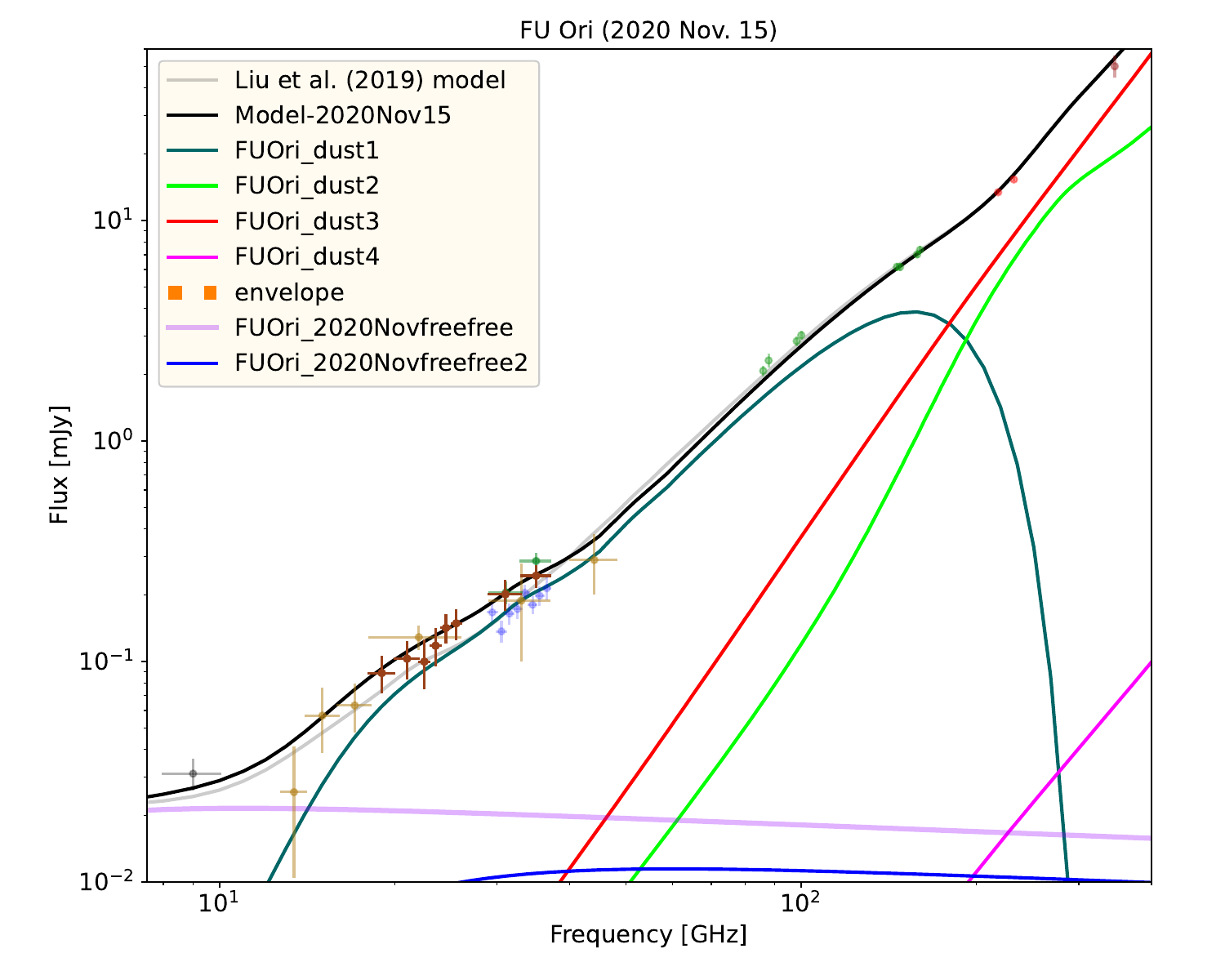} &  
        \includegraphics[width=9.8cm]{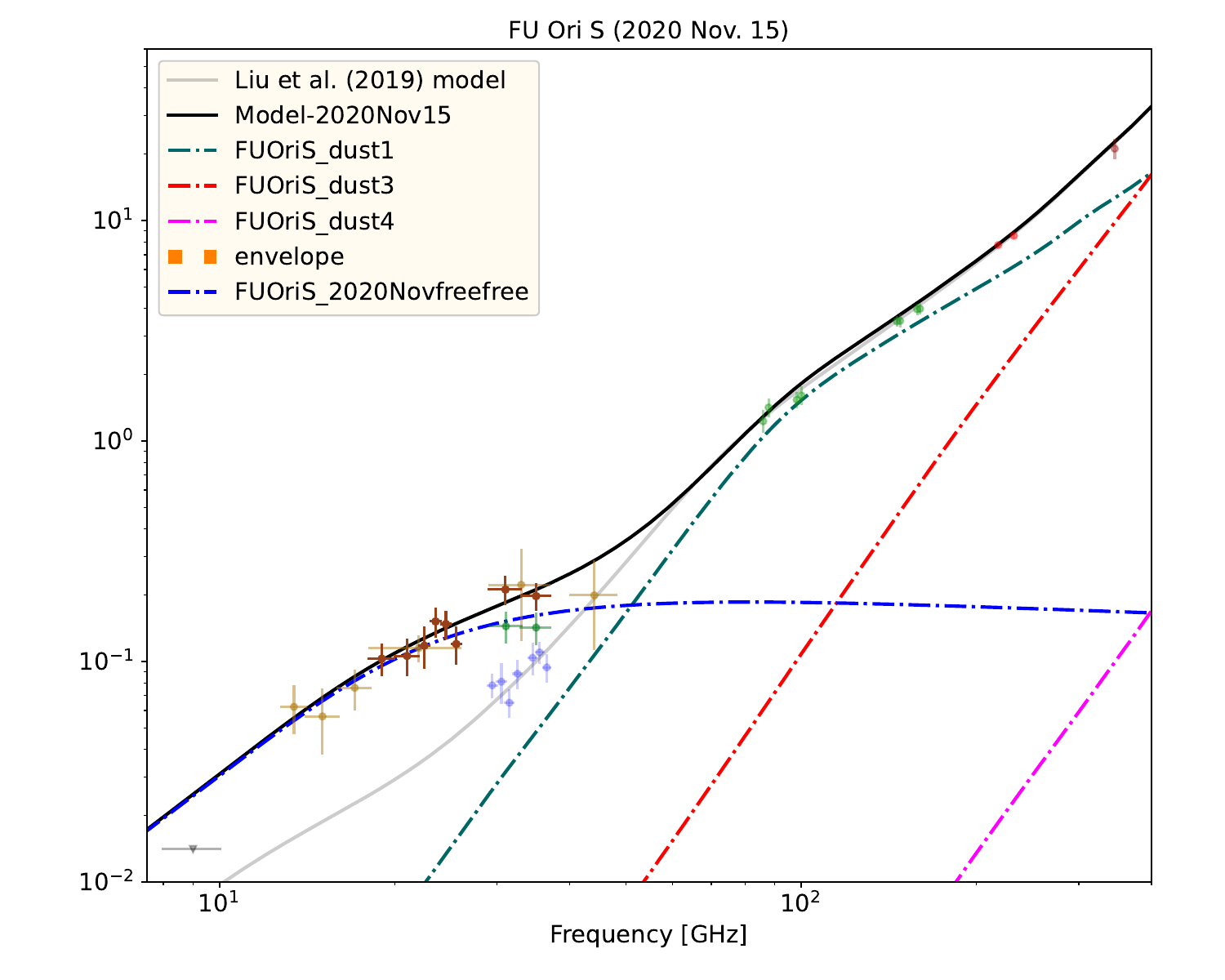} \\
    \end{tabular}
    \caption{
    The bottom panels of Figure \ref{fig:fuoriFlux} in a log-log scale, which additionally include the ALMA Bands 3, 4, 6 and 7 data reported by \citet{Hales2015ApJ,Liu2019ApJ...884...97L,Perez2020ApJ}.
    }
    \label{fig:fig2loglog}
\end{figure}

Figure \ref{fig:fig2loglog} shows radio and (sub)millimeter observations on FU\,Ori and FU\,Ori\,S taken with the JVLA and ALMA, which are over-plotted with our best fit models (Section \ref{sub:sed}, \ref{sub:sedresult}).

\section{Hypothesis about viscous heating and turbulence in the FU\,Ori hot inner disk} \label{appendix:turbulent}

In this section, we provide the rationales of why we hypothesize that the hot inner disk of FUOri\_dust1 may be turbulent (e.g., characterized by an $\alpha_{t}$ that is $>$0.01).
Although we understand that our estimates are very uncertain, we still think that this scenario may be possible.
The uncertainties are mainly caused by lacking some observational constraints on the essential physical parameters (mean and UV opacity, disk geometry/morphology, gas-to-dust ratio, etc), such that we have to base on various assumptions that are likely oversimplified. 

\begin{figure}
    \hspace{-2cm}
    \begin{tabular}{p{9cm} p{9cm} }
         \includegraphics[width=10cm]{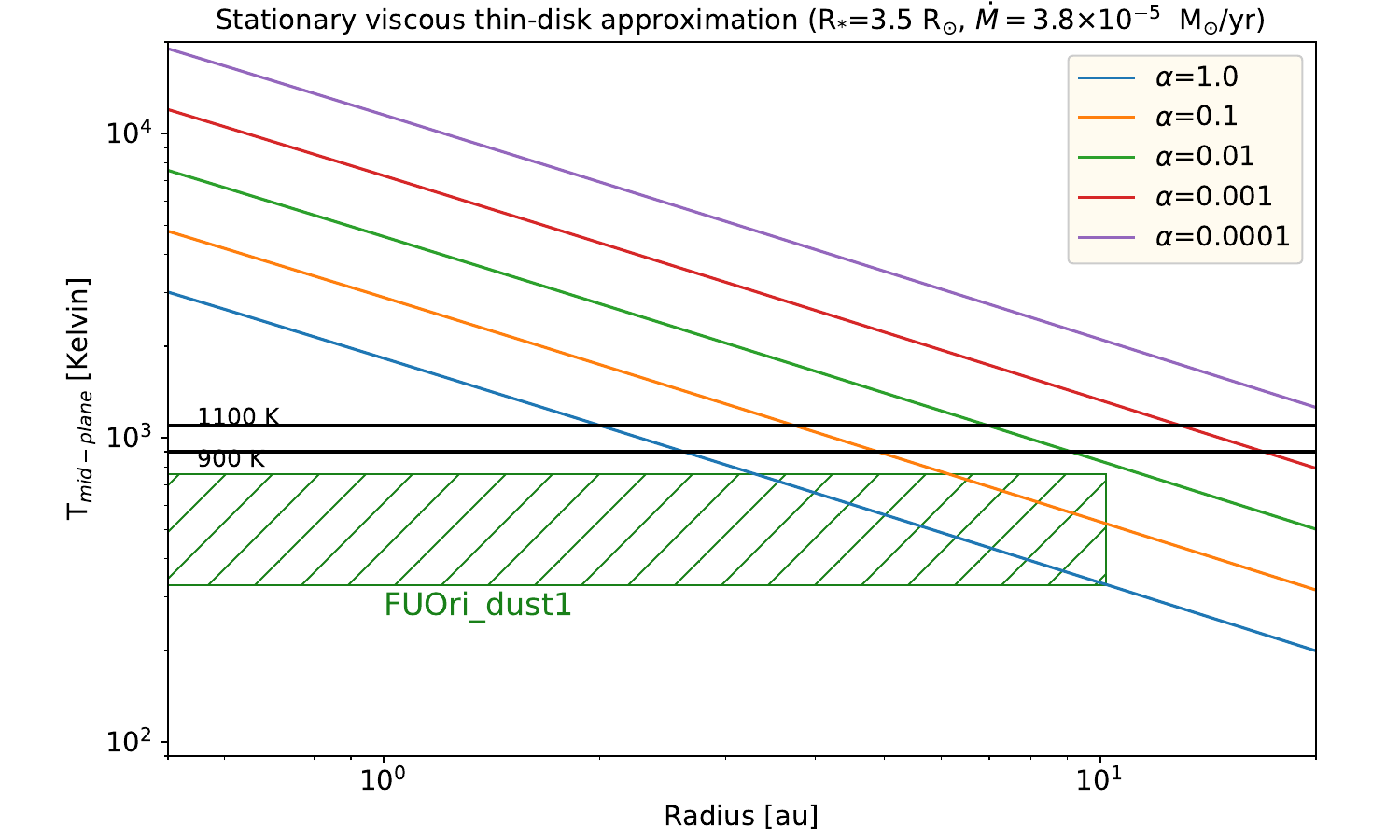} &
         \includegraphics[width=10cm]{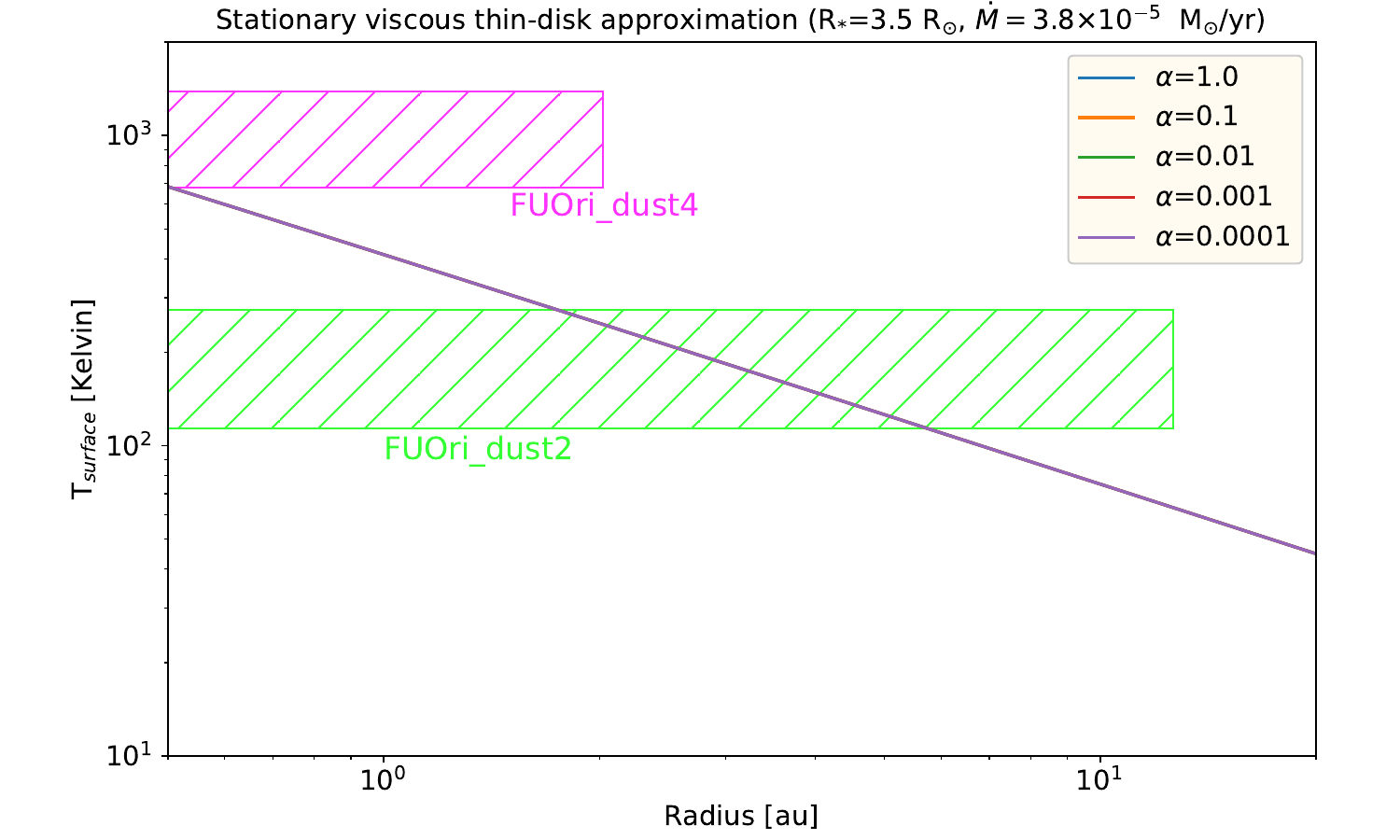} \\
    \end{tabular}
    \caption{Mid-plane ({\it left}) and surface ({\it right}) temperature profiles of a geometrically thin viscous accretion disk evaluated for $\alpha$=0.0001--1.0, which are compared with the measurements from the inner $\sim$10 au FU\,Ori disk. In the left panel, the 900-1100 K temperature range is labeled.}
    \label{fig:temperature}
\end{figure}

First, we assume that FUOri\_dust1 (see Figure \ref{fig:schematic}; Table \ref{tab:dustmodel}) is a standard steady Sunyaev \& Shakura viscous accretion thin-disk (\citealt{Shakura1973A&A....24..337S}) of which the disk accretion rate is a constant over the radius and is identical to the protostellar accretion rate.
The constant accretion rate helps understand why FU\,Ori is approximately stationary over a 10$^{2}$ years timescale (i.e., the mass at any radius is always sufficiently replenished).
Otherwise, we might expect significant variability on the timescales that are shorter than the dynamic (e.g., orbital, $\sim$50 years) timescales at the $\sim$10 au radius, which would be contradictory with the observations.
The viscosity is characterized by the variable $\alpha_{t}$ that is smaller than 1 (e.g., equal to 1 in the most MRI turbulent cases; \citealt{Shakura1973A&A....24..337S}).
These assumptions are not necessarily realistic. But without them, we are not able to proceed.
Here we quote the stellar radius and accretion rate from \citet{Perez2020ApJ} which might also have some uncertainties.
The derived mid-plane and surface temperature profiles for such models are presented in Figure \ref{fig:temperature}.

In these models, the surface temperature profile has no dependence on $\alpha_{t}$ given that it is merely determined by a balance between the radiative cooling rate and the rate to convert gravitational potential energy to heat.
The latter only depends on the accretion rate (i.e., there is no explicit dependence on the micro-physics which are assisting the accretion or inducing viscosity). 
This surface temperature is reasonably well consistent with the measured dust temperature from FUOri\_dust2 (which is not only heated by viscous heat generation but also radiation).
For a $\alpha_{t}=$0.01 disk, at $\sim$10 au radii, the gas temperature at the disk mid-plane is around 1000 K.
Under our assumptions, the detected lower ($\sim$400 K) temperature of FUOri\_dust1, which is a mass-weighted average from the inner $\sim$10 au region, appears to favor the $\alpha_{t}$ values that are well above 0.01.
Qualitatively, this can be understood since the mid-plane temperature is determined by the surface temperature and the optical depth in the vertical direction: the the higher optical depth, the higher the mid-plane temperature.
When the viscosity is stronger (i.e., $\alpha_{t}$ is larger), it requires a lower column density thus lower optical depth to achieve the same accretion rate, thereby yields the lower mid-plane temperature.

The origin of $\alpha_{t}$ is very uncertain.
It can be partly or largely due to MRI.
Although in this specific target source, the region which can thermally ionize the gas yet does not sublimate dust is not big.
However, in an environment that dust (in the mid-plane) can be sublimated, the UV opacity distribution is very uncertain and it may not be sufficient to ignore UV and X-ray photo-ionization.
We presently do not have the development to tackle this issue self-consistently.
Finally, in such a dynamically active region, there might (easily) be other sources of turbulence.

\bibliography{main}{}
\bibliographystyle{aasjournal}



\end{document}